\documentclass[11pt]{article}
\usepackage{rpmacros}
\usepackage[dvipsnames,svgnames,x11names,hyperref]{xcolor}
\usepackage{amsthm}
\usepackage{xfrac}
\usepackage{mathpazo}
\usepackage{gitinfo}
\usepackage{enumitem}
\usepackage{fullpage}
\usepackage{thmtools}
\usepackage{textgreek}
\usepackage[LGR,T1]{fontenc} 
\usepackage[colorlinks]{hyperref}
\hypersetup{
  linkcolor=[rgb]{0,0,0.4},
  citecolor=[rgb]{0, 0.4, 0},
  urlcolor=[rgb]{0.6, 0, 0}
}
\usepackage{lipsum}
\usepackage{mathrsfs,bbm}
\usepackage{bm}
\usepackage{todonotes}
\usepackage{tikz}
\usetikzlibrary{decorations.pathreplacing,backgrounds}
\usepackage{multirow}
\usepackage{setspace}
\usepackage{amsmath}

\usepackage[linesnumbered,vlined,ruled]{algorithm2e}
\newcommand{\algwidth}{\linewidth}

\usepackage{amsthm}
\usepackage{thmtools,thm-restate}
\usepackage[nameinlink,capitalise]{cleveref}

\numberwithin{equation}{section}
\declaretheoremstyle[bodyfont=\it,qed=\qedsymbol]{noproofstyle}

\declaretheorem[name=Observation,numberlike=equation]{observation}

\declaretheorem[name=Observation,numbered=no]{observation*}

\declaretheorem[numberlike=equation]{theorem}

\declaretheorem[name=Theorem,numbered=no]{theorem*}

\declaretheorem[numberlike=equation]{lemma}
\declaretheorem[name=Lemma,numbered=no]{lemma*}

\declaretheorem[numberlike=equation]{corollary}
\declaretheorem[name=Corollary,numbered=no]{corollary*}

\declaretheorem[name=Proposition,numbered=no]{proposition*}

\declaretheorem[numberlike=equation]{claim}
\declaretheorem[name=Claim,numbered=no]{claim*}

\declaretheorem[name=Conjecture,numbered=no]{conjecture*}

\declaretheorem[name=Question,numbered=no]{question*}

\declaretheoremstyle[bodyfont=\it,qed=$\lozenge$]{defstyle} 

\declaretheorem[numberlike=equation,style=defstyle]{definition}
\declaretheorem[unnumbered,name=Definition,style=defstyle]{definition*}

\declaretheorem[numberlike=equation,style=defstyle]{example}
\declaretheorem[unnumbered,name=Example,style=defstyle]{example*}

\declaretheorem[unnumbered,name=Notation=defstyle]{notation*}

\declaretheorem[unnumbered,name=Construction,style=defstyle]{construction*}

\declaretheorem[numberlike=equation,style=defstyle]{remark}
\declaretheorem[unnumbered,name=Remark,style=defstyle]{remark*}

\renewcommand{\phi}{\varphi}
\renewcommand{\epsilon}{\varepsilon}

\newcommand{\SP}{\Sigma\Pi}
\newcommand{\SPsize}[1]{(\SP)^k\operatorname{-size}}

\usepackage{nth}
\usepackage{intcalc}
\usepackage{etoolbox}
\usepackage{xstring}

\usepackage{ifpdf}
\ifpdf
\else
\usepackage[quadpoints=false]{hypdvips}
\fi

\newcommand{\ECCCurl}[2]{http://eccc.hpi-web.de/report/\ifnumcomp{#1}{>}{93}{19}{20}#1/#2/}

\newcommand{\shortECCC}[2]{\texttt{\href{http://eccc.hpi-web.de/report/\ifnumcomp{#1}{>}{93}{19}{20}#1/#2/}{eccc:TR#1-#2}}}

\newcommand{\parseECCC}[1]{
\StrSubstitute{#1}{TR}{}[\tmpstring]%
\IfSubStr{\tmpstring}{/}{ 
\StrBefore{\tmpstring}{/}[\ecccyear]%
\StrBehind{\tmpstring}{/}[\ecccreport]%
}{
\StrBefore{\tmpstring}{-}[\ecccyear]%
\StrBehind{\tmpstring}{-}[\ecccreport]%
}%
\shortECCC{\ecccyear}{\ecccreport}}

\newif\ifblind

\newif\ifdraft
\drafttrue
\ifdraft
\newcommand{\SKnote}[1]{\textcolor{BrickRed}{\guillemotleft SK: #1 \guillemotright}}
\newcommand{\MKnote}[1]{\textcolor{Green}{\guillemotleft MK: #1 \guillemotright}}
\newcommand{\HSnote}[1]{\textcolor{Blue}{\guillemotleft HS: #1 \guillemotright}}
\else
\newcommand{\SKnote}[1]{}
\newcommand{\MKnote}[1]{}
\newcommand{\HSnote}[1]{}
\fi

\newcommand{\RS}{\mathrm{RS}}
\newcommand{\1}{\mathbb{1}}
\newcommand{\supbrak}[2]{#1^{(#2)}}
\newcommand{\wt}{\mathrm{wt}}
\newcommand{\RM}{\mathrm{RM}}

\newcommand{\imult}{\mathbf{IMult}}
\newcommand{\OO}{{\mathcal{O}}}

\newcommand{\length}{{\mathrm{length}}}

\newcommand{\CAP}[1]{\mathrm{CAP}_{#1}}
\newcommand{\GAP}[1]{\mathrm{GAP}_{#1}}
\newcommand{\Simplex}[1]{\Lambda_{#1}}
\newcommand{\vecX}{\mathbf{X}}

\newcommand{\vecZ}{\mathbf{Z}}

\title{High Rate Multivariate Polynomial Evaluation Codes}

\ifblind
\else
\author{
	{Swastik Kopparty \thanks{Department of Mathematics and Department of Computer Science, University of Toronto, Toronto, Canada. Email: \texttt{swastik.kopparty@utoronto.ca}. Research supported by an NSERC grant.}}
	\and
    {Mrinal Kumar \thanks{Tata Institute of Fundamental Research, Mumbai, India. Email: \texttt{mrinal@tifr.res.in}. Research supported by the Department of Atomic Energy, Government of India, under project number RTI400112 and in part by a Google Research grant and SERB grants.}}
    \and
    {Harry Sha \thanks{Department of Computer Science, University of Toronto, Toronto, Canada. Email: \texttt{shaharry@cs.toronto.edu}.}}
}
\fi
\date{}

\begin{document}

\maketitle

\begin{abstract}
The classical Reed-Muller codes over a finite field
$\F_q$ are based on evaluations of $m$-variate polynomials
of degree at most $d$ over a product set $U^m$, for some $d < |U|$.
Because of their good distance properties, as well as the ubiquity
and expressive power of polynomials, these codes have played an influential role in coding theory and
complexity theory. This is especially so in the setting of $U$ being ${\F}_q$ where they possess deep locality properties. However, these Reed-Muller codes have a significant limitation in terms of the rate achievable --- the rate cannot be more than $\frac{1}{m{!}} = \exp(-m \log m)$.

In this work, we give the first constructions of multivariate polynomial evaluation codes which overcome the rate limitation -- concretely, we give explicit evaluation domains $S \subseteq \F_q^m$ on which evaluating $m$-variate polynomials of degree at most $d$ gives a good code. For $m= O(1)$, these new codes have relative distance $\Omega(1)$ and rate $1 - \epsilon$ for any $\epsilon > 0$. In fact, we give two quite different constructions, and for both we develop efficient decoding algorithms for these codes that can decode from half the minimum distance.

The first of these codes is based on evaluating multivariate polynomials on simplex-like sets. The distance of this code is proved via a generalized Schwartz-Zippel lemma on the probability of non-zeroness
when evaluating polynomials on sparser subsets of $U^m$ -- the final bound only depends on the ``shape'' of the set, and recovers the Schwartz-Zippel bound for the case of the full $U^m$, while still being $\Omega(1)$ for much sparser simplex-like subsets of $U^m$.

The second of these codes is more algebraic and, surprisingly (to us), has some strong locality properties. It is based on evaluating multivariate polynomials at the intersection points of hyperplanes in general position. 
It turns out that these evaluation points have many large subsets of collinear points. These subsets form the basis of a simple local characterization, and using some deeper algebraic tools generalizing ideas from Polischuk-Spielman~\cite{PS}, Raz-Safra~\cite{razSubconstantErrorprobabilityLowdegree1997} and Ben-Sasson-Sudan~\cite{ben-sassonRobustLocallyTestable2006}, we show that this gives a local test for these codes.
Interestingly, the set of evaluation points for these  locally testable multivariate polynomial evaluation codes can be as small as $O(d^m)$, and need not occupy a constant or even noticeable fraction of the full space $\F_q^m$.
\end{abstract}

\tableofcontents

\section{Introduction}
\label{sec:intro}

Polynomial evaluation codes, such as the Reed-Solomon and the Reed-Muller codes, have played a central and outsized role in the classical theory of error-correcting codes. In recent decades, they have also been key to fundamental advances in complexity theory, pseudorandomness, cryptography and extremal combinatorics, for example in
PCPs, interactive proofs, randomness extractors, pseudorandom generators and algebraic complexity theory. Furthermore, the pursuit of algorithms for and understanding of these codes has generated deep ideas in algebraic algorithms, property testing, and pseudorandom constructions.



A polynomial evaluation code is an error-correcting code obtained by evaluating $m$-variate polynomials of degree\footnote{Throughout this paper, degree refers to the total degree of a polynomial. Thus the degree of $X^3Y^7 + X^6Y^6$ equals $12$.} at most $d$ with coefficients in a finite field $\F$ at some subset $S$ of $\F^m$.
$$ \mathcal C =  \left\{ (P(\mathbf a) )_{\mathbf a \in S}  \mid P(X_1, \ldots, X_m) \in \F[X_1, \ldots, X_m], \deg(P) \leq d \right\} \subseteq \F_q^S.$$
The rate of this code (which is the ratio of the dimension of the code as a linear space to the length of the codewords) is easily calculated to be $R=\frac{ {d+m \choose m} }{|S|}$. The relative distance of this code (which is the largest fraction of coordinates on which any two codewords are guaranteed to be different) equals the largest $\delta$ such that any polynomial of degree at most $d$ has at least $\delta |S|$ non-zero evaluations on $S$, and is more complicated to determine. We are interested in families of codes, with their length $|S|$ going to infinity, and the rate $R$ and the relative distance $\delta$ are both $\Omega(1)$.

The most basic and well-known polynomial evaluation code is the $m = 1$ case: the Reed-Solomon code. Regardless of the choice of $S \subseteq \F$ of size $n$, this code has relative distance $\delta = 1-\frac{d}{n}$: since distinct polynomials of degree $d$ can agree on at most $d$ points. 
Thus, Reed-Solomon codes achieve a rate-distance tradeoff of $R+ \delta = 1$, which turns out to be optimal among all codes.

For the multivariate case, $m \geq 2$, the standard code is called the Reed-Muller code, and is based on choosing the evaluation domain $S$ to be a product set $U^m$, where\footnote{Reed-Muller codes also include the code of degree $d$ polynomials evaluated on $U^m$ with the degree $d \geq |U|$ (and individual degree $< |U|$), but these codes cannot achieve constant rate and relative distance.} $|U| > d$. In this case, the Schwartz-Zippel lemma provides the necessary bound on the minimum distance: it says that this code has relative distance at least $1- \frac{d}{|U|}$. This gives a worse rate-distance tradeoff of
$$R  = \frac{(1-\delta)^m}{m!}.$$
For constant $m$, this gives constant rate and constant relative distance, albeit with $R < 1/2$ unless $m = 1$.

Reed-Muller codes with $U$ equal to the whole finite field $\F$ are especially interesting because of their amazing locality properties.
Indeed, $\F^m$ contains within it $m-1$ and lower dimensional spaces, and $m$-variate polynomials of degree $d$ stay degree $d$ when restricted to these lower dimensional spaces. This is the basis for a natural local characterization of codewords -- and leads to the extremely useful local testability and local decodability of these codes.


This rate limitation of multivariate polynomial codes is what our paper addresses -- we find evaluation domains for multivariate polynomials of total degree $d$ so that the resulting code has rate close to $1$ while still having constant relative distance. This is a derandomization question -- a random such evaluation domain of the right size has this property with high probability.

We give two kinds of explicit such evaluation domains, along with polynomial time decoding algorithms for the corresponding codes. 
The first code is more combinatorial, and its analysis and decoding involves a generalized Schwartz-Zippel lemma bounding the probability that a polynomial evaluates to zero on a general ``shape''. The second kind is more geometric, based on intersections of hyperplanes in general position, and its analysis and decoding comes from the presence of the many intersecting lines contained within the evaluation domain. The rich geometric structure that comes from these lines, to our surprise, leads to the {\em local testability} of these codes, something that we did not expect was possible from such a drastic sparsification of the Reed-Muller code.

\subsection{Our results}

We give explicit multivariate polynomial evaluation codes with rate close to $1$ and constant relative distance for the setting of $m = O(1)$.\\

\noindent {\bf Theorem A (Informal):} {\em For $m = O(1)$ and any $\epsilon > 0$, there are explicit polynomial evaluation codes of $m$-variate polynomials that have rate $R \geq 1 - \epsilon$ and relative distance $\Omega(1)$ (see \Cref{thm:code-construction-simplexes,thm:code-construction-geometric}).}\\

We give three proofs of this theorem. 

The fastest (but least informative) proof is obtained via a general result, proved using an application of the polynomial method, showing that {\em interpolating sets} for $m$-variate polynomials of degree $(1 + \Omega(1))d$ are themselves automatically good evaluation domains for polynomials of degree $d$;  any polynomial of degree $d$ has many non-zero evaluations on such a set. This construction gives the tradeoff:
\begin{align}
\label{eq:Rvsdelta}
R^{1/m} + \delta^{1/m} = 1,
\end{align}
which for constant $m$ gives us the statement of Theorem A.

Instantiating this with two known constructions of interpolating sets from the literature\footnote{See~\cite{interpolation-survey} for a detailed history of interpolating sets for multivariate polynomials.} (and elegantly presented in the paper of Bl\"aser and Pandey~\cite{BP20}), we get the two high rate multivariate  polynomial evaluation codes which this paper is about. We give a small taste of these codes now and in a later subsection we describe them in full generality.

\begin{itemize}
 \item {\bf CAP (Combinatorial Arrays for Polynomials) codes:} These codes are obtained by evaluating multivariate polynomials on simplex-like sets.
 
 A simple example of such a set for the case $m = 3$ is given below.
 Assume $\F$ has characteristic at least $1.1 d$.
 Then taking $S \subseteq \F^3$ to be the ``simplex'' given by:
 $$ S = \{ (i, j, k ) : 0 \leq  i,j,k \leq 1.1d,  i+j+k \leq 1.1d \},$$
 the code obtained by evaluating $3$-variate polynomials of degree at most $d$ on $S$ turns out to have rate $R = \left(\frac{1}{1.1}\right)^3$ and relative distance $\delta = \left(\frac{0.1}{1.1}\right)^3$.
 
 This example seems somewhat related to the Reed-Muller code with evaluation domain $[1.1d]^3$. Indeed it is very closely related, and our most general CAP codes come from a unified viewpoint which captures both of these and many more kinds of evaluation domains.
 
 \item {\bf GAP (Geometric Arrays for Polynomials) codes:} These codes are obtained by evaluating multivariate polynomials on intersection points of hyperplanes in general position.
 
 A simple example of such a set for the case $m = 3$ is given below.
 Let $U \subseteq \F$ with $|U| = 1.1 d$.
 Then we take the set $S \subseteq \F^3$ by evaluating the elementary symmetric polynomials in $3$ variables on distinct $3$-tuples of $U$:
 $$ S = \{ (a+b+c, ab+bc+ca, abc ) :   a,b,c \in U, \mbox{ distinct } \},$$
 The code obtained by evaluating $3$-variate polynomials of degree at most $d$ on $S$ turns out to have rate $R = \left(\frac{1}{1.1}\right)^3$ and relative distance $\delta = \left(\frac{0.1}{1.1}\right)^3$.
 
 It may not be clear why the example above has anything to do with hyperplanes in general position, but it does, and it in fact contains many large collinear subsets of points. This leads to the (very surprising to us) local testability of these codes (Theorem C below).
  
\end{itemize}

Our second and third proofs of Theorem A are by a direct analysis 
of the distance of CAP and GAP codes. This yields much more information, and in particular,
allows us to prove our next main result: that there are efficient unique decoding algorithms for both CAP codes and GAP codes up to half their minimum distance. \\

\noindent {\bf Theorem B (Informal):} {\em There are polynomial time algorithms for unique decoding CAP codes and GAP codes from a fraction of errors that equals half their minimum distance (see \Cref{thm:decode-mvariate-simplex,thm:algo-mD-hperplane-decoder-analysis}).
}\\

A unique decoding algorithm that decodes a code unto half its minimum distance is effectively providing an ``algorithmic proof'' of the minimum distance of the code. To prove Theorem B, we develop alternate and deeper understandings of these codes, and then identify and algorithmize the core combinatorial and geometric phenomena that give CAP codes and GAP codes their distance.

\subsubsection{CAP codes}
\label{sec:capintro}

The CAP code described in the example above is based on evaluating polynomials at a small subset of the product set $[1.1d]^3$, thus improving upon the standard Reed-Muller codes, which require the whole product set.

In \autoref{sec:zpp}, we develop a second, more transparent, proof that CAP codes have distance, based on a new generalization of the Schwartz-Zippel Lemma. 
This generalization comes from an observation that the standard proof of the Schwartz-Zippel lemma not only bounds the number of zeros in the grid, also gives information about the shape in which the zeros may appear. 
This information is quite powerful and underlies the ability of CAP codes to achieve rate close to $1$.

To state this generalized Schwartz-Zippel Lemma, we introduce some notation to specify ``shapes''.
Let $t$ be a natural number.  Let $J$ be a subset of $[t]^m$ which is downward closed under the coordinatewise order $\preceq$: namely, if $\mathbf j, \mathbf{j'} \in [t]^m$ with $\mathbf{j'} \preceq \mathbf j$ (which means we have coordinatewise inequality), then $\mathbf{j} \in J$ implies $\mathbf{j'} \in J$.


We say that a downward-closed set $J \subseteq [t]^m$ has $d$-robustness at least $B$ if, whenever we delete $d$ axis parallel $(m-1)$-dimensional hyperplanes from $J$, at least $B$ points of $J$ remain.

We can now state our generalized Schwartz-Zippel lemma.\\

\noindent{\bf Generalized Schwartz-Zippel Lemma}\ \ {\em
 Let $J$ be a subset of $[t]^m$ with $d$-robustness at least $B$.
 
 \noindent Let $U \subseteq \F$ with $|U| = t$, and let $U = \{u_1, \ldots, u_t\}$.
 Define $U[J]$ to be the set $S \subseteq U^m$ given by:
 $$ S = \{ (u_{j_1}, u_{j_2}, \ldots, u_{j_m}) \mid \mathbf j \in J \}.$$
 
 \noindent Let $P(X_1, \ldots, X_m) \in \F[X_1, \ldots, X_m]$ be a nonzero polynomial
 of degree at most $d$.
 Then:
 $$ | \{ \mathbf a \in S \mid P(\mathbf a) \neq 0 \} | \geq B.$$
}\\

Thus if we find a subset $J$ with size $n$ and $d$-robustness $\geq \delta n$, the polynomial evaluation code with evaluation set $U[J]$ will have relative distance $\delta$ (and rate $\frac{{d+m. \choose m}}{n}$).

When we take $J = [t]^m$, then we recover the standard Schwartz-Zippel lemma.
When we take $J = \{ \mathbf j \in [t]^m \mid \sum_{i} j_i \leq t \}$, we recover the bound for simplex-like CAP codes described earlier (but in much greater generality -- since we do not need the regular-spacedness of the set $U$).

Other $J$ can be taken, and they can give other rate vs distance tradeoffs. For given constants $R, \delta \in [0,1]$ and $m = O(1)$, the question of whether there is a downward-closed $J \subseteq [t]^m$ (for some large $t$) of a given size $n = (1+o(1)) \cdot \frac{1}{R} \cdot \frac{d^m}{m!}$ with $d$-robustness $\delta n$ is an interesting combinatorial question. Interestingly, there are choices of $J$ which go beyond (even in the asymptotic setting) both the full product set and the simplex constructions, even though both are ``locally optimal''.

The proof of the Generalized Schwartz-Zippel Lemma is by induction on a stronger statement -- this statement not only bounds the size of, but also gives structural information about, the ``shape'' of the zero set of any multivariate polynomial. This refined information also turns out to be useful for our decoding algorithms, which we discuss next.\footnote{It was pointed to us by S Venkitesh \cite{V24} that this generalized Schwartz-Zippel lemma can also be shown as a consequence of a classic theorem of Macaulay \cite{Mac27} using a slight adaptation of the techniques in \cite{STV20}. }

\paragraph{Decoding:} Based on the ideas above, we show that CAP codes can be decoded in polynomial time from half their minimum distance. 
We do this by developing a suitable generalization of a peeling-based algorithm of Kim and Kopparty \cite{KimK2017}    for decoding Reed-Muller codes on a product set. 
Suppose we are trying to decode the bivariate Reed-Muller code over a grid $U^2$. 
At a high level, \cite{KimK2017} write $f(X, Y) = \sum_{i=0}^df_i(X)Y^i$. 
They then decode the univariate polynomials $f(x, Y)$ on each column. 
From these decodings, they have estimates of $f_d(x)$ for each $x \in U$ and use these to decode for $f_d$ using the GMD algorithm -- the classical algorithm for decoding concatenated codes that
uses ``soft'' information. Peeling off the polynomial $f_d(X)Y^d$ from $f(X,Y)$, and iteratively applying this argument again gives the final algorithm. 
The main challenge for adapting this algorithm to CAP codes is that each column has a variable length (and thus, each univariate polynomial $f(x, Y)$ is evaluated on a different number of points). 
We address this by introducing a variant of the GMD algorithm that allows for varying inner code lengths and inner code distances. This uses a recent new combinatorial analysis of the GMD algorithm by~\cite{BHKS-2023}.

\subsubsection{GAP codes}

GAP codes are based on intersections of hyperplanes in general position.
Suppose we take a collection $\mathcal H$ of $t$ such hyperplanes in $\F^m$.
Let $\mathcal{S}$ be the set of all $m$-wise intersections of hyperplanes from $\mathcal H$. This set $\mathcal{S}$, of size ${t \choose m}$, is the set of evaluation points for GAP codes\footnote{The example GAP code presented above was just an instantiation of this framework with the ``moment-curve construction'' of hyperplanes in general position.}.\\

Combinatorially these sets look quite complex, but there is rich structure sitting inside them. Indeed, if we take the intersection of some $(m-1)$ hyperplanes from $\mathcal H$, we get a line $L$ in $\F^m$. It is easy to see that this line contains $t-(m-1)$ points of $\mathcal{S}$. If $t- (m-1)$ is at least $(1 + \Omega(1))$ times $d$, then polynomials of degree $d$ evaluated on $\mathcal{S}$ will look like Reed-Solomon codes of constant relative distance when restricted $\mathcal{S} \cap L$.

\paragraph{Decoding:} We can use these observations to get a decoding algorithm for GAP codes up to half the minimum distance.
The algorithm is based on decoding along each line $\mathcal{S} \cap L$ to the nearest Reed-Solomon codeword. How do we stitch together all these Reed-Solomon codewords? Miraculously, viewing things the right way, the $m$-variate GAP code turns out to be a {\em concatenated code}, where the outer codes are Reed-Solomon codes over the function field $\F(X_1, \ldots, X_{m-1})$, and the inner codes are 
$(m-1)$-variate GAP codes. From this point of view, the way to stitch together all these Reed-Solomon codewords is to simply use the classical GMD decoder of Forney for concatenated codes!

Compared to the big struggle that happened within~\cite{KimK2017} for decoding Reed-Muller codes, which involved slowly correcting errors in tiny steps via a  back and forth between the inner Reed-Solomon decoder and the GMD decoder,
and a refined version of that struggle that we just did for decoding CAP codes, this clean decoder for GAP codes is a source of great joy (and relief).

\paragraph{Local Testing:} These local Reed-Solomon codes present in GAP codes also form the basis for a local characterization -- we show in~\Cref{thm:local-characterization-mD} that a function defined on $S$ is a codeword of the GAP code if and only if its restriction to any line $L$ gives a codeword of the Reed-Solomon code of degree $d$ univariate polynomials.


Using (and generalizing) the ideas of Polischuk-Spielman~\cite{PS} and Raz-Safra~\cite{razSubconstantErrorprobabilityLowdegree1997}, we upgrade this local characterization to a {\em robust} local characterization (involving planes), and thereby show local testability of these codes.\\

\noindent {\bf Theorem C (Informal):} {\em 
    For $m = O(1)$, GAP codes of length $n$ with constant relative distance are locally testable with $O(n^{2/m})$ queries. In particular, for any $\epsilon > 0$, there exist GAP codes of length $n$ and rate $(1-\epsilon)$ that are locally testable with $O(n^{2/m})$ queries (see \Cref{thm:local-test-main-theorem-plane-point}).
}\\


The above result is proved using a version of the plane-vs-point test for standard Reed-Muller codes of Raz-Safra~\cite{razSubconstantErrorprobabilityLowdegree1997} (see also its abstraction to tensor codes by Ben-Sasson-Sudan~\cite{ben-sassonRobustLocallyTestable2006} and Viderman~\cite{V15}). The rich geometric structure within GAP codes turns out to be sufficient to implement this proof strategy, despite the drastically smaller number of planes available.

In fact, the proof turns out to be substantially {\em easier} than both the proofs of Raz-Safra and Ben-Sasson-Sudan. The underlying reason is the following fact (\autoref{lem:local-consistency-to-global-consistency}) about gluing polynomials defined on hyperplanes {\em in general position}: if we are given $(m-1)$-dimensional hyperplanes $H_1, \ldots, H_\ell$ in  $\F^m$ in general position for some $\ell \geq d$, and degree $d$ polynomials $P_i: H_i \to \F$ defined on them which are mutually consistent on the intersections $H_i \cap H_j$, then
there is a unique degree $d$ polynomial $P(X_1, \ldots, X_m)$ consistent with all of them; namely $P|_{H_i} = P_i$ for all $i$. Without the general position assumption, this no longer holds (for example, because of parallel hyperplanes), and this complicates the analysis of the hyperplane consistency graph in~\cite{razSubconstantErrorprobabilityLowdegree1997} and~\cite{ben-sassonRobustLocallyTestable2006}. We think this explains why GAP codes admit a much simpler analysis, and also why the analysis directly works in the low-distance/high-rate regime\footnote{For the Ben-Sasson-Sudan setting of tensor codes, the original proof only worked when the base code had large distance. Testability when the base code has low distance was only later discovered by Viderman~\cite{V15} using some further clever ideas.}.


We also analyze the natural line-vs-point test for GAP codes, but this ends up being quantitatively weaker -- it only gives nontrivial testability when the rate is at most $ (1-\epsilon) 2^{-m}$ (which is still higher than the rate achievable by the classical Reed-Muller code). Thus, unlike for the plane-vs-point test, our analysis for the line-vs-point test does not work all the way up to rate approaching 1. 

Along the way, we prove an independently interesting statement (\autoref{lem:ps-style-lemma-md}) about a local-to-global phenomenon for divisibility. If we have two polynomials $A(X_1, \ldots, X_m)$ and $B(X_1, \ldots, X_m)$ of degree at most $d$ such that  for at least $3d$ hyperplanes $H$ in general position, the $(m-1)$-variate polynomial 
$A|_H$  divides the $(m-1)$-variate polynomial $B|_H$, then
in fact $A$ divides $B$. Such statements with a $O(d^2)$ bound on the number of hyperplanes needed were implicitly proved in a work of Forbes~\cite{Forbes} and a work of Harsha, Kumar, Saptharishi and Sudan~\cite{HKSS23}. Our work is the first to get an $O(d)$ bound for this setting. For $m=2$ our proof is based on an adaptation of an argument of Polischuk-Spielman, using Bezout's theorem and a basic claim about intersection multiplicities of two curves and with a common tangent. For larger $m$ it involves much more nontrivial algebraic machinery involving intersection multiplicities of varieties, but the final statement is still quite clean.

Finally, we remark that GAP codes can be viewed as Tanner codes (or HDX-Tanner codes) on the complete graph (for $m=2$) or the complete $m$-dimensional complex (for larger $m$) on $t$ vertices, where the inner codes are Reed-Solomon codes. That these codes have high rate is a surprise; it goes beyond what pure dimension counting implies. That these codes are in fact multivariate polynomial evaluation codes is another surprise. This viewpoint is related to the recent constructions of Dinur, Liu and Zhang~\cite{DLZ}.

\subsubsection{Comparison to Reed-Muller codes}

How do CAP codes and GAP codes compare to Reed-Muller codes?

At a very high level, CAP codes take the principles underlying Reed-Muller codes and squeeze everything that could possibly be squeezed out of it. The proof of distance of CAP codes generalizes the Schwartz-Zippel lemma, and our decoding algorithms for CAP codes generalize the decoding algorithms for Reed-Muller codes from~\cite{KimK2017}. We feel that our generalized Schwartz-Zippel lemma as well as our decoding algorithm for CAP codes sheds light on the true content of the Schwartz-Zippel lemma and the reason for the efficient decodability of Reed-Muller codes, while simultaneously highlighting the route to modifying Reed-Muller codes to achieve higher rate.

On the other hand, GAP codes look nothing like Reed-Muller codes, and in many ways they are much cleaner and more natural than Reed-Muller codes. The proof of distance is much simpler, and in fact the decoding algorithm just follows directly from viewing it as a concatenated code in the right way. Even more interestingly, GAP codes are locally testable (in part because of the many lower dimension subspaces present in the evaluation domain), while Reed-Muller codes are not unless the evaluation domain is all of $\F_q^m$. 

Reed-Muller codes on general product sets $U^m$ also have many local constraints, with $m$ axis-parallel lines through each evaluation point.
The local constraint says that the restriction to these lines should be degree at most $d$.
But the key difference with GAP codes is that these local constraints do not characterize Reed-Muller codes -- they only characterize evaluation codes of multivariate polynomials of {\em individual degree} at most $d$ -- and this is a significantly bigger code than the Reed-Muller code which has a total degree condition.

Finally we remark that since the evaluation domains for CAP and GAP codes do not essentially depend on the degree $d$, CAP and GAP codes automatically are {\em multiplication codes}~\cite{meir13} -- the pointwise product of codewords is a codeword of another good code -- one of the key properties of Reed-Muller codes which is important for applications in complexity theory. 

\subsection{Other Related Work}

There have been many influential works studying the problem of finding small evaluation sets for $m$-variate degree-$d$ polynomials which preserve their good distance. Our results are focused on $m$ constant and $d$ big (where we get codes rate nearly $1$), while the existing literature addresses large $m$. The known results give good-distance polynomial evaluation codes with inverse polynomial rate
when $m$ and $d$ are both large, and rate nearly $1$ when $d$ is constant and $m$ is large.

Recall the setup. Let $k = {m + d \choose d}$, which is the dimension of the space of $m$-variate polynomials of degree at most $d$. We are seeking evaluation domains where such polynomials still have constant fraction distance on this domain, with the size of the evaluation domain being as small a function of $k$ as possible. 

There is a long line of work (Chen-Kao~\cite{CK97}, Lewin-Vadhan~\cite{LV98}, Agrawal-Biswas~\cite{AB98}, Klivans-Spielman~\cite{KS02}) studying pseudorandom generators against low degree polynomials over large fields using few random bits. The fundamental (and very general) result of Klivans and Spielman on fooling sparse polynomials from this line of work yields explicit evaluation sets
for $m$-variate degree-$d$ polynomials of size ${m + d \choose d }^{O(1)} = k^{O(1)}$. When $m = d$ for example, this gives a multivariate polynomial evaluation code with block-length $k^{O(1)}$, while the evaluation domains coming from the Schwartz-Zippel lemma would have to have size $d^m = k^{\Omega(\log k)}$. These results require the field size to be very large (at least ${m + d \choose d }$).

Another line of work exploiting deeper algebraic geometric properties of polynomials over large fields gives interesting evaluation sets over fields of moderately large size. These works by Bogdanov~\cite{bogdanov05}, Derksen-Viola~\cite{DV} and Dwivedi-Guo-Volk~\cite{DGV} lead to explicit evaluation sets of size $k^{O(1)}$ when the field size is $\poly(d,n)$
(this can also work over slightly smaller fields, giving slightly larger explicit evaluation sets).

In the other extreme setting of parameters, when $d = O(1)$, $m$ is large and the field is small (e.g. $\F_2$), there is another approach to this problem\footnote{To be precise, here we look for evaluation domains in $\F_q^m$ for polynomials of total degree $d$ and {\em individual degree} at most $q-1$: the individual degree constraint is natural because points in $\F_q^m$ cannot witness the nonzeroness of the polynomials $X_i^q - X_i$. For this setting, $k$ is taken to be the dimension of this space of polynomials, and may be smaller than ${m + d \choose d}$.}  using sumsets of epsilon-biased sets. Here the result of Dvir-Shiplka~\cite{DS11} and Viola~\cite{viola08} (using a key construction of Bogdanov-Viola~\cite{BV07}, see also Lovett~\cite{Lovett08}) give explicit evaluation sets of size $(1+ \epsilon) k$. There is also an elegant decoding algorithm for these codes, given by Dvir and Shpilka~\cite{DS11}.

We also mention a classical coding theory approach to constructing evaluation sets for the setting $m =2$, based on Algebraic Geometry codes. If we take the $\F$-points of an irreducible degree $d+1$ curve in the plane, the number of points where a bivariate polynomial $Q(X_1, X_2)$ of degree at most $d$ can vanish on $C$ is at most $d \cdot (d+1)$ (by Bezout's theorem). Thus,
if we take $n = (1+ \Omega(1))d^2$ points on $C$ as our evaluation domain, 
we guarantee that polynomials of degree $d$ have positive fraction distance on this domain. Note, however, that this argument cannot achieve rates $\geq 1/2$, and is thus comparable to the Schwartz-Zippel approach for $m= 2$.

A very interesting recent work by Dinur, Liu and Zhang~\cite{DLZ} gave new constructions of constant query locally testable codes  of subconstant rate using the explicit description of high dimensional expanders using finite field geometry, and with Reed-Solomon ``inner codes'' in a Tanner code like construction. Our GAP codes can be viewed through a similar lens, but with the complete $m$-dimensional complex in place of the HDX.

Another interesting recent line of work is by Bafna, Srinivasan and Sudan~\cite{BSS} and Amireddy, Srinivasan and Sudan~\cite{ASS},  who studied the local properties of multivariate polynomial codes evaluated on product sets. These gave the first examples of nontrivial locality for multivariate polynomial evaluation codes which are not evaluated on the full $\F^m$. Our GAP codes give another such example (but only for local testability).

We close by mentioning two nice open problems.
\begin{enumerate}
\item Are there explicit polynomial evaluation codes of rate and relative distance both $\Omega(1)$ when $m$ and $d$ are both large?
 \item By a small variant of the argument of Saraf-Yekhanin~\cite{saraf-yekhanin}, a random choice of evaluation points in $\F^m$ for the space of polynomials of degree at most $d$ will achieve near-optimal tradeoff of $R + \delta = 1 - o(1)$, provided $|\F| > \omega(d)$. Thus, there are better multivariate polynomial evaluation codes out there, even for $m = 2$. We think finding an explicit one of them for which there is efficient decoding (and possibly locally decoding and locally testing) is a very interesting problem for future work.

\end{enumerate}

\paragraph*{Organization: }
The rest of this paper is organized as follows. We set up the necessary notation and prelims in \autoref{sec:prelims}. We describe our construction of high rate codes in \autoref{sec:highrate-construction}, followed by a more detailed discussion on the ideas underlying the construction and properties of CAP codes in \autoref{sec:zpp}. We describe the decoding algorithms for CAP codes in \autoref{sec:cap-decoding} and GAP codes in \autoref{sec:decoding-hyperplane-intersection-codes}. We prove the local testability of GAP codes in \autoref{sec:testing}.


 

\section{Notation and preliminaries}\label{sec:prelims}
\subsection*{Notation}
Throughout the paper, we use $\F$ to denote a field. We use $X, Y, Z$ etc in capitals to denote formal variables, and $x, y, z$ etc in small to denote field elements in $\F$. We use bold for vectors and subscripts to denote the entries. For example, $\mathbf{x} \in \N^m$, and the entries of $\mathbf{x}$ are $x_1,...,x_m$. Let $\N_{<k} = \{0,1,...,k-1\}$. The arity of bold letters will be clear from the context. For a polynomial $f \in \F[X_1,...,X_m]$,  $Z(f), N(f)$ be the set of zeros and non-zeros of $f$ in $\F$. 

For any alphabet $\Sigma$, natural number $n \in \N$, and strings $\veca, \vecb \in \Sigma^n$, $\Delta(\veca, \vecb)$ denotes the Hamming distance between $\veca$ and $\vecb$, i.e. the number of coordinates on which $\veca$ and $\vecb$ disagree with each other, and $\delta(\veca, \vecb)$ denotes their fractional Hamming distance, i.e. $\delta(\veca, \vecb) = \Delta(\veca, \vecb)/n$.


\subsection*{Error-correcting codes}

The following are standard definitions in coding theory. A set $C \subset \Sigma^n$ is called an error-correcting code of distance at least $d$ if for any $c, c' \in C$, $\Delta(c, c') \geq d$. The relative distance, $d / n$, typically denoted as $\delta(C)$ or simply $\delta$ where $C$ is clear from the context. The dimension of the code is $\log_{|\Sigma|}(C)$. The rate of the code, typically $R$, is equal to $\log_{|\Sigma|}(C) / N$.

\subsection*{Codes based on polynomial evaluation}

This paper is focused on error-correcting codes obtained from polynomial evaluations. The message space is the space of $m$-variate polynomials of total degree at most $d$, and a polynomial, $f$, is encoded by recording the evaluations of $f$ on a subset $S \subset \F^m$. We think of the list of evaluations $(f(\mathbf{x}))_{\mathbf{x} \in S}$ as a function from $S$ to $\F$. More formally, we have the following definition. 

\begin{definition}[Polynomial Evaluation Code]\label{def:polynomial-eval-code}
    The $m$-variate Polynomial Evaluation Code of degree $d$ on a set $S$, denote $E_{m, d, S}$ is the set of evaluations of (total) degree $d$, $m$-variate polynomials on the set $S$. That is 
    $$
    E_{m, d, S} = \left\{ f: S \to \F : f \in \F[X_1,...,X_m], \deg(f) \leq d \right\}
    $$
\end{definition}

Important examples of polynomial evaluation codes include Reed-Solomon codes, where $m = 1$, and Reed-Muller Codes, where $S$ is a $m$-dimensional grid. 

\begin{definition}[Reed-Solomon Code]\label{def:RS}
    For any $d, n \in \N$ with $d<n$, let $\RS_{d, n} = E_{1, d, U}$ where $U$ is some subset of $\F$ of size $n$.
\end{definition}

\begin{definition}[Reed-Muller Code]\label{def:RM}
    For any $d, n, m \in \N$ with $d<n$, $m \geq 1$,  let $\RM(m, d, n) = E_{m, d, U^m}$, where $U$ is some subset of $\F$ of size $n$.
\end{definition}

Let $C$ be a polynomial evaluation code where the evaluation set is $S$. 
Define an error and erasure pattern to be a function $e: S \to  (\F \cup \{?\})^N$. 
When added to a codeword, the symbol $?$ corresponds to erasures, and non-zero entries of $e$ correspond to errors. 
Define the weight of an error and erasure pattern to be $\wt(e) = 2|\{x \in S: e(x) \neq 0, e(x) \neq ?\}| + |\{x \in S: e(x) = ?\}|$;
i.e., the weight of an error and erasure pattern is twice the number of errors plus the number of erasures. 
We call $e$ an error pattern when there are no erasures. 

Let $A$ be an algorithm and $C$ be a code. 
We say that $A$ decodes $C$ from error (and erasure) patterns of weight at most $D$ if for any error (and erasure) pattern $e$ of weight at most $D$, and any codeword $c \in C$, $A$ returns $c$ on input $c + e$.


\subsection*{Hitting sets and interpolating sets}

\begin{definition}[Hitting sets]\label{def:hitting-sets}
    A subset $S \subseteq \F^m$ is called a hitting set for $m$-variate polynomials of degree $d$ if every non-zero polynomial $f$ of degree at most $d$ in $\F[X_1,...,X_m]$, there is some point $\mathbf{x} \in S$ such that $f(\mathbf{x}) \neq 0$.
\end{definition}

\begin{definition}[Interpolating sets]\label{def:interpolating-sets}
    A set $S \subseteq F^m$ is an interpolating set for $m$-variate polynomials of degree $d$ if for all functions $f : S\to \F$, there is a unique polynomial, $g \in \F[X_1,...,X_m]$ of degree at most $d$ such that for all $a \in S, f(a) = g(a)$. 
\end{definition}

\subsection*{Concatenated Codes and GMD decoding}  
A key step in our decoding algorithms is to express CAP and CAP codes as concatenated codes. 
This allows us to leverage known algorithms for decoding concatenated codes such GMD algorithm of Forney \cite{forneyGeneralizedMinimumDistance1966}.

\begin{definition}\label{defn:concatenated-codes}
Let $C_{out} : \Sigma_1^K \to \Sigma_2^N$ be a code with block length $N$ over an alphabet $\Sigma_2$ of size $Q$. Let $\mathcal{C}_{in} = (C_{1}, C_{2}, \ldots, C_{N})$ be an $N$-tuple of codes where for each $i$, the code $C_{i}$ is of the form $C_{i}:\Sigma_2 \to \Sigma_3^n$, i.e. the message space is identified with the alphabet $\Sigma_2$ of $C_{out}$ and $C_{i}$ maps each such alphabet to a vector of length $n$ over a new alphabet $\Sigma_3$. The concatenation of $C_{out}$ with $\mathcal{C}_{in}$, denoted as $\mathcal{C}_{in} \circ C_{out}$ is a code that maps messages in the space $\Sigma_1^K$ to vectors of length $(Nn)$ over the alphabet $\Sigma_3$, and the encoding maps a message $\vecm \in \Sigma_1^K$ to the codeword 
\[
\mathcal{C}_{in} \circ C_{out}(\vecm) = \left(C_{1}(C_{out}(\vecm)_1), C_{2}(C_{out}(\vecm)_2), \ldots, C_{N}(C_{out}(\vecm)_N) \right) ,
\]
where, 
\[
C_{out}(\vecm) = \left(C_{out}(\vecm)_1, C_{out}(\vecm)_2, \ldots, C_{out}(\vecm)_N\right) \in \Sigma_2^N .
\] 
\end{definition}
We now state a result of Forney for decoding concatenated codes up to half their minimum distance that will be crucially used in our proofs. 
\begin{theorem}[Forney \cite{forneyGeneralizedMinimumDistance1966}]\label{thm:GMD-decoding}
Let $C_{out}, \mathcal{C}_{in} = (C_{1}, C_{2}, \ldots, C_{N})$ be codes  and $\mathcal{C}_{in} \circ C_{out}$ be their concatenation as in \autoref{defn:concatenated-codes}. 
Suppose each inner code $C_i$ has distance $d$, and $C_{out}$ has distance $D$.
Suppose there exists algorithms $A_1,...,A_N, A_{out}$ such that 
\begin{itemize}
\item For every $i$, $A_{i}$ decodes $C_{i}$ from error patterns of weight less than $d$.
\item $A_{out}$ decodes $C_{out}$ from errors and erasures patterns of weight less than $D$
\end{itemize}
Then, there is a deterministic algorithm for (unique) decoding of the concatenated code $\mathcal{C}_{in} \circ C_{out}$ from error patterns of weight less than $dD$. 
Furthermore, the algorithm makes at most one function call to each $A_{i}$, at most $O(N)$ function calls to the algorithm $A_{out}$, and performs at most $\poly(N, n, \log (|\Sigma_1|\cdot |\Sigma_2|\cdot |\Sigma_3|))$ additional computation. 
\end{theorem}

\section{Explicit High Rate Polynomial Evaluation Codes}\label{sec:highrate-construction}
The following technical lemma gives a general connection between robust hitting sets for polynomials of degree $d$ and hitting sets for polynomials of slightly higher degree. 
The lemma will give us a clean and direct way of analyzing our constructions of high-rate polynomial evaluation codes. 

\begin{lemma}\label{lem:hitting-sets-to-robust-interpolating-sets}
Let $m, d, D$ be natural numbers with $D > d$ and $\F$ be any field. Let $S$ be a subset of $\F^m$ that is a hitting set for polynomials of degree $D$. Then, for any non-zero $m$-variate polynomial $f$ of degree at most $d$, the number of non-zeroes of $f$ in the set $S$ is at least $\binom{D-d+m}{m}$.
\end{lemma}
\begin{proof}
The proof of the lemma is via an instantiation of the polynomial method. 

Let $f$ be an arbitrary non-zero polynomial on $m$ variables and degree $d$ and $T_f \subseteq S$ denote the set of non-zeroes of $f$ in the set $S$. Suppose $|T_f| < \binom{D-d+m}{m}$.
The next general interpolation claim exploits this assumption.
\begin{claim}\label{clm:poly-vanishing-on-non-zeroes}
If $|T_f| < \binom{D-d + m}{m}$, then there is a non-zero polynomial $g$ on $m$ variables and degree at most $(D-d)$ that vanishes everywhere on $T_f$. 
\end{claim}
\begin{proof}[Proof of \autoref{clm:poly-vanishing-on-non-zeroes}]
For the proof of the claim, we think of the coefficients of $g$ as formal variables and consider the system of homogeneous linear equations obtained by imposing the constraints that $g$ vanishes at every point in $T_f$. The number of variables in this linear system equals the number of monomials of degree at most $(D-d)$ on $m$ variables, i.e., $\binom{D-d+m}{m}$ and the number of homogeneous linear constraints equals the size of $T_f$, which by our assumption is strictly less than $\binom{D-d+m}{m}$. Thus, the number of variables in this homogeneous linear system is strictly more than the number of constraints imposed. Hence, the system has a non-zero solution.
\end{proof}

With the $g$ given by the above claim in hand, consider the polynomial $h = f\times g$. Clearly, the degree of $h$ is at most $D$, and $h$ is non-zero since $g$ and $f$ are non-zero. Moreover, since $f$ vanishes on $S\setminus T_f$ (by definition of $T_f$) and $g$ vanishes on $T_f$ (by construction of $g$), we get that $h$ vanishes on all points in the set $S$, which contradicts the fact that $S$ is a hitting set for $m$-variate polynomials of degree at most $D$. Thus, the size of the set $T_f$ must be at least $\binom{D-d+m}{m}$. 
\end{proof}
We now discuss the construction of two sets of evaluation points that give us polynomial evaluation codes of high rate and constant relative distance  (when the number of variables is constant). From \autoref{lem:hitting-sets-to-robust-interpolating-sets}, we get that one way of doing this is to construct hitting sets of small size. Two such constructions of hitting sets were given by Bl{\"{a}}ser \& Pandey \cite{BP20}. In fact, their hitting sets are optimal, in the sense that their hitting set for $m$-variate degree $D$ polynomials has size \emph{equal} to $\binom{m+D}{D}$ (in other words, these hitting sets are interpolating sets). In the next section, we discuss these constructions and invoke them with \autoref{lem:hitting-sets-to-robust-interpolating-sets} to get our codes. 

\subsection{A Geometric Construction}
To describe this construction, we rely on the notion of hyperplanes in general position that we now define. Recall that a hyperplane $H$ in $\F^m$ is just an affine subspace of co-dimension one. In other words, there is a non-constant linear polynomial $L(X_1, X_2, \ldots, X_m) := \alpha_0 + \sum_{i = 1}^m \alpha_i X_i$, where each $\alpha_i$ is a field element such that 
\[
H = \left\{(b_1, \ldots, b_m) \in \F^m : L(b_1, \ldots, b_m) = 0 \right\} \, .
\]
We say that a set $\mathcal{H} = \{H_1, H_2, \ldots, H_t\}$ of hyperplanes in $\F^m$ are in \emph{general position} if for every subset of hyperplanes in $\mathcal{H}$ of size $m$ intersect in one point, and no subset of hyperplanes in $\mathcal{H}$ of size $m+1$ intersect.

Given any collection of $D + m$ hyperplanes in general position in $\F^m$, Bl{\"{a}}ser \& Pandey gave a construction of hitting sets of optimal size for $m$-variate degree $D$ polynomials. It is not hard to show that this set of points is, in fact, an interpolating set (and not just a hitting set) for degree $D$ polynomials on $m$ variables. This stronger property turns out to be useful for our proofs in this paper. 
\begin{theorem}[Bl{\"{a}}ser \& Pandey \cite{BP20}]\label{thm:geometric-hitting-set}
    Let $m, D$ be natural numbers, $\F$ be a field, and let $\mathcal{H}$  be a set of size $(m+D)$  hyperplanes in general position in $\F^m$. Let $T \subseteq \F^m$ be the set of all points obtained by intersecting some $m$ elements of $\mathcal{H}$. Then, $T$ is of size $\binom{m+D}{m}$ and $T$ is an interpolating set for $m$-variate polynomials of degree $D$ over $\F$. 
\end{theorem}
We give a quick proof in \autoref{appendix:GAPsets}, which is slightly simpler than the already simple proof in~\cite{BP20}.

We refer to codes obtained from hitting sets obtained from \autoref{thm:geometric-hitting-set} as Geometric Arrays for Polynomials (GAP). 
\begin{definition}[GAP]\label{def:gap-codes}
    Let $t > 0$, and $\F$ be a finite field of size at least $t$.
    Define $\GAP{m, d, t} = E_{m, d, T}$, where $T$ is the set of $m$-wise intersections of some $t$ hyperplanes in general position. 
    I.e., $\GAP{m, d, t}$ is the set of evaluations of $m$-variate polynomials of total degree at most $d$ on $T$.
    To specify a specific set of $t$ hyperplanes, $\mathcal{H}$, we write $\mathrm{GAP}_{m, d, t}^{\mathcal{H}}$
\end{definition}

We now combine \autoref{thm:geometric-hitting-set} and \autoref{lem:hitting-sets-to-robust-interpolating-sets} to obtain the following theorem. 
\begin{theorem}\label{thm:code-construction-geometric}
    Let $m, d, t \in \N$ with $t > m + d$. 
    Then $\GAP{m, d, t}$ has distance at least $\binom{t - d}{m}$. 
    Furthermore, for any $\epsilon > 0$, if $t = m + d + \epsilon d - 1$, then rate of $\GAP{m, d, t}$ is least $\left(\frac{1}{1 + \epsilon}\right)^m$ and relative distance is at least $\left( \frac{\epsilon}{1 + \epsilon}\right)^m$.  
\end{theorem}

\begin{proof}
    Let $C = \GAP{m, d, t}$. Let $\mathcal{H}$ be a set of $t$ hyperplanes in general position, and let $T$ be the set of their $m$-wise intersections.

    By \Cref{thm:geometric-hitting-set}, $T$ is a hitting set for polynomials of degree $t - m$. 
    Let $f$ be any $m$-variate polynomial of degree at most $d$. 
    By \autoref{lem:hitting-sets-to-robust-interpolating-sets}, the number of non-zeros of $f$ in $T$ is at least $\binom{t - m - d + m}{m} = \binom{t - d}{m}$.
    Therefore, since $C$ is linear, $C$ has distance at least $\binom{t - d}{m}$.

    Now let $t = m + d + \epsilon d$. 
    We will now compute the rate, $R$, and relative distance, $\delta$. 
    Recall that $C$ has dimension $\binom{m + d}{m}$, and block length $\binom{t}{m}$. 
    Thus, 
    \begin{align*}
    R &= \frac{\binom{m+d}{m}}{\binom{m+d+\epsilon d}{m}} \\
    &= \frac{(m+d)!}{m!d!} \cdot \frac{m! (d + \epsilon d)!}{(m + d + \epsilon d)!}\\
    &=  \prod_{i = 0}^{m-1} \frac{(m + d - i)}{(m + d + \epsilon d - i)} \,.
    \end{align*}
    Now, using the observation that for all $i \leq  m$, $\frac{(m + d - i)}{(m + d + \epsilon d - i)}$ is at least $1/(1 + \epsilon)$, we get that $R \geq \left(\frac{1}{1+\epsilon}\right)^m$. 
    Then, 
    \begin{align*}
    \delta &= \frac{\binom{m+\epsilon d}{m}}{\binom{m+d+\epsilon d}{m}} \\
    &=\frac{(m+\epsilon d)!}{m!(\epsilon d)!} \cdot \frac{m! (d + \epsilon d)!}{(m + d + \epsilon d)!}\\
    &=\prod_{i = 0}^{m-1} \frac{(m + \epsilon d - i)}{(m + d + \epsilon d - i)} \, .
    \end{align*}
    We now observe that for all $i \leq  m$, $\frac{(m + \epsilon d - i)}{(m + d + \epsilon d - i)}$ is at least $\epsilon /(1 + \epsilon)$. 
    Thus, $\delta \geq \left( \frac{\epsilon}{1 + \epsilon}\right)^m$, as required.
\end{proof}

\begin{remark}
    The distance and rate trade-off for GAP codes is $R^{1/m} + \delta^{1/m} \geq 1$.
\end{remark}

In comparison, the trade-off obtained by the standard Reed-Muller code, $\RM_{m, d, \ell}$ is $R = (1 - \delta)^m / m!$. 
Note that the rate of such codes are always bounded by $1/m!$. 
By contrast, GAP codes can have rate arbitrarily close to 1, while having constant relative distance.

This construction can be instantiated using any explicit collection of hyperplanes in general position\footnote{which are equivalent to MDS codes.}. For example, here is a classical construction of hyperplanes in general position coming from Vandermonde matrices, which works over any sufficiently large field.
\begin{observation}[The Vandermonde hyperplane family in general position, see eg.\ Bl{\"{a}}ser \& Pandey \cite{BP20}]\label{obs:example-hype-general-position}
Let $t, m$ be natural numbers with $t \geq m$ and let $\F$ be a field of size at least $t$.
Let $\alpha_1, \alpha_2, \ldots, \alpha_t$ be $t$ distinct elements from the field $\F$.

For $i \in [t]$, let $H_i \subseteq \F^m$ be the hyperplane given by:
\[
H_i := \left\{(b_1, \ldots, b_m) \in \F^m : L_{\alpha_i}(b_1, \ldots, b_m) = 0 \right\} , 
\]
where for $\alpha \in \F$, $L_\alpha(X_1, \ldots, X_m)$ is the linear function given by:
\[
L_\alpha(\vecX) :=  \alpha^m - \alpha^{m-1} X_1 + \alpha^{m-2}  X_2 +  \cdots + (-1)^{m-1} \alpha  X_{m-1} + (-1)^m X_m.
\]

Then, the hyperplanes $H_1, H_2, \ldots, H_t$ are in general position in $\F^m$. 
\end{observation}

The nonstandard choice of signs in the above construction is to enable the following elegant description of the evaluation points of the associated GAP code. Viewing $\F^m$ as the space of monic univariate polynomials of degree $m$ (by associating the point $(b_1, \ldots, b_m)$ with the polynomial $T^m - b_1 T^{m-1} + \ldots + (-1)^m b_m \in \F[T]$), we get that the hyperplane $L_{\alpha}$ is simply the set of monic polynomials that vanish at $\alpha$. Then the intersection of $m$ such hyperplanes $\{ L_{\alpha}: \alpha \in J\}$ (where $|J| = m$) is simply the unique polynomial of degree $m$ whose set of roots equals $J$ -- its coefficients are thus the (signed) elementary symmetric polynomials evaluated at the elements of $J$.
Instantiating this, we get the following very clean description of the evaluation domain for a GAP code.

\begin{definition}[Vandermonde GAP code evaluation domain]
Let $A\subseteq \F$ be a set of size $t$. Define the evaluation domain:
$$S_{A, m} =  \{  (e_1(\mathbf a), e_2(\mathbf a), \ldots, e_m(\mathbf a)) \mid \mathbf a = (a_1, \ldots, a_m), \mbox{ the }a_i \mbox{ are pairwise distinct elements of } A \}.$$
\end{definition}
Note that $|S_{A,m}| = {t \choose m}$. By the previous discussion, evaluating polynomials of degree $d$ on $S_{A,m}$ yields a code of distance 
${t-d \choose m}$.

\subsection{A Construction Based on Simplices}

We now give another construction of a polynomial evaluation code with parameters matching those of the construction in \autoref{thm:code-construction-geometric}. 
The overall framework of analysis will also remain the same and rely on the use of \autoref{lem:hitting-sets-to-robust-interpolating-sets}. 
The main difference is that we instantiate this framework using an alternative construction of hitting sets for polynomials based on taking the hitting set to be all points in an appropriate simplex. 
Let 
$$\Simplex{m, D} = \{(x_1,...,x_m) \in \N^m: x_i \in \N \text{ and }x_1 + x_2 + ... 
+ x_m < D\}$$ be the $m$-dimensional simplex of side length $D$. 
This construction of hitting sets is also described in the work of Bl{\"{a}}ser \& Pandey \cite{BP20}. 
We start with a statement of their result.

\begin{theorem}[Bl{\"{a}}ser \& Pandey \cite{BP20}]\label{thm:simplex-hitting-set}
Let $m, D$ be natural numbers and $\F$ be a field of characteristic greater than $D$ or zero. Then, $\Simplex{m,D+1}$ is of size $\binom{m + D}{m}$ and is a hitting set for $m$-variate polynomials of degree $D$ over the field $\F$. 
\end{theorem}

Since the characteristic of $\F$ is either zero or larger than $D$, we can view the set $\{0, 1, \ldots, D\}$ as a subset of $\F$ in a natural sense. 
Thus, the set $\Simplex{m, D+1}$ as constructed above is indeed a subset of $\F^m$.

Instantiating this hitting/interpolating set with \autoref{lem:hitting-sets-to-robust-interpolating-sets}, we immediately get another family of multivariate polynomial evaluation codes with high rate. 
In fact, investigating the combination of the proof of \autoref{thm:simplex-hitting-set} from~\cite{BP20} and \autoref{lem:hitting-sets-to-robust-interpolating-sets} together, we develop a new ``Schwartz-Zippel lemma theory'', which is able to analyze the distance of polynomial codes evaluated over a wide family of combinatorial shapes -- simultaneously generalizing the original Schwartz-Zippel Lemma for general product sets and the distance bound that we just saw for the above rigid $\{0,1,\ldots, D\}$ simplex. See the statement in \autoref{sec:capintro}.

For now, we will simply state a special case: combinatorial simplices. 
\begin{definition}[Combinatorial Simplex]
     Let $A \subseteq \F$ with $|A| = \ell$. 
     Let $A = \{a_0, a_1, \ldots, a_{\ell-1} \}$ be an arbitrary ordering of $A$.
     Define:
     $$\Simplex{m, A} = \{(a_{x_1}, a_{x_2}, \ldots, a_{x_m}) \mid x_i \in \mathbb N \mbox{ with } \sum_{i} x_i < \ell \}.$$
\end{definition}
In \autoref{sec:zpp} we show that combinatorial simplices are good evaluation domains for polynomials. We refer to the codes obtained from combinatorial simplices as Combinatorial Arrays for Polynomials (CAP) codes.

\begin{definition}[CAP]\label{def:cap-codes}
    Define $\CAP{m, d, \ell} = E_{m, d, S}$, where $S = \Simplex{m, A}$ for some set $A$ of size $\ell$. That is, $\CAP{m, d, \ell}$ is the set of evaluations of $m$-variate polynomials of degree at most $d$ on the combinatorial simplex of $A$, where $|A| = \ell$.
\end{definition}

We again have that these codes have good rate and distance, with rate able to approach $1$ when $m = O(1)$.
\begin{theorem}\label{thm:code-construction-simplexes}
    Let $m, d, \ell \in \N$ with $\ell > d$. 
    Then $\CAP{m, d, \ell}$ has distance at least $\binom{\ell - d + m - 1}{m}$. 
    Furthermore, for any $\epsilon > 0$, if $\ell = d + \epsilon d$, then rate of $\CAP{m, d, \ell}$ is least $\left(\frac{1}{1 + \epsilon}\right)^m$ and relative distance is at least $\left( \frac{\epsilon}{1 + \epsilon}\right)^m$.  
\end{theorem}

The proof of the distance and the calculations of rate and relative distance are the same as that of the proof of \autoref{thm:code-construction-geometric}. 

\section{Zero Patterns of Polynomials}
\label{sec:zpp}

In this section we develop an alternate approach to the distance of CAP codes by understanding the zero patterns of polynomials. 
Though this method requires a bit more effort, the technique we develop leads to a generalized Schwartz-Zippel lemma, and is able to prove the distance properties of a wider variety of evaluation sets. It will also be the foundation for our efficient decoding algorithms for CAP codes.

As a particular application of this more general theory, we give an example in \autoref{sec:step-eval} of an evaluation set in 2-dimensions obtaining better rate versus distance trade-offs than both CAP/GAP codes and traditional RM codes for some settings of rate and distance.

Since polynomial evaluation codes are linear, finding the distance of a polynomial evaluation code is equivalent to finding lower bounds on the minimum number of non-zeros in $S$ of any non-zero polynomial of total degree $d$. 
Alternatively, one can find upper bounds on the number of zeros in a set $S$. 
In the univariate case, the degree mantra states that univariate polynomials of degree at most $d$ have at most $d$ zeros. 
When the evaluation set is the grid, we have the Schwartz-Zippel lemma. 

Thus, in order to prove the distance properties of polynomial evaluation codes on more general sets, we will study the zero patterns of polynomials and prove a generalization of the Schwartz-Zippel lemma. 
We state the The Schwartz-Zippel Lemma below for reference.

\begin{lemma}[Schwartz-Zippel]\label{lem:SZ}
    Let $f \in \F[X_1,...,X_m]$ be any non-zero polynomial of total degree at most $d$, let $S \subset \F$. Then $|Z(f) \cap S^m| \leq d|S|^{m-1}$
\end{lemma}

Throughout this section, we identify $\F = \F_q$ and $\N$ in the following way. Suppose $\left\{ \alpha_0,...,\alpha_{q-1} \right\}$ are the elements of $\F$, think of $k \in \N$ as $\alpha_k$. Note that the mapping of $\alpha_i$ to elements of $\F$ is irrelevant since we will only be working on the indices. We will always assume $\F$ is large enough so that when we write $\N_{<\ell}$, there are indeed at least $\ell$ distinct elements in $\F$. Furthermore, we will write $\N = \N_{<q} = \F$.

\subsection{The Bivariate Case}

We start by considering the bivariate case. 
The standard proof of the Schwartz-Zippel Lemma proceeds by induction on $m$. 
We will present the first inductive step (proving the bivariate case from the univariate case), and observe that it not only gives bounds on the number of zeros but also provides restrictions on their locations in the grid.

Let $d_Y$ be the $Y$ degree of $f$ (i.e., the maximum power of $Y$ in any monomial with non-zero coefficient). 
Then we can write $f(X, Y) = \sum_{i = 0}^{d_Y} f_i(X)Y^i$, where $f_i$ is a univariate polynomial in $X$ of degree at most $d - i$. 
Then, consider the number of zeros in each column. 
I.e., for each $x$, how many zeros are there in $\{(x, y): y \in S\}$? 
Let $x \in S$, there are two cases. 

\begin{enumerate}
    \item If $f_{d_Y}(x) \neq 0$, then the univariate polynomials $f(x, Y)$ is non-zero (since the coefficient for $Y^{d_Y}$ is non-zero). Furthermore, $f(x, Y)$ has degree at most $d_Y$, so there are at most $d_Y$ zeros in the $x$th column. 
    \item $f_{d_Y}(x) = 0$, then $f(x, Y)$ may be identically zero, so the entire $x$th with the column could be zero. 
\end{enumerate}

The proof of the Schwartz-Zippel Lemma then proceeds to bound the number of zeros as follows. 
Since $f_{d_Y}$ has degree at most $d - d_Y$, case 2 happens for at most $d - d_Y$ values of $x$, and thus the total number of zeros is at most 
$$
|S|(d - d_Y) + d_Y(|S| - d + d_Y) =  d|S| + d_Y(d_Y - d) \leq d|S|.
$$

\begin{figure} 
    \vspace{-5em}
    \begin{center}
        \includegraphics[width=0.65\textwidth]{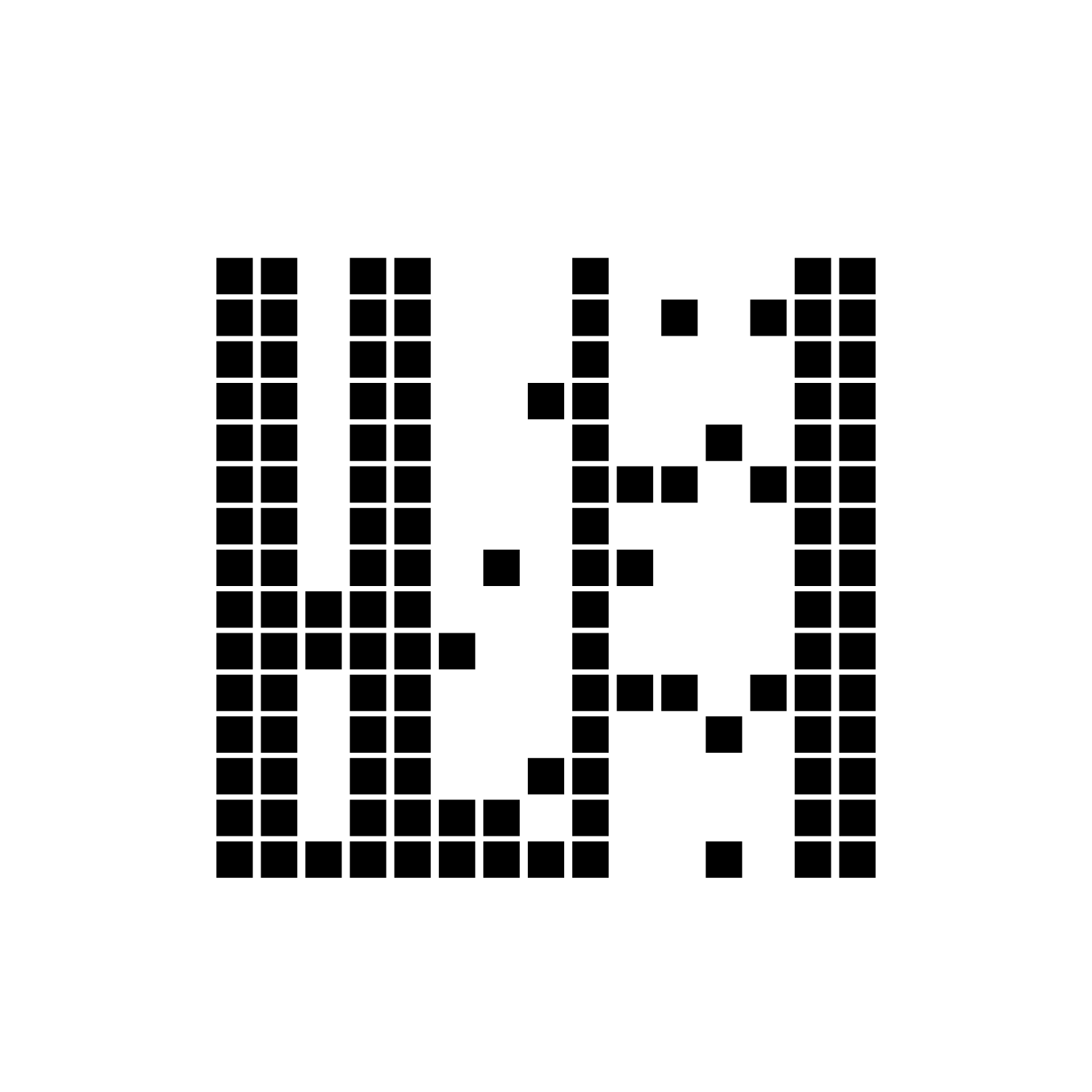}
    \end{center}
    \vspace{-4em}
    \caption{
        Possible locations of zeros of a bivariate polynomial, $f(X, Y)$ of degree $10$, $Y$-degree $3$, evaluated on $\N_{< 15} \times \N_{< 15}$. Zeros are denoted with black squares.
    }\label{fig:zero-pattern}
\end{figure}

Upon close inspection, the proof not only bounds the number of zeros but also proves that the zeros can only occur in a certain pattern. 
In particular, there exists $a \in \{0,1,2,...,d\}$ (which is the $Y$ degree of $f$), such that at most $d - a$ columns are entirely zero, and the rest of the column have at most $a$ zeros. 
\Cref{fig:zero-pattern} shows an example.

To demonstrate the power of this structure, consider the following set.

\begin{example}[Triangular Evaluation Set]\label{example:triangle}
    $T = \Simplex{2, \frac{3}{2}d} = \{(x, y) \in \N \times \N : x+y < \frac{3}{2}d\}$. 
\end{example}

Let $f \in \F[X, Y]$ be a polynomial of total degree at most $d$, how many points in $T$ can $f$ vanish? Note that $T$ is contained in the grid $(\N_{<\frac{3}{2}d})^2$. 
Applying the Schwartz-Zippel Lemma directly to this grid, we get that there are at most $\frac{3}{2}d^2$ zeros. 
By a stars and bars argument, $|T| = \binom{\frac{3}{2}d}{2} \approx \frac{9}{8}d^2 \leq \frac{3}{2}d^2$. 
Thus, the Schwartz-Zippel lemma does not rule out the possibility that $f$ vanishes entirely on $T$. 

\begin{figure}
    \vspace{-5em}
    \begin{center}
        \includegraphics[width=0.65\textwidth]{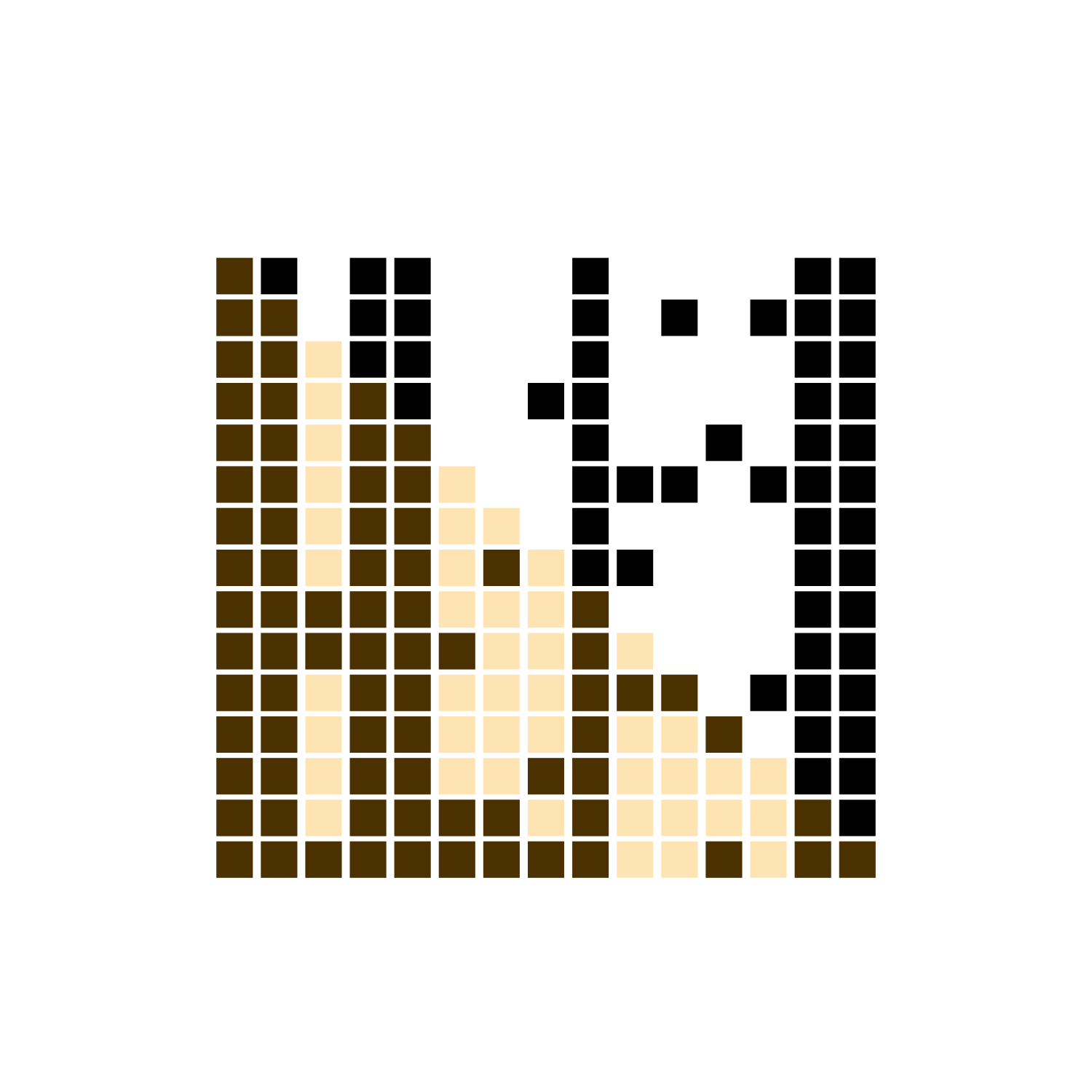}
    \end{center}
    \vspace{-4em}
    \caption{
        A possible zero pattern (same as \Cref{fig:zero-pattern}) overlayed with the triangle evaluation set from \Cref{example:triangle} in orange.
    }
    \label{fig:triangle}
\end{figure}

Now, let's consider the structure of the zeros. 
From the previous discussion, we know that there exists some $a \in \{0,1,...,d\}$ such that $f$ vanishes on at most $d - a$ of the columns, and has at most $a$ zeros on the remaining columns. 
Let $C \subset S$ be the set of columns on which $f$ vanishes. 
Then the smallest value of $x$ for which the $x$th column does not vanish is at most $d-a$ (in the case that $C = \left\{ 0,1,...,d-a-1 \right\}$). 
However, this column has at least $\frac{3}{2}d - (d - a) = d/2 + a$ points in $T$. 
Since $f$ can vanish on at most $a$ of these points, $f$ must be non-zero on at least $d/2$ points in this column. 
Extending this argument to the remaining columns, $f$ must be non-zero on at least $(d/2) + (d/2-1) + (d/2-2) + ... + 1 \approx d^2/8$ points in $T$. 
This example is illustrated in \Cref{fig:triangle}.

Slightly more generally, if we fix any $\epsilon > 0$, and take $T = \{(x, y) \in S: x + y \leq (1 + \epsilon)d\}$, we have that any degree $d$ polynomial is non-zero on at least $\epsilon^2d^2/2$ points. 
As a polynomial evaluation code, $E_{2, d, T}$ has rate $\frac{1}{(1 + \epsilon)^2}$, and relative distance $\frac{\epsilon^2}{(1 + \epsilon)^2}$. 
Thus, taking $\epsilon$ to be small, we can get codes with rate arbitrarily close to 1, with constant distance. 
Notice that this recovers \Cref{thm:code-construction-simplexes} for $m = 2$.

In the next section, we formalize the idea of zero patterns and generalize it to $m$ variables. 

\subsection{The General Case}

\paragraph{Slicing.} To facilitate working over high-dimensional grids, we will introduce some notation for slices. 
Let $S \subset \N^m$. 
For $\mathbf{x} \in \N^{m-1}$, and $i \in J$, define $S_{\mathbf{x}, i} = \left\{ a \in \N: (x_1,...,x_{i-1},a,x_i,....,x_{m-1}) \in S \right\}$, and call $S_{\mathbf{x}, i}$ the slice of $S$ at $\mathbf{x}$ in the $i$ direction. 
Note that to define a subset of $\N^m$, it suffices to specify all the slices in the $i$ direction for some $i \in [m]$.

We define a set of zero patterns, $\mathcal{Z}_{m, d}$.
\begin{definition}[$\mathcal{Z}_{m, d}$]
$\mathcal{Z}_{m, d}$ is a subset of $\wp(\N^m)$ and is defined recursively as follows. 
$\mathcal{Z}_{0, d} = \{\emptyset\}$, and $\forall E \subset \N^m$,  $E \in \mathcal{Z}_{m, d}$ if and only if there exists $a \in \{0,1,...,d\}$, and $D \in \mathcal{Z}_{m-1, d-a}$ such that for all $\mathbf{x} \in \N^{d-1}$, if $\mathbf{x} \in D$, then $E_{\mathbf{x}, m} = \N$, and if $\mathbf{x} \notin D$, $|E_{\mathbf{x}, m}| = a$\footnote[1]{Note that we could have had replaced this condition with $|E_{\mathbf{x}, m}| \leq a$, but having $|E_{\mathbf{x}, m}|$ exactly equal to $a$ will make some later proofs slightly cleaner.}.
\end{definition}

The following lemma shows that that for every non-zero polynomial $f$ on $m$ variables with total degree at most $d$, the zeros of $f$ are a contained in some set in $\mathcal{Z}_{m, d}$. 
\begin{lemma}[Zero Patterns]\label{lem:zero-patterns}
    Let $f$ be a non-zero polynomial on $m$ variables with total degree at most $d$. 
    Let $Z(f)$ be the set of zeros of $f$ on $\N^m$. 
    Then there is some $E \in \mathcal{Z}_{m, d}$ such that $Z(f) \subset E$.
\end{lemma}

\begin{proof}
    By induction on $m$. 

    \textbf{Base case.} For $m = 0$, $f$ is a polynomial on zero variables and thus a constant. Since $f$ is non-zero, it has no zeros; hence, $Z(f) = \emptyset$.

    \textbf{Inductive step.} Assume the claim is true for $m-1$. 
    Let $f \in \F[X_1,..., X_m]$ be a non-zero polynomial with total degree at most $d$. 
    Let $d_m \in \{0,1,...,d\}$ be the $X_m$ degree of $f$, and write $f = \sum_{i=0}^{d_m}f_i(X_1,...,X_{m-1})X_m^i$. 
    Note that $f_{d_m}$ is a non-zero polynomial on $m-1$ variables of total degree at most $d-d_m$. 
    Thus, the induction hypothesis applies, and there exists some $D \in \mathcal{Z}_{m-1, d-d_m}$ such that $Z(f_{d_m}) \subset D$. 

    We'll construct $E$ by its $m$ directional slices as follows. 
    For $\mathbf{x} \in D$, we take everything - i.e., $E_{\mathbf{x}, m} = \N$. 
    For $\mathbf{x} \notin D$, $f(\mathbf{x}, X_m)$ is a non-zero univariate polynomial of degree $d_m$, and thus has at most $d_m$ zeros. 
    Thus, $|Z(f)_{\mathbf{x}, m}| \leq d_m$. 
    So for each $\mathbf{x} \notin D$, we set $E_{\mathbf{x}, m}$ to some subset containing $Z(f)_{\mathbf{x}, m}$ of size exactly $d_m$.

    Then, $E \in \mathcal{Z}_{m, d}$ and $E$ contains $Z(f)$, so we're done.
\end{proof}

\subsection{Shifting}

To get a better handle on $\mathcal{Z}_{m, d}$, we will define a shifting operation and find that, after shifting, elements of $\mathcal{Z}_{m, d}$ are very easy to describe. 

\begin{figure}
    \vspace{-5em}
    \begin{center}
        \includegraphics[width=\textwidth]{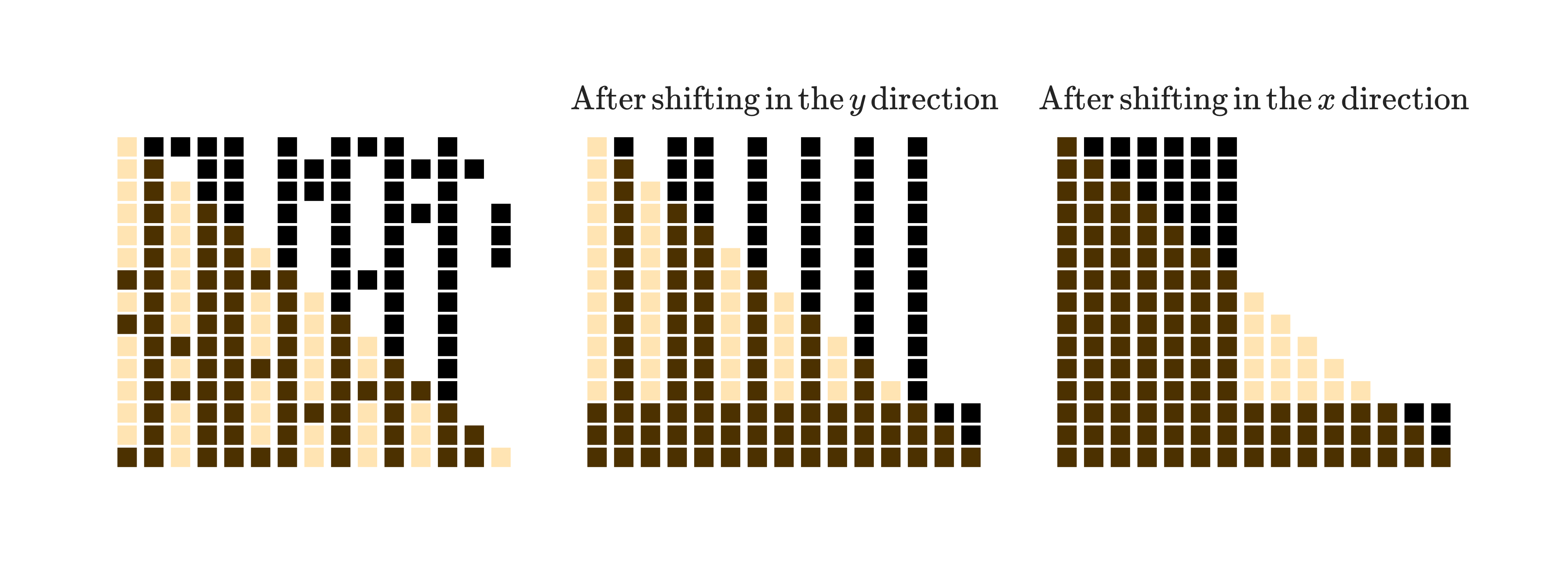}
    \end{center}
    \vspace{-4em}
    \caption{
        Example of shifting the zeros from \Cref{fig:triangle}
    }
    \label{fig:shifting}
\end{figure}

Let's again think back to the bivariate case. 
Suppose $f \in \F[X, Y]$ is any polynomial of total degree at most $d$ and $Y$ degree $d_Y$. 
We earlier found that there are at most $d - d_Y$ columns where $f$ was completely zero, and the remaining columns had at most $d_Y$ zeros each. 
Let $C \subset \N$ be the subset of entirely zero columns. 
Consider an operation that first pushes zeros in each column as far down as possible. 
This has no effect for columns what were already completely zeros, but now, for each other column, the zeros are now contained within the first $d_Y$ rows. 
Next, for each row, push all the zeros as far left as possible. 
Since there are now at most $d-d_Y$ columns that have non-zero values above the $d_Y$th row, we have that all of the zeros are contained in the first $d_Y$ rows and $d-d_Y$ columns. 
This shifting procedure is shown in \Cref{fig:shifting}. 

We now formally define the $i$th shifting operator, $\sigma_i(Z)$, that pushes points down in the $i$ direction. 
Define $\sigma_i(Z)$ by its slices in the $i$ direction. For any $\mathbf{x} \in \N^{m-1}$, 
$$
(\sigma_i(Z))_{\mathbf{x}, i} = \N_{<|Z_{\mathbf{x}, i}|}.
$$
That is, we replace every slice in the $i$ direction with an initial segment of the same size as the original slice.

We now prove some basic facts about $\sigma_i$.
\begin{lemma}[Properties of $\sigma_i$]
    \crefalias{enumi}{lemma}
    \begin{enumerate}[before=\leavevmode,label=\upshape(\Roman*),ref=\thetheorem (\Roman*)]
        \item \label{lem:pushing-decrease} $\sigma_i(Z)$ can be obtained by decreasing the $i$ coordinate of zero or more elements of $Z$. 
        \item \label{lem:push-full} For any $i \in [m]$. $\sigma_i(\N^{m}) = \N^{m}$
        \item \label{lem:pushing-disjoint} Let $i, j \in [m]$ with $i \neq j$, and $A, B \subset \N^m$, and suppose $A$ and $B$ are disjoint in a coordinate $i$, that is $\{a_i: \mathbf{a} \in A\} \cap \{b_i: \mathbf{b} \in B\} = \emptyset$, then if $j \neq i$, $\sigma_j(A \cup B) = \sigma_j(A) \cup \sigma_j(B)$.
        \item \label{lem:pushing-product} Let $j \in [m-1]$, and $A \subset \N^{m-1}$ and $C \subset \N$. Then $\sigma_{j}(A \times C) = \sigma_{j}(A) \times C$.
    \end{enumerate}
\end{lemma}

\begin{proof}
    \begin{enumerate}[before=\leavevmode,label=\upshape(\Roman*),ref=\thetheorem (\Roman*)]
        \item For any $\mathbf{x} \in \N^{m-1}$, $(\sigma_i(Z))_{\mathbf{x}, i}$ is the initial segment of $\N$ of size $|Z_{\mathbf{x}, i}|$. 
            Let $z_0,z_1,...$ be the elements of $Z_{\mathbf{x}, i}$ in ascending order. 
            Then $f: Z_{\mathbf{x}, i} \to \sigma_i(Z)_{\mathbf{x}, i}$ defined by $f(z_j) = j$ is always such that $f(x) \leq x$.
        \item For every $\mathbf{x} \in \N^{m-1}$, $(\N^m)_{\mathbf{x}, i} = \N$. 
            Thus, $(\sigma_i(\N^m))_{\mathbf{x}, i} = \N$, and $\sigma_i(\N^m) = \N^m$.
        \item Let $\mathbf{x} \in \N^{m-1}$. 
            Since $A$ and $B$ are disjoint in some coordinate $i$, at most one of $A_{\mathbf{x}, j}$, and $B_{\mathbf{x}, j}$ can be non-empty (otherwise, there would be elements in $A$ and $B$ that both have $i$ coordinate $x_i$). 
            Thus, $A_{\mathbf{x}, j} \cup B_{\mathbf{x}, j}$ is either $A_{\mathbf{x}, j}$ or $B_{\mathbf{x}, j}$, and consequently, $|A_{\mathbf{x}, j} \cup B_{\mathbf{x}, j}| = \max(|A_{\mathbf{x}, j}|, |B_{\mathbf{x}, j}|)$. 
            We now show $\sigma_j(A \cup B)$ and $\sigma_j(A) \cup \sigma_j(B)$ are equal by comparing their $j$-directional slices.
    \begin{align*}
        (\sigma_j(A) \cup \sigma_j(B))_{\mathbf{x},j} &= \N_{<|A_{\mathbf{x}, j}|} \cup \N_{<|B_{\mathbf{x}, j}|}\\
                                                      &= \N_{<\max(|A_{\mathbf{x}, j}|, |B_{\mathbf{x}, j}|)}\\
                                                      &= \N_{<|A_{\mathbf{x}, j} \cup B_{\mathbf{x}, j}|}\\
                                                      &= (\sigma_j(A \cup B))_{\mathbf{x}, j}
    \end{align*}
\item Let $c \in \N$, and $\mathbf{x} \in \N^{m-2}$. 
    Note that the slice at $\mathbf{x}$ in the $j$th direction for $A$ is the same as the slice at $[x_1,...,x_{m-2}, c]$ in the $j$th direction for $A \times \{c\}$ since all we changed was to set the last coordinate to $c$ for each element of $A$. 
    Thus, $\sigma_{j}(A \times \{c\}) = \sigma_j(A) \times \{c\}$. 
    Then note that $A \times C = \bigcup_{c \in C} A \times \{c\}$. 
    Each of these sets in the union is disjoint last coordinate, so by \Cref{lem:pushing-disjoint},
    $$
    \sigma_{j}(A \times C) = \bigcup_{c \in C} \sigma_{j}(A \times \{c\})  = \bigcup_{c \in C} \sigma_{j}(A) \times \{c\} = \sigma_{j}(A) \times C
    $$
    \end{enumerate}
\end{proof}

Let $\mathbf{d} \in \N^m$, and define $L(\mathbf{d}) = \{\mathbf{x} \in \N^m: x_1 < d_1 \lor x_2 < d_2 \lor ... 
\lor x_m < d_m\}$. 
We use $L$ to denote this set since, in the two-dimensional case, it has an ``L'' shape. 
We now claim that every element of $\mathcal{Z}_{m, d}$ looks like $L(\mathbf{d})$ for some $\mathbf{d}$ after shifting in every coordinate.

\begin{lemma}\label{lem:pushing-zero-patterns}
    Let $\sigma_{1,...,m} = \sigma_1 \circ \sigma_2 \circ ... \circ \sigma_m$. For every $E \in \mathcal{Z}_{m, d}$, $\sigma_{1,...,m} (E) = L(\mathbf{d})$ for some $\mathbf{d} \in \{0,1,...,d\}^m$ with $d_1 + d_2 +...+ d_m = d$.
\end{lemma}

\begin{proof}
    By induction on $m$. 
    To make the base case work nicely, define the empty concatenation of operators to be the identity, and define $L([]) = \emptyset$

    \textbf{Base case.} For $m = 0$, $E = \emptyset = L([])$

    \textbf{Inductive Step.} Suppose the claim is true for $m-1$, and let $E \in \mathcal{Z}_{m, d}$. 
    Then by the recursive definition of $\mathcal{Z}$, there exists some $d_m \leq d$, and $D \in \mathcal{Z}_{m-1, d-d_m}$ such that $E_{\mathbf{x}, m} = \N$ for $\mathbf{x} \in D$, and $|E_{\mathbf{x}, m}| =  d_m$ for $\mathbf{x} \notin D$. 
    Then, $$\sigma_m(E) = D \times \N \cup \overline{D} \times \N_{<d_m}$$
    Since $D$ and $\overline{D}$ partition $\N^{m-1}$, we have that every $\mathbf{x} \in \N^m$ with $x_m < d_m$ is in $\sigma_m(E)$. 
    Thus, we can rewrite the set as $D\times \N_{\geq d_m} \cup \N^{m-1} \times \N_{<d_m}$. 
    Plugging this in, we get 
    \begin{align*}
        \sigma_{1,...,m}(E) &= \sigma_{1,...,m-1}(D\times\N_{\geq d_m} \cup \N^{m-1} \times \N_{<d_m})\\
        &= \sigma_{1,...,m-1}(D\times\N_{\geq d_m}) \cup \sigma_{1,...,m-1}(\N^{m-1} \times \N_{<d_m})\\
        &= \sigma_{1,...,m-1}(D)\times\N_{\geq d_m} \cup \sigma_{1,...,m-1}(\N^{m-1})\times \N_{<d_m}\\
        &= \sigma_{1,...,m-1}(D)\times\N_{\geq d_m} \cup \N^{m-1}\times \N_{<d_m}.
    \end{align*}
    The second line follows from the fact that $D\times\N_{\geq d_m}$, and $\N^{m-1} \times \N_{<d_m}$ are disjoint in the $m$ coordinate, and $\sigma_i$ for $i < m$ does not change the $m$ coordinate, and applying \Cref{lem:pushing-disjoint} repeatedly. 
    The third line follows from \Cref{lem:pushing-product}, and the fourth from \Cref{lem:push-full}. 

    From the inductive hypothesis, there exists $\mathbf{d} \in \{0,1,...,d\}^{m-1}$ such that $d_1 + d_2 + ... + d_{m-1} = d - d_m$, and $\sigma_{1,...,m-1}(D) = L(d_1,...,d_{m-1})$. 
    Substituting this back, we finally get
    $$
    \sigma_{1,...,m}(E) = L(d_1,...,d_{m-1}) \times \N_{\geq d_m} \cup \N^{m-1} \times \N_{<d_m}  = L(d_1,...,d_m).
    $$
    Also note that $\sum_{i=1}^m d_i = d_m + \sum_{i=1}^{m-1} d_i = d_m + d-d_m = d$, and this completes the proof.
\end{proof}

\subsection{d-robustness}\label{sec:d-robust}
Since we want to prove upper bounds on the number of zeros in $S$, we require that $S$ satisfies some property to ensure that the shifting operation does not move zeros out of the set. 
Shifting can be viewed as decreasing the values of certain coordinates (\Cref{lem:pushing-decrease}), so a very natural definition is the following.

\begin{definition}[Downward Closed]
    $S \subset \N^m$ is downward closed if for all $\mathbf{x}, \mathbf{y}\in \N^m$, with $x_1 \leq y_1$, $x_2\leq y_2$,...,$x_m \leq y_m$, if $\mathbf{y} \in S$, then $\mathbf{x} \in S$. In other words, if $\mathbf{x}$ is at most $\mathbf{y}$ in every coordinate, and $\mathbf{y} \in S$, then $\mathbf{x} \in S$.
\end{definition}

Given, this definition, we see that shifting cannot decrease the size of the intersection with a downward closed set.

\begin{lemma}\label{lem:shift-downward-closed}
    Let $S, Z \subset \N^m$ such that $S$ is downward closed. Then for any $i \in [m]$, 
    $$
    |S \cap Z| \leq |S \cap \sigma_i(Z)|
    $$
\end{lemma}

Our next lemma bounds the number of zeros of an $m$-variate polynomial can have on a downward closed set.

\begin{lemma}\label{lem:zeros-inL}
    Let $f \in \F[X_1,..,X_m]$ be of total degree $d$, and $S$ be a downward closed set. Then there exists $\mathbf{d} \in \{0,1,...,d\}^m$ such that $\sum_{i=1}^m d_i = d$, and 
     $$
     |Z(f) \cap S| \leq |L(\mathbf{d}) \cap S|
     $$
\end{lemma}

\begin{proof}
    By \Cref{lem:zero-patterns}, $Z(f) \subset E$ for some $E \in \mathcal{Z}_{m, d}$. By \Cref{lem:pushing-zero-patterns}, we have that there is some $\mathbf{d} \in \{0,1,...,d\}^m$, with $\sum_{i=1}^m d_i = d$, such that $\sigma_{1,...,m}(E) = L(\mathbf{d})$. Then, we have
    \begin{align*}
        |Z(f) \cap S| &\leq |E \cap S| & (Z(f) \subset E)\\
                      &\leq |\sigma_{1,...,m}(E)\cap S| &(\text{Lemma \ref{lem:shift-downward-closed}})\\
                      &= |L(\mathbf{d}) \cap S|
    \end{align*}
\end{proof}

We also state the corresponding bound for non-zeros, a direct result of the previous lemma. 
For $\mathbf{d} \in \{0,1,...,d\}^m$, define $H(\mathbf{d}) = \overline{L(\mathbf{d})}$. 
I.e., $H(\mathbf{d})$ is the set of points $\mathbf{x}$ in which every coordinate $x_i$ is at least $d_i$.

\begin{lemma}\label{lem:nonzeros-inL}
Let $f \in \F[X_1,..,X_m]$ be of total degree $d$, and $S$ be a downward closed set. 
Then, there exists $\mathbf{d} \in \{0,1,...,d\}^m$ such that $\sum_{i=1}^m d_i = d$, and 
$$
|N(f) \cap S| \geq |H(\mathbf{d}) \cap S|
$$
\end{lemma}

Now, we define a combinatorial property of a set $S$ that guarantees that any $m$-variate polynomial of total degree at most $d$ will have at least some number of non-zeros on $S$.

\begin{definition}[$d$-robustness of a downward closed set]
    Let $S \subset \N^m$ be downward closed.
    Define the $d$-robustness of $S$, as follows.
    $$
        \Pi_d(S) = \min \left\{ 
            |H(\mathbf{d}) \cap S| : {\mathbf{d} \in \N^m , \sum_{i=1}^m d_i = d}.
        \right\}
    $$
    The relative $d$-robustness is then $\pi_d(S) = \Pi_d(S) / |S|$.
\end{definition}

In words, $\Pi_d(S) \geq B$ if for any choice of $\mathbf{d}$ such that $\sum_{i=1}^m d_i = d$, at least $B$ elements $\mathbf{x}$ in $S$ are such that $x_i \geq d_i$ for all $i \in [m]$.

Here are several useful facts about $d$-robustness
\begin{remark}
    For any downward closed $S$. $\Pi_d(S) = |S| - \max\left\{|L(\mathbf{d}) \cap S| : {\mathbf{d} \in \N^m , \sum_{i=1}^m d_i = d} \right\}$. Additionally, the following conditions are equivalent 
        \begin{itemize}
        \item $\pi_d(S) \geq \delta$
        \item For every $\mathbf{d}$ with $\sum_{i=1}^m d_i = d$, $\Pr_{\mathbf{x} \sim S}[\mathbf{x} \in H(\mathbf{d})] \geq \delta$
        \item For every $\mathbf{d}$ with $\sum_{i=1}^m d_i = d$, $\Pr_{\mathbf{x} \sim S}[\mathbf{x} \in L(\mathbf{d})] \leq 1-\delta$
    \end{itemize}
\end{remark}

Following the definition of $d$-robustness, and \Cref{lem:nonzeros-inL}, we have the following bounds on the number of zeros/non-zeros in sets with $d$-robustness $\delta n$.

\begin{lemma}[Zeros in $d$ robust sets]\label{lem:zeros-in-robust}
    Let $f \in \F[X_1,...,X_m]$ be a polynomial of total degree $d$. 
    Let $S$ be downward closed of size $n$ of size $n$ such that $\Pi_d(S) \geq \delta n$. 
    Then, we get the following equivalent results:
    \begin{enumerate}
        \item $|N(f) \cap S| \geq \delta n$.
        \item $\Pr_{\mathbf{x} \sim S} [\mathbf{x} \in N(f)] \geq \delta$.
        \item $|Z(f) \cap S| \leq (1-\delta) |S|$.
        \item $\Pr_{\mathbf{x} \sim S} [\mathbf{x} \in Z(f)] \leq 1-\delta$.
    \end{enumerate}
\end{lemma}

A direct consequence of \Cref{lem:zeros-in-robust} (versions 1, 2) is that polynomial evaluation codes on downward closed sets have distance.

\begin{corollary}[Distance of polynomial evaluation codes]\label{cor:distance-rms}
    Let $m, d \in \N$ with $m \geq 1$, and let $S \subset \N^m$.
    If $\pi_d(S) \geq \delta$, then $E_{m, d, S}$ has relative distance at least $\delta$.
\end{corollary}

Thus, to compute the distance of a polynomial evaluation code where the evaluation set is downward closed, we just need to compute the $d$-robustness of the set. 
A direct consequence of \Cref{lem:zeros-in-robust} (versions 3, 4) is a generalization of the Schwartz-Zippel Lemma.

\begin{corollary}[Schwartz-Zippel for downward closed sets]\label{cor:sz}
    Let $m, d \in \N$ with $m \geq 1$, and let $S \subset \N^m$.
    Let $f \in [X_1,...,X_m]$ be a polynomial of total degree $d$, and suppose $\pi_d(S) \geq \delta$.
    Then $\Pr_{\mathbf{x} \sim S}[f(\mathbf{x} = 0)] \leq 1 - \delta$, equivalently, $|Z(f) \cap S| \leq (1 - \delta)|S|$.
\end{corollary}

To see that \Cref{cor:sz} is indeed a generalization of the Schwartz-Zippel lemma, calculate $d$-robustness of the grid of side length $\ell$,  $(\N_{< \ell})^m$. 
Let $\mathbf{d} \in \{0,1,...,d\}^m$ such that $\sum_{i=1}^m d_i = d$. 
We have
$$
\Pr[\mathbf{x} \in L(\mathbf{d})] = \Pr[\exists i \in [m]. (x_i < d_i)] \leq \sum_{i=1}^m \Pr[x_i < d_i] = \sum_{i=1}^m d_i/\ell = d/\ell \, .
$$
Thus, $(\N_{< \ell})^d$ has relative $d$-robustness $1 - d/\ell$, and we have $\Pr[f(x) = 0] \leq d/\ell$, which is exactly what is given by Schwartz-Zippel.

As an alternate proof that CAP codes have distance (\Cref{thm:code-construction-simplexes}), we can calculate the $d$-robustness of simplex, $\Simplex{m, \ell}$.
Let $\mathbf{d} \in \{0,1,...,d-1\}^m$ be such that $\sum_{i=1}^m d_i = d$. 
Then 
$$
\Simplex{m, \ell} \cap H(\mathbf{d}) = \{\mathbf{x} \in \N^m : x_i \geq d_i, \sum_{i=1}^m x_i < \ell\}
$$
Then, $f(\mathbf{x}) = \mathbf{x} - \mathbf{d}$ is a bijection from $\Simplex{m, \ell} \cap H(\mathbf{d})$ to $\Simplex{m, \ell - d}$. 
Thus, the $d$-robustness of $\Simplex{m, \ell}$ is 
$$
\binom{m + \ell - d - 1}{m}.
$$
Combining this with \Cref{lem:zeros-in-robust}, we get that $\CAP{m,d,\ell}$ has distance a least $\binom{m + \ell - d -1}{m}$, recovering \Cref{thm:code-construction-simplexes}.
Note that we get the same result for $\Simplex{m, A}$, where $A \subset \F$ has size $\ell$, by identifying the elements of $A$ with $\{0,1,...,\ell - 1\}$ using any ordering of the elements in $A$.

\subsection{The step evaluation set}\label{sec:step-eval}

\newcommand{\Step}{\mathrm{Step}}

We can also use \Cref{lem:zeros-in-robust} to prove that other evaluations sets provide codes with good distance.
For example, in this section, we show an evaluation set beating both the grid and the simplex for certain values of $\delta$. 
The evaluation set is $\Step(\ell) = \{(x, y) \in \N_{<\ell} \times \N_{< \ell}: x < \ell/2 \lor y < \ell/2\}$.

\begin{claim}[Step Evaluation Set]
    $E_{2, d, \Step(\ell)}$ has relative distance $\frac{2}{3}(1 - d/\ell)$, and rate at least $\frac{2d^2}{3\ell^2}$. 
\end{claim}

\begin{proof}
    First we calculate the rate. 
    Note that $|\Step(\ell)| = \frac{3}{4}\ell^2$, so the rate of the code is $\frac{\binom{d + 2}{2}}{\frac{3}{4}\ell^2} \geq \frac{2d^2}{3\ell^2}$. 
    To calculate the distance, we find the $d$-robustness of $\Step(\ell)$, and apply \Cref{cor:distance-rms}. 
    To find the $\pi_d(\Step(\ell))$, we study the size of $\Step(\ell) \cap L(a, d-a)$ for each $a \in \{0,1,...,d\}$. 
    WLOG, suppose $a \geq d - a$. 
    Consider two cases. 
    \begin{itemize}
        \item $a \leq \ell/2$. 
            Since $a \geq d -a$, we also have $d - \ell/2 \leq a$. 
            Then, $|\Step(\ell) \cap L(a, d-a)| = a\ell + (d - a)\ell - a(d-a)$. 
            Since this is a quadratic expression in $a$ with a positive coefficient, for $a \in [d - \ell/2, \ell/2]$, this function is maximized at $a = \ell/2$ (or equally $a = d - \ell/2$), and we have $|\Step(\ell) \cap L(a, d-a)| \leq d\ell/2 + \ell^2/4$.
        \item $a \geq \ell/2$. 
            In this case, $|\Step(\ell) \cap L(a, d-a)| = \ell a - (a - \ell/2)\ell/2 + (d - a)(\ell-a)$. 
            This function another quadratic in $a$ with a positive coefficient of $a^2$. 
            Therefore, it is again maximized at the boundaries $a = \ell/2$, and $a = d$, and $|\Step(\ell) \cap L(a, d-a)| \leq d\ell/2 + \ell^2 / 4$.
    \end{itemize}

    Rearranging, we find that $\Step(\ell)$ has relative $d$-robustness equal to $\frac{2}{3}(1 - d/\ell)$, as required. 
\end{proof}
\begin{remark}
    The rate distance trade-off of $\delta = 2/3-\sqrt{2R/3}$.
\end{remark}

\begin{figure}
    \vspace{-5em}
    \begin{center}
        \includegraphics[width=0.85\textwidth]{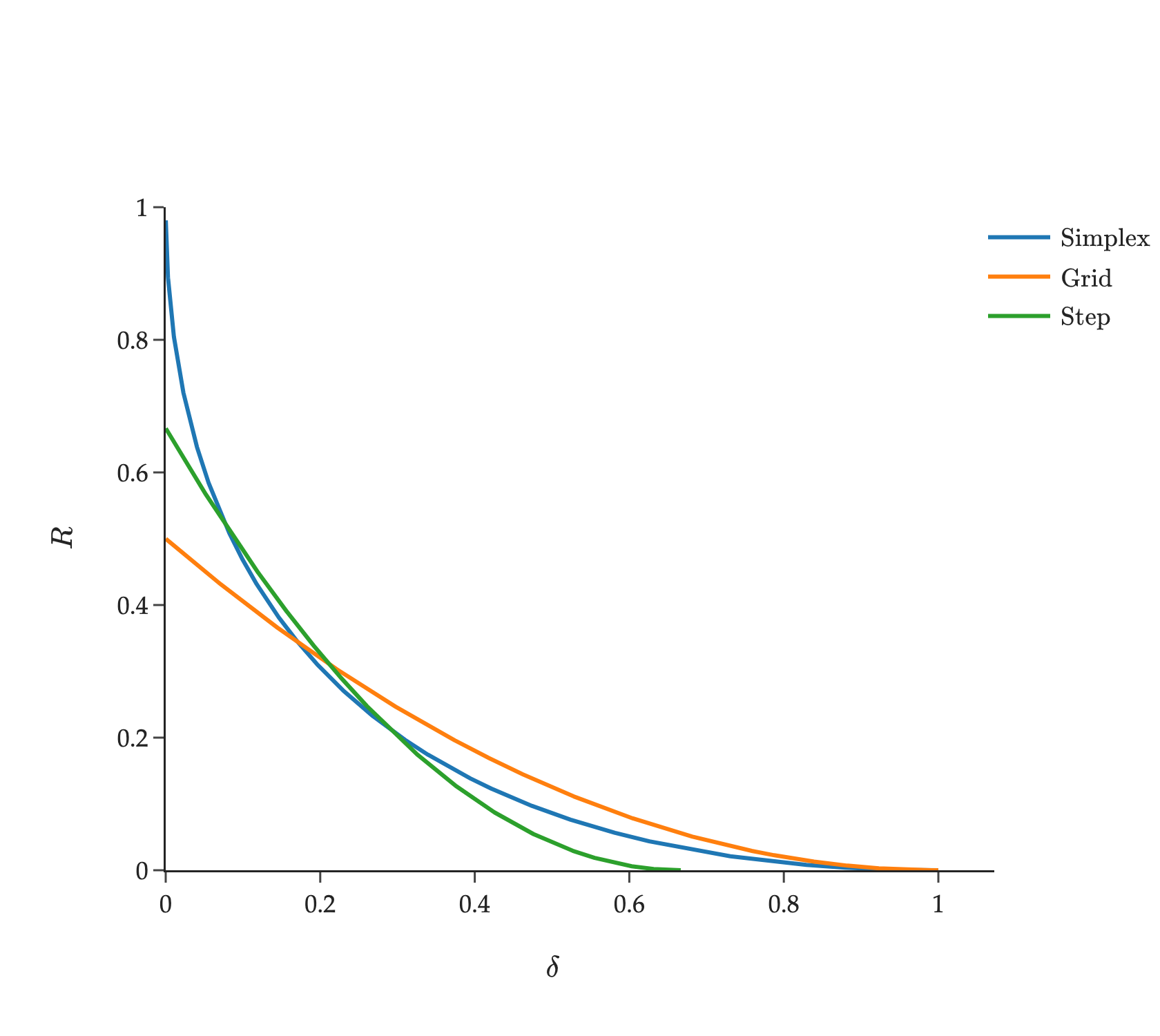}
    \end{center}
    \vspace{-3em}
    \caption{Comparing rate vs. distance trade-offs for the grid, simplex and step evaluation sets}\label{fig:trade-offs}
\end{figure}

\Cref{fig:trade-offs} compares the rate versus distance trade-offs for the simplex, grid, and step. 
In particular, we see that for $\delta \in [0.078, 0.21]$, the step construction beats both the simplex and the grid. 
\footnote{The exact range for when the step construction beats the grid is $\delta \in [\frac{1}{3}(3 - \sqrt{6} - \sqrt{5 - 2 \sqrt{6}}), \frac{1}{2} - \frac{1}{6}\sqrt{3}]$.}

The following is a natural question. For a fixed $\delta$, what is the smallest set $S$ with relative $d$-robustness equal to $\delta$? We explore this question in \Cref{appendix:d-robustness-bounds}, where we prove several lower bounds in two dimensions. In particular, we show that for $\delta > 1/2$, the grid is optimal, and for $\delta \to 0$, the simplex has the right asymptotic behavior.

\section{Unique Decoding of CAP Codes}\label{sec:cap-decoding}

In this section, we give an efficient decoding algorithm for CAP codes. 
The algorithm is based on the decoding algorithm for Reed Muller codes given by Kim and Kopparty \cite{KimK2017}.

\subsection{Uneven GMD}\label{sec:unevengmd}

In this section we describe a variant of the GMD decoding algorithm of Forney \cite{forneyGeneralizedMinimumDistance1966} that allows for inner codes of varying distances. 
This variant of GMD decoding, \Cref{alg:GMD}, will be crucial in our decoding algorithm for the simplex evaluation set.
As it turns out, the only change we need to make to the standard (deterministic) GMD algorithm is the setting of the weights. 
For a reference on the GMD algorithm, see section 14.3 of \cite{essentialcoding}.

\begin{algorithm}
    \caption{$\mathrm{GMD}(r, A_{out}, A_1,...,A_N)$}\label{alg:GMD}
    Write $r$ as $N$ blocks $r_1,...,r_N$ corresponding to $N$ inner codewords\\
    \For{$i \in [N]$}{
        $y_i = A_i(r_i)$\\
        $s(i) \gets $ number of erasures in $r_i$.\\
        $x(i) \gets $ number of non-erasure coordinates in which $y_i$ and $r_i$ differ\\
        $w(i) \gets d_i - 2x(i) - s(i)$\\
    }
    \For{$i \in [N]$}{
        $t \gets w(i)$\\
        \For{$j \in [N]$}{
            \eIf{$w(j) < t$}{
                $z_j \gets ?$ 
            }{
                $z_j \gets r_j$.
            }
        }
        $c \gets A_{out}(\mathbf{z})$\\
        \If{$\wt(C(\supbrak{c}{i}) - r) < dD$}{
            \KwRet c
        }
    }
    \KwRet{$\bot$}
\end{algorithm}

\begin{theorem}[Uneven GMD Decoding]\label{thm:uneven-gmd}
    Let $C_{out}$ code with block length $N$ and distance $D$. For $i \in [N]$, let $C_i$ be a code with distance $d_i$. Let $C = (C_1,...,C_N) \circ C_{out}$ be the concatenated code as in \Cref{defn:concatenated-codes}. Then, $C$ has minimum distance at least $\min_{S\subseteq [N]: |S| = D}\sum_{i \in S} d_i$, i.e. the sum of the $D$ smallest inner distances.

    Furthermore, suppose there exist decoding algorithms
    $A_{out}, A_1, A_2,..., A_N$ with corresponding time complexities $T_{out}, T_1,..., T_N,$ such that $A_{out}$ decodes $C_{out}$ from error and erasure patterns of weight less than $D$, and for each $i \in [N]$, $A_i$ decodes $C_i$ from error (and erasure) patterns of weight less than $d_i$.

    Then \Cref{alg:GMD} instantiated with the $A_{out}$, $A_1,..., A_N$ decodes the concatenated code $C$ from error (and erasure) patterns of weight less than $\min_{S\subseteq [N]: |S| = D}\sum_{i \in S} d_i$. Furthermore, the time complexity of the algorithm is $\mathrm{poly}(T_{out}, T_1,...,T_N, N, n)$.
\end{theorem}

\begin{proof}
    The claim about the distance of the concatenated code is simple. WLOG suppose $d_1 \leq d_2 \leq ... \leq d_N$ so that $\min_{S\subseteq [N]: |S| = D}\sum_{i \in S} d_i = \sum_{i = 1}^D d_i$. Let $m, m'$ be distinct messages with $m \neq m'$. Let $X \subseteq [N]$ be the set of indices where the encodings of $m$ and $m'$ differ in the outer code. Since the distance of the outer code is at least $D$, $|X| \geq D$. For each $i \in X$, the inner encoding of $i$th symbol of the outer codeword must differ in at least $d_i$ coordinates. Thus, the total number of coordinates in which the codewords differ is at least $\sum_{i \in X}d_i$, which is at least $\sum_{i = 1}^D d_i$.

    We now show the correctness of \Cref{alg:GMD}. The proof follows the combinatorial proof of Forney's algorithm given in \cite{BHKS-2023}. Let $e$ be an error pattern of weight at most $\sum_{i = 1}^Dd_i$. Let $E$ be the total number of errors and $S$ be the total number of erasures. Recall $\wt(e) = 2E + S$. Let $e(i)$ be the number of errors in the $i$th block, and $s(i)$ be the number of erasures in the $i$th block.

    Consider the situation after decoding the inner codes. Let $x(i)$ be the Hamming distance between the closest (inner) codeword and the received word in the $i$th block not counting erasures. Define $w(i) = d_i - 2x(i) - s(i)$ to be the weight of each block. Intuitively, the $w(i)$ captures the `confidence' in the $i$th block. We'll now show that there exists a threshold, $t$, such that setting blocks with weight less than $t$ to erasures allows the outer decoder to decode correctly.
        Let $A\subseteq [N]$ be the set of correctly decoded blocks. Let $B\subseteq [N]$ be the set of incorrectly decoded blocks. Note that these are unknown to the algorithm and are purely for analysis. Let $A_{\geq t} = \left\{ i \in A: w(i) \geq t \right\}$. Define $B_{\geq t}$ similarly. The number of errors remaining after erasing blocks with weight $< t$ is $|B_{\geq t}|$. The number of blocks we set to erasures is $N - |B_{\geq t}| - |A_{\geq t}|$.
    Thus, the outer decoder can decode so long as 
    $$
    2|B_{\geq t}| + N - |B_{\geq t}| - |A_{\geq t}| < D.
    $$
    I.e., 
    $$
    |B_{\geq t}| + N - D < |A_{\geq t}|.
    $$

    \begin{claim}
        There exists a threshold $t \in \{w(i): i \in [N]\}$ such that the above holds.
    \end{claim}

    If the claim holds, we can decode by trying each weight as a threshold. The rest of this proof will show that the claim is true.

    By contradiction, suppose the claim is false, i.e. 
    \begin{equation}
        |B_{\geq t}| + N - D \geq |A_{\geq t}|,
        \label{eq:threshold-contradiction}
    \end{equation}
    for every choice in $t \in \{w(i): i \in [N]\}$. The contradiction we will eventually obtain is that the weight of $e$, i.e., $2E + S$, is too large.

    As a first step, we bound the size of $|B|$. Since the inner codes are decodable as long as $2e(i) + s(i) < d_i$, we have
    $$
    \sum_{i \in [B]}2e(i) + s(i) \geq \sum_{i \in B}d_i \geq \sum_{i =1}^{|B|}d_i,
    $$
    where the second inequality holds because $d_1 \leq d_2 \leq ... \leq d_N$.
    Thus, since we have $2E + S < \sum_{i=1}^D d_i$, $|B| < D$. In particular, let $|B| = D - u$ for some positive integer $u$, and $|A|= N - D + u$.

    Let $a_1 \geq a_2 \geq ... \geq a_{N-D+u}$ be an enumeration of $A$, and. Let $b_1 \geq b_2 \geq ... \geq b_{D - u}$ be an enumeration of $B$. 

    We show that for every $r \in [u]$, $w(a_{N - D + r}) \leq w(b_r)$. By contradiction, suppose $w(a_{N - D + r}) > w(b_r)$. Then consider setting $t = w(a_{N - D +r})$, we have $|A_{\geq t}|=N - D + r$, and $|B_{\geq t}| < r$, which contradicts \Cref{eq:threshold-contradiction}. 

    We now use this relationship to bound the weight of the errors in blocks $a_{N - D + r}$ and $b_r$. For simplicity, let's abbreviate $a = a_{N - D +r}$, and $b =b_r$. Substituting the definition of the weights, we get
    \begin{align*}
        d_a - 2x(a) - s(a) &\leq  d_b - 2x(b) - s(b) \iff \\
        d_a - d_b &\leq  2x(a) + s(a) - 2x(b) - s(b)\\
    \end{align*}
    By the triangle inequality on block $b$, we have that $x(b) \geq d_b - s(b) - e(b)$. Additionally, since $a \in A$, the $a$th block was correctly decoded, thus $x(a) = e(a)$. Substituting these facts into the above and rearranging, we get
    \begin{align*}
        d_a + d_b &\leq  2e(a) + s(a) + 2e(b) + s(b)
    \end{align*}
    For the remaining blocks in $i \in B$, we know that the weight of the errors in block $i$ is at least $d_i$ since otherwise, the inner codeword would have been correctly decoded. Summing over the weights of the errors in each of these blocks, we have 
    \begin{align*}
        2E + S &\geq \sum_{r \in [u]} d_{a_{N - D + r}} + d_{b_r} + \sum_{i = u+1}^{D - u} 2e(b_{i}) + s(b_i)\\
        &\geq \sum_{r \in [u]} d_{a_{N - D + r}} + d_{b_r} + \sum_{i = u+1}^{D - u} d_{b_{i}}.
    \end{align*}
    There are $2u + D - 2u =  D$ unique inner distances in the sum, which is at least the sum of the $D$ smallest inner distances, $\sum_{i=1}^D d_i$, which is a contradiction since $2E + S$ was supposed to be less than that quantity.
\end{proof}

\begin{remark}
    Note that if the inner code can only be decoded from errors, then the theorem is still true if there are only errors in the received word (and no erasures). However, since we are setting certain blocks to erasures for the outer decoding, the outer decoder must be able to correct errors and erasures even if there are no erasures in the received word.
\end{remark}

\subsection{Decoding in two-dimensions}
\newcommand{\Coeff}{\mathrm{Coeff}}

Let $f \in \F[X, Y]$ be any bivariate polynomial of total degree at most $d$. We can write it as $$f(X, Y) = \sum_{i = 0}^d c_i(X)Y^i,$$ where $c_i$, a univariate polynomial in $X$, is the coefficient of $Y^i$.
Let $G = \N_{< \ell} \times \N_{< \ell}$ be the two dimensional $\ell \times \ell$ grid. 
View the grid as having $\ell$ columns, each of size $\ell$. 
Let $f: G \to \F$ be a codeword and $e: G \to \F$ be an error pattern. 
The received word is $r = f + e$. 
Kim and Kopparty combine the following observations to decode polynomial evaluation codes on the grid. 
Firstly, the $x$th column contains $\ell$ evaluations of the univariate polynomial $f(x, Y)$, and thus, we can decode each column using standard RS decoding. 
Secondly, for any fixed $i$ and $x$, the coefficient of $Y^i$ in $f(x, Y)$ is the evaluation of $c_i$ at $x$. 
Thus, the decodings of each of the $\ell$ columns combine to give $\ell$ evaluations of $c_i$, and we can decode for $c_i$ using standard RS decoding. 
Their algorithm is iterative, where the first iteration uses this procedure to find $c_d$, the second finds $c_{d-1}$, and so on.

We first present an alternate analysis. 
In particular, to retrieve the coefficient $c_d$, we view the codeword $f$ as the encoding of $c_d$ in some concatenated code and directly apply \Cref{alg:GMD}. 
The decoding algorithm is formally defined in \Cref{decode-simplex-2}. 
In \Cref{decode-simplex-2}, and in the rest of the paper, let $\Coeff_{Y^k}$ be a function that maps a polynomial to its $Y^k$ coefficient, and let $\mathrm{RSDecode}_{d, n}$ be the decoding algorithm for $\RS_{d, n}$ correcting error and erasure patterns of weight less than $n - d$. 
Note that $\mathrm{RSDecode}$ is the standard Berlekamp-Welch algorithm \cite{BW86}.

\begin{algorithm}
    \caption{$\mathrm{DecodeBivariate}_{d, \ell}(r, k)$}\label{decode-simplex-2}
    \eIf{$k < 0$}{
        \KwRet{[]}
    }{
        $A_{out} \gets \mathrm{RSDecode}_{d - k, \ell}$\\
        $A_x \gets \Coeff_{Y^{k}} \circ \mathrm{RSDecode}_{k, \ell - x}$ for each $x \in \{0,1,...,\ell-1\}$\\
        $c_k \gets \mathrm{GMD}(A_{out}, A_0,...,A_{\ell-1}, r)$\\
        \KwRet{$\mathrm{DecodeBivariate}_{d, \ell}(r - c_kY^k, k-1) + [c_k]$}
    }
\end{algorithm}

Recall that the distance of $\CAP{m, d, \ell}$ is $\binom{\ell - d + m - 1}{m} = |\Simplex{m, \ell - d}|$.
\begin{theorem}\label{thm:decode-bivariate}
    Let $d, \ell \in \N$ with $\ell > d$. 
    Let $f \in \CAP{2, d, \ell}$, be any codeword with $Y$-degree at most $k$ (and total degree at most $d$), $e$ be an error pattern with weight less than $\binom{\ell - d + 1}{2}$, and let $r = f + e$. 
    Write $f = \sum_{i=0}^k c_i(X)Y^i$, Then $\mathrm{DecodeBivariate}_{d, \ell}(r, k)$ (\Cref{decode-simplex-2}) returns $f$ as a list of coefficients $[c_0,...,c_{k}]$.
\end{theorem}

\begin{proof}
    By induction. Suppose the claim is true for $k - 1$. We'll show the claim for $k$.

    Let $f$ be a polynomial with $Y$-degree at most $k$. Thus, $f = \sum_{i = 0}^{k}c_i(X)Y^i$, where $c_i$ is a univariate polynomial in $X$ of degree at most $d - i$. 

    We claim that the evaluations of $f$ on $S$ is in one-to-one correspondence with the encoding of $c_k$ under the following concatenated code.

    Let $C_{out}$ be the code that encodes $c_k$ by evaluating it on $\N_{< \ell}$. For each $x \in \{0,...,\ell-1\}$, define the inner code $C_x$ as the code that encodes elements of $\alpha \in \F$ in the following way. For any $\alpha \in \F$, map it to the univariate polynomial $\alpha Y^k + \sum_{i=0}^{k-1}c_i(x)Y^i$, and evaluate this polynomial on $\N_{< \ell - x}$.  Let $C = (C_0,...,C_{\ell - 1})\circ C_{out}$.

    Let $(x, y) \in \Simplex{2, \ell}$. Then $f(x, y)$ is the $y$th element of the $x$th block in the encoding of $c_k$ under $C$. Indeed, the $x$th block of the outer code is $c_k(x)$, and the $y$th element of the encoding of $c_k(x)$ under the inner code is $c_k(x)y^k + \sum_{i=0}^{k-1}c_i(x)y^i = f(x, y)$. Furthermore, it is not hard to check that if $(x, y) \in \Simplex{2, \ell}$, then $C_x$ indeed contains the evaluation at $y$: In $C_x$, the polynomial is evaluated on $\N_{<\ell -x}$, and $y < \ell - x$ since $(x, y) \in \Simplex{2, \ell}$. This establishes the one-to-one correspondence.

    We now show that the concatenated code can be decoded to the desired distance using \Cref{alg:GMD}. $C_{out} = \RS_{d - k, \ell}$, can thus be decoded from errors and erasures up to weight $\ell - d + k$ using $\mathrm{RSDecode}_{d - k, \ell}$. For each $x \in \N_{<\ell}$, $C_x$ is a subset of $\RS_{k, \ell - x}$, and thus one can decode for the polynomial $\alpha Y^k + \sum_{i=0}^{k-1}c_i(x)Y^i$ using $\mathrm{RSDecode}_{k, \ell - x}$ for error and erasure pattens with weight less than $\max(\ell - x -k, 0)$. One can then extract $\alpha$ as the coefficient of $Y^k$.

    The distance of the concatenated code is then the sum of the $\ell - d + k$ smallest inner distances, which is the distances of the last $\ell - d + k$ inner codes. Note that the distance of $C_x$ is $0$ for each $x \in \{\ell - k, \ell - k + 1,...,\ell - 1\}$ (the last $k$ codes). For $x = \ell - k - i$, the distance is $i$. Thus, the sum of the $\ell - d + k$ smallest inner distances is just $1 + 2 + ... + \ell - d = \binom{\ell - d + 1}{2}$. Since the weight of $e$ is less than $\binom{\ell - d + 1}{2}$, the GMD algorithm \Cref{alg:GMD} correctly decodes $c_k$.

     Then, $r - c_kY^k = (f - c_kY^k) + e$. This is now a received word with $Y$-degree at most $k-1$. By the inductive hypothesis, the recursive call correctly returns $[c_{0},...,c_{k-1}]$, and the final return value of the function is the full list of coefficients $[c_0,c_1,...,c_{k}]$ as required.
\end{proof}

\begin{corollary}
    $\CAP{2, d, \ell}$ can be decoded from error patterns of weight less than the minimum distance.
\end{corollary}

\begin{proof}
    For any $f \in \CAP{2, d, \ell}$, the total degree of $f$ is at most $d$. Therefore, the $Y$-degree of $f$ is at most $d$, and by \Cref{thm:decode-bivariate},
    $\mathrm{DecodeBivariate}_{d, \ell}(r, d)$ correctly decodes so long as the weight of the error pattern is less than the minimum distance.
\end{proof}

\subsection{Decoding on the \textit{m}-dimensional simplex}
\newcommand{\prev}{\mathrm{prev}}

We now generalize to $m$ dimensions. For brevity, for any $\mathbf{v} \in \N^{m-1}$, let $\mathbf{X}^\mathbf{v} = X_1^{e_1}X_2^{e_2}\cdot ... \cdot X_{m-1}^{e_{m-1}}$, and write $\deg(v) = v_1 + v_2 + ... + v_{m-1}$. Write an $m$-variate polynomial $f$ as an element of $(\F[X_m])[X_1,...,X_{m-1}]$. In particular, write 
$$
f = \sum_{\mathbf{v}: \deg(\mathbf{v})\leq d}c_{\mathbf{v}}(X_m)\mathbf{X}^{\mathbf{v}},
$$
where $c_{\mathbf{v}}$ is a univariate polynomial of degree at most $d - \deg(\mathbf{v})$. Let $\prec$ be an ordering over $\N^{m-1}$, corresponding to the graded lexicographical ordering of monomials. That is, $\mathbf{v} \prec \mathbf{v'}$ iff either $\deg(\mathbf{v}) < \deg(\mathbf{v'})$ or $\deg(\mathbf{v}) = \deg(\mathbf{v'})$ and $\mathbf{v}$ comes before $\mathbf{v'}$ in lexicographical order. For example,
$$
[0,0] \prec [0,1] \prec [1, 0] \prec [0,2] \prec [1, 1] \prec [2, 0].
$$
Notice that $\prec$ is a total order. Our algorithm will recover $c_{\mathbf{v}}$ in descending order with respect to $\prec$. Define the function $\mathrm{\prev} : \N^{m-1} \to \N^{m-1} \cup \{\bot \}$, where $\prev(\mathbf{v})$ returns the direct predecessor of $\mathbf{v}$ with respect to the order $\prec$. I.e. $\prev(\mathbf{v}) = \max(\{\mathbf{v}' : \mathbf{v}' \prec \mathbf{v} \})$. As an edge case, define $\prev([0,0,...,0]) = \bot$ to specify an invalid monomial. The algorithm for decoding on the $m$-variable simplex is described in \Cref{decode-simplex}.

\begin{algorithm}
    \caption{$\mathrm{DecodeSimplex}_{m, d, \ell}(r, \mathbf{v})$}\label{decode-simplex}
    \eIf{$\mathbf{v}= \bot$}{
        \KwRet{[]}
    }{
        $k \gets \deg(\mathbf{v})$\\
        $A_{out} \gets \mathrm{RSDecode}(d - k, \ell)$\\
        \For{ $i \in \{0,...,\ell - 1\}$ }{
            $A_{i} \gets \Coeff_{\mathbf{X^v}} \circ \mathrm{DecodeSimplex}_{m-1, k, \ell - i}(\cdot, [k,0,...,0])$\\
        }
        $c_{\mathbf{v}} \gets \mathrm{GMD}(A_{out}, A_0,...,A_{\ell -1}, r)$\\
        $\mathbf{v'}=\prev(\mathbf{v})$\\
        \KwRet{$\mathrm{DecodeSimplex}_{m, d, \ell}(r - c_{\mathbf{v}}\mathbf{X^v}, \mathbf{v'}) + [c_{\mathbf{v}}]$}
    }
\end{algorithm}

\begin{theorem}\label{thm:decode-mvariate-simplex}
    Let $m, d, \ell \in \N$, such that $\ell > d$, and $m \geq 1$. 
    Let $f \in \CAP{m, d, \ell}$, and write $f = \sum_{\mathbf{u}: \deg(\mathbf{u})\leq d}c_{\mathbf{u}}(X_m)\mathbf{X}^{\mathbf{u}}$. 
    Let $\mathbf{v}$ be an exponent vector of degree at most $d$ and suppose for all $\mathbf{w}$ with $\mathbf{v}\prec \mathbf{w}$, $c_{\mathbf{w}} = 0$. 
    Let $e$ be an error pattern with weight less than $|\Simplex{m, \ell - d}|$, and let $r = f + e$. 
    Then $\mathrm{DecodeSimplex}_{m, d, \ell}(r, \mathbf{v})$ (\Cref{decode-simplex}) returns $f$ as a list of coefficients $(c_{\mathbf{v}})_{\mathbf{v}: \deg(\mathbf{v}) \leq d}$.
\end{theorem}

\begin{proof}
    Let $\ell, d \in \N$ with $\ell > d$. 
    For notational simplicity, suppose the code is the CAP code on the standard simplex of side length $\ell$, $\Simplex{m, ell}$.
    The proof works exactly the same on $\Simplex{m, A}$ for any $A \subset F$ with $|A| = \ell$ by replacing elements of $A^m$ with the vector of their indices in $\left\{ 0,1,...,\ell - 1 \right\}$.

    By induction on $m$. 

    For the base case $m = 1$, observe that $E_{m, d, \Simplex{m, \ell}}$ is a Reed-Solomon code, so $\mathrm{DecodeSimplex}_{1, d, \ell}$ is $\mathrm{RSDecode}_{d, \ell}$.

    Now, suppose the theorem is true for $\mathrm{DecodeSimplex}_{m-1, d', \ell'}$ for any $d'$ and $\ell'$ with $\ell' > d'$. 
    We'll show the theorem for $\mathrm{DecodeSimplex}_{m, d, \ell}$. 
    The proof is similar to that of \Cref{thm:decode-bivariate}. 
    By induction on $\mathbf{v}$ with respect to $\prec$. 

    For the inductive hypothesis, suppose that for any polynomial $g$ where the maximum monomial with a non-zero coefficient is $\mathbf{X}^{\prev(\mathbf{v})}$, and error pattern $e$ with weight less than $|\Simplex{m, \ell - d}|$, $\mathrm{DecodeSimplex}_{m, d, \ell}(g + e, \prev(\mathbf{v}))$ returns $g$.

    Let $f$ be any polynomial where the maximum monomial with a non-zero coefficient is $\mathbf{X}^{\mathbf{v}}$, and let $e$ be an error pattern of weight less than $|\Simplex{m, \ell - d}|$, we'll show  $\mathrm{DecodeSimplex}_{m, d, \ell}(f + e, \prev(\mathbf{v}))$ returns $f$. Let $k = \deg(\mathbf{v})$. We first show that $f$ is also the encoding of $c_{\mathbf{v}}$ under the following concatenated code. 

    Let $C_{out}$ be the code that encodes the univariate polynomial $c_{\mathbf{v }}$ by evaluating it on $\N_{< \ell}$. 
    For $i \in \N_{< \ell}$, define the inner code $C_{i}$ to be be the code that encodes symbols $\alpha \in \F$ as a $m-1$ variate polynomial 
    $$f_{\alpha}(X_1,...,X_{m-1}) = \alpha \mathbf{X}^{\mathbf{v}} + \sum_{\mathbf{u}: \mathbf{u} \prec \mathbf{v}}c_{\mathbf{u}}(x_m)\mathbf{X^{u}},$$ 
    and evaluates it on $\Simplex{m-1, \ell - i}.$
    Consider the code $C = (C_0,C_1, \ldots ,C_{\ell - 1}) \circ C_{out}$.

    We claim that the evaluation of $f$ on $\Simplex{m, \ell}$ is the encoding of $c_{\mathbf{v}}$ under $C$. 
    Let $(x_1,...,x_{m - 1}, i) \in \Simplex{m, \ell}$. 
    Then, this corresponds to the element indexed by $(x_1,...,x_{m-1})$ in the $i$th block. 
    Since the $i$th element in the outer encoding of $c_{\mathbf{v}}$ is $c_{\mathbf{v}}(i)$, and the coordinate of the $i$th inner code indexed by $(x_1,...,x_{m-1})$ is the evaluation $f_{c_{\mathbf{i}}}(x_1,...,x_{m-1}) = f(x_1,...,x_m)$. 
    Since $x_1 + x_2 + ... + x_{m-1} + i < \ell$, we have $x_1 + x_2 + ... + x_{m-1} < \ell - i$, so $(x_1,...,x_{m-1}) \in \Simplex{m, \ell - i}$ and hence is indeed an evaluation point in $C_{i}$. 

    The degree of $c_{\mathbf{v}}$ is at most $d - k$, thus $C_{out} = RS(d-k, \ell)$, and can be decoded from error and erasure patterns with weight less than $\ell - d + k$. 
    For each $i \in \N_{< \ell}$, $C_i$ is a subcode of $E_{m-1, k, \Simplex{m-1, \ell - i}}$ since $f_\alpha$ has degree at most $\deg(\mathbf{v}) = k$. 
    Thus, for any $\alpha$, $f_\alpha$ can be decoded from error and erasure patterns with weight less than $|\Simplex{m-1, \ell - i - k}|$ using the (outer) inductive hypothesis. 
    To retrieve $\alpha$, we just extract the coefficient of $\mathbf{X^v}$.

    Instantiating the GMD decoder (\Cref{alg:GMD}) with these outer and inner decoders, we get that we can decode so long as the number of errors is less than the sum of the $\ell - d + k$ smallest inner distances. 
    Note that for $x \in \{\ell - k, \ell - k + 1,..., \ell - 1\}$, the last $k$ codes, the distance is zero. 
    For $x = \ell - k - i$, the distance is $|\Simplex{m-1, i}|$. 
    Thus, the sum of the $\ell - d + k$ smallest inner distance is $\sum_{i = 1}^{\ell - d} |\Simplex{m-1, i}|= |\Simplex{m, \ell - d}|$. 
    Since the weight of $e$ is less than $|\Simplex{m, \ell - d}|$, GMD instantiated with these outer and inner decoders correctly decodes $c_{\mathbf{v}}$.

    Then, $r - c_{\mathbf{v}}Y^k = (f - c_{\mathbf{v}}Y^k) + e$. 
    Notice that this is now a received word where the maximum monomial present with respect to $\prec$ is at most $\prev(\mathbf{v})$. 
    By the inductive hypothesis, the recursive call correctly returns $[c_{\mathbf{u}}]_{\mathbf{u} \prec \mathbf{v}}$, and the final return value of the function is the full list of coefficients.
\end{proof}

\begin{corollary}
    $\CAP{m, d, \ell}$ can be decoded from error patterns of weight less than the minimum distance.
\end{corollary}
\begin{proof}
    Note that for any $f \in \CAP{m, d, \ell}$, the total degree is at most $d$. 
    Viewing $f \in (\F[X_m])[X_1,...,X_{m-1}]$, the greatest monomial with respect to $\prec$ is $X_1^d$, corresponding to the exponent vector $[d,0,0,...,0]$. 
    Thus, by \Cref{thm:decode-mvariate-simplex}, $\mathrm{DecodeSimplex}$ with the $\mathbf{v}$ parameter as $[d, 0,...,0]$ decodes any received word so long as the weight of the error patten is less than the minimum distance.
\end{proof}

\section{Unique Decoding of GAP codes}\label{sec:decoding-hyperplane-intersection-codes}
This section gives efficient algorithms for decoding GAP codes \autoref{thm:code-construction-geometric} up to half their minimum distance. 
At a high level, for our decoding algorithms, we essentially view these codes as a specific instance of decoding concatenated codes where the outer code is a Reed-Solomon code, and the inner codes are GAP codes in a smaller dimension, up to some (crucial) technical differences. 
We massage this point of view sufficiently to be in a position to invoke the framework of generalized minimum distance (GMD) decoding for concatenated codes of Forney \cite{forneyGeneralizedMinimumDistance1966}. 
While this appears to be a natural approach, we note that apriori, it is quite unclear whether this strategy can decode up to half the minimum distance of the code. 
For instance, we do not know if this strategy can give a simple unique decoding algorithm for decoding polynomial evaluation codes on arbitrary product sets (a problem that was solved only relatively recently by Kim and Kopparty \cite{KimK2017}) or our second construction of such codes in \autoref{thm:code-construction-simplexes}, both of which, though relying on GMD decoding seem to require further ideas and a more complex algorithm overall. 
Perhaps this innate simplicity of the decoding algorithm is further evidence that the codes in \autoref{thm:code-construction-geometric} are quite natural and merit further investigation.

In this section, we work with concatenated codes where the outer code is a Reed-Solomon code, and the inner codes are all polynomial evaluation codes with the same block length and minimum distance, with potentially different evaluation points. 
The Reed-Solomon codes in the outer codes here are in a slightly unusual setting, and we formally define them now.  
\begin{definition}[Reed-Solomon codes over a polynomial ring]\label{defn:RS-over-poly-rings}
Let $\F$ be a finite field, $\vecX = (X_1,  \ldots, X_{m-1})$ be an $(m-1)$-tuple of variables, $S \subseteq \F[\vecX]$ be a set of $t$ affine linear forms and $d \in \N$ be a natural number. Then, Reed-Solomon codes $\mathcal{RS}(m,t,d, S)$ are defined as follows.  

The message space of the code is identified with the set of $m$-variate polynomials $f(\vecX,Y)$ of total degree at most $d$. To encode any such message $f(\vecX,Y)$, we view it as a degree $d$ univariate in $Y$ with coefficients in the field $\F(\vecX)$ and the codeword associated to $f$ equals $\left(f(\vecX, L_1(\vecX)), \ldots, f(\vecX, L_t(\vecX)) \right)$, i.e. the \emph{evaluation} of $f(\vecX,Y) \in (\F(\vecX))[Y]$ on the $t$ inputs $L_1(\vecX), \ldots, L_t(\vecX) \in S$.  
\end{definition}
The following observation summarises some of the basic properties of this code. 
\begin{observation}\label{obs:props-of-RS-over-polyring}
Let $\mathcal{RS}(m,t,d,S)$ be the codes defined in \autoref{defn:RS-over-poly-rings}. Then, the following are true. 
\begin{itemize}
\item The distance of the code is $(t-d)$.
\item For any message polynomial $f(\vecX,Y)$ of total degree at most $d$, every coordinate of its encoding $\left(f(\vecX, L_1(\vecX)), \ldots, f(\vecX, L_t(\vecX)) \right)$ is an $(m-1)$-variate polynomial of degree $d$ in $\F[\vecX]$. Thus, the alphabet of the code can be naturally identified with ${\F}^{\binom{d + m-1}{m-1}}$. 
\end{itemize}
\end{observation}
Reed-Solomon codes are very well studied from the point of view of decoding algorithms and, in particular, are known to be efficiently decodable up to half their minimum distance via the well-known decoding algorithm of Berlekamp \& Welch \cite{BW86}. 
However, typically, we study Reed-Solomon codes where the underlying alphabet is a finite field, in contrast to the Reed-Solomon codes in \autoref{defn:RS-over-poly-rings} where the underlying alphabet is the set of $(m-1)$-variate polynomials of degree $d$ over the field $\F$. 
The standard Berlekamp-Welch algorithm can be naturally extended to this setting with two additional technical ingredients. 
The first ingredient is an efficient algorithm that, when given a rank deficient $N \times N$ matrix $M$ whose entries are $m$-variate polynomials in $\F[\vecX]$ of degree at most $D$ (for some parameter $D$) outputs a non-zero vector with entries in $\F[\vecX]$ that is in the kernel of $M$. 
Lemma $3$ in \cite{BHKS-2023} addresses precisely this problem and shows that there is a deterministic algorithm for this problem that runs in time $\poly(N^m, D^m)$. 
Moreover, the degree of every coordinate of the output vector of the algorithm is at most $ND$. 
The second technical ingredient is an efficient algorithm that takes any two $m$-variate polynomials $F$ and $G$ of degree at most $D$ as input and outputs the quotient $G/F$ if $F$ divides $G$ and a bot $(\bot)$ otherwise. 
One of the ways of doing this is to fix any valid monomial ordering (e.g., the graded lexicographic order) and just do a standard long division. 
This gives a deterministic algorithm for this problem with time complexity $\poly(D^m)$.   

With these two ingredients in place, the Berlekamp-Welch decoder for Reed-Solomon codes naturally extends to the codes in \autoref{defn:RS-over-poly-rings}. 
For our applications here, we need to decode Reed-Solomon codes from errors and erasures, which the Berlekamp-Welch algorithm handles without any further issues. 

\begin{theorem}[Berlekamp-Welch \cite{BW86}]\label{thm:BW}
Let $\F$ be any finite field. There is a deterministic algorithm, that for all $d, m, t \in \N$ and subsets $S \subseteq \F[\vecX]$ of size $t$ of affine linear forms, decodes the code $\mathcal{RS}(m,t,d,S)$ in time $\poly(t^m, d^m, \log |\F|)$ from errors and erasures with weight less than the distance of the code.
\end{theorem}

We start with the bivariate case and then rely on induction on the number of variables to lift these ideas to the multivariate case. 

\subsection{Decoding in two dimensions}
Before moving on to the decoding algorithm, we first view $\GAP{m, d, t}$ as a concatenated code. 
We begin with the $m = 2$ case, where the details are cleaner, and which will also serve as the base case for our induction based argument. 

We have the following set up. 
Let $\mathcal{H} = \{\ell_1, \ell_2, \ldots, \ell_t\}$ be a set of $t$ lines in two dimensions that are in general position, i.e., every pair of these lines intersect at a point, and no three of them intersect anywhere. 
For every $i$, let $\ell_i$ be parameterized as \[
\ell_i := \{(x, a_ix+ b_i) : x \in \F \} ,
\]
where $a_i, b_i$ elements of $\F$. 
Let $T$ be the set of pairwise intersections of lines in $\mathcal{H}$.
We will decode $\GAP{2, d, t}^{\mathcal{H}}$. Recall that $\GAP{2, d, t}^{\mathcal{H}}$ has distance $\binom{t-d}{2}$.

\paragraph*{An intermediate concatenated code: }
For describing our decoding algorithm, we will rely on another family of codes that are closely related to $\GAP{2, d, t}^{\mathcal{H}}$. These intermediate codes are denoted by $\mathcal{I}_{2}^\mathcal{H}$ and are defined as follows. 

\begin{definition}\label{defn:intermediate-code-2D}
The message space of $\mathcal{I}_{2}^\mathcal{H}$ is the space of bivariate polynomials of total degree at most $d$ in $\F[X,Y]$. The block length of the code is $t(t-1)$ and the alphabet is $\F$. We index the coordinates of the codewords by pairs $(i,j)$, where $i, j \in \{1, 2, \ldots, t\}$, $i \neq j$, and the $(i,j)$th coordinate of the codeword corresponding to a message polynomial $f$ is the evaluation of $f$ at the intersection of the lines $\ell_i$ and $\ell_j$. 
\end{definition}

From the above definition, it follows that the main difference between the encoding of $f$ under the codes $\mathcal{I}_{2}^\mathcal{H}$ and $\GAP{2, d, t}^{\mathcal{H}}$ is that in the former, the evaluation of $f$ on the intersection point of any two distinct lines $\ell_i$ and $\ell_j$ appears on two separate coordinates, namely, $(i,j)$ and $(j, i)$, whereas in the latter, this evaluation appears exactly once, at a coordinate indexed by the set $\{i,j\}$. Moreover, given the encoding of any message $f$ under one of these codes, we can efficiently compute its encoding as a codeword of the other code.

The following lemma summarises some more interesting properties of this new code $\mathcal{I}_{2}^\mathcal{H}$ that will be useful for our algorithm. Again, we follow the notation set up earlier in this section.
\begin{lemma}\label{lem:props-intermediate-code-2D}
Let $C_{out}$ be the code $\mathcal{RS}(2,t,d,S)$, where $S$ is the set $\{a_iX + b_i : i \in \{1, 2, \ldots, t\}\}$ of $t$ distinct affine forms. 

For $i \in \{1, 2, \ldots, t\}$, let $C_{i}$ be the Reed-Solomon code of dimension $(d+1)$ and block length $(t-1)$ with the evaluation points being the set $S_i = \{\alpha_j : (\alpha_j, a_i\alpha_j + b_i) \in \ell_i \cap \ell_j \text{ for some } j \neq i\}$. 

Then, the following are true. 
\begin{itemize}
\item \label{item:props-intermediate-cdoe-2D-concat} $\mathcal{I}_{2}^\mathcal{H}$ can be viewed as the code obtained by the concatenation of a code $C_{out}$ and $\mathcal{C}_{in} = (C_{1}, \ldots, C_{t})$. 
\item \label{item:props-intermediate-cdoe-2D-dist} $\delta(\mathcal{I}_{2}^\mathcal{H}) = \delta(\GAP{2, d, t}^{\mathcal{H}}) = \delta(C_{out}) \cdot \delta(C_{i})$, for every $i \in \{1, 2, \ldots, t\}$.  

\end{itemize}
\end{lemma}
We now sketch the proof of the lemma.
\begin{proof}[Proof Sketch]
The first item immediately follows from the definitions of the codes involved. 

For the second item, we note that $\mathcal{I}_{2}^\mathcal{H}$, as defined,  is a linear code, and hence its distance equals the minimum of the weights of non-zero codewords in it. As discussed earlier in this section, there is a bijection between  $\mathcal{I}_{2}^\mathcal{H}$ and $\GAP{2, d, t}^{\mathcal{H}}$, where every codeword of the former is obtained by repeating twice every symbol in a codeword of the latter (up to an appropriate indexing of coordinates). Thus, the Hamming weight of any codeword of minimum weight in $\mathcal{I}_{2}^\mathcal{H}$ is twice that of  $\GAP{2, d, t}^{\mathcal{H}}$, and hence equals $2\cdot \binom{t-d}{2} = (t-d)(t-d-1)$. Since the block length of $\mathcal{I}_{2}^\mathcal{H}$ is also twice the block length of $\GAP{2, d, t}^{\mathcal{H}}$, we get that their fractional distances must be the same. The last equality immediately follows from the first item of the lemma and by observing that the distance of $C_{out}$, which is really a Reed-Solomon code over the field $\F(X)$ equals $(t-d)$. The distance of every inner code $C_{i}$ equals $(t-1 -d)$, and hence their product equals $(t-d)(t-d-1)$. 
\end{proof}
From \autoref{lem:props-intermediate-code-2D}, we have that $\mathcal{I}_{2}^\mathcal{H}$ is a concatenated code. Thus, from  \autoref{thm:GMD-decoding} and \autoref{thm:BW}, we have the following corollary. 

\begin{corollary}\label{cor:unique-decode-intermediate-2D}
There is a deterministic algorithm, denoted by $\mathrm{DecodeIntermediateBivariate}$, that decodes the code $\mathcal{I}_2^\mathcal{H}$ from error patterns of weight less than the minimum distance in time  $\poly(t,d, \log |\F|)$.

\end{corollary}

\paragraph*{The decoding algorithm : }We now describe our decoding algorithm for two dimensional $\GAP{}$ codes. 

\begin{minipage}{\algwidth}
\begin{algorithm}[H]
    \caption{$\mathrm{DecodeBivariateGeometric}\left(t, d, \mathcal{H}, \vecr\right)$}
    \label{alg:decoder-hyperplane-2D}
    \KwIn{$t, d \in \N$ with $t \geq d$, a set $\mathcal{H} = \{\ell_1, \ldots, \ell_t\}$ of lines in $\F^2$ in general position and received word $\vecr \in \F^{\binom{t}{2}}$}
    \KwOut{The unique polynomial $f \in \F[X,Y]$ of degree at most $d$ such that $\Delta\left(\GAP{2, d, t}^{\mathcal{H}}(f),\vecr\right) < \binom{t-d}{2}/2$ or, $\bot$ if  no such polynomial exists.}
    \For{$i\in \{1, \ldots, t\}$}{
        \For{$j \in \{1, \ldots, t\}\setminus \{i\}$}{
        	$\tilde{\vecr}(i,j) \gets \vecr(\{i,j\})$ \\
        }
        }
 Output $\mathrm{DecodeIntermediateBivariate}(t,d,\mathcal{H},\tilde{\vecr})$ (\autoref{cor:unique-decode-intermediate-2D}). 
\end{algorithm}
 \end{minipage}

Given the discussion leading up to this point, the analysis of \autoref{alg:decoder-hyperplane-2D} follows almost immediately and is summarised in the following theorem. 

\begin{theorem}\label{thm:algo-2D-hperplane-decoder-analysis}
    Let $t, d, \in \N$ with $t \geq d$, and $\mathcal{H} = \{\ell_1, \ldots, \ell_t\}$ be a set of $t$ lines in $\F \times \F$ in general position. 
    Then \autoref{alg:decoder-hyperplane-2D} decodes $\GAP{2, d, t}^{\mathcal{H}}$ from error patterns with weight less than the minimum distance. I.e., when the number of errors is less than half the minimum distance.
    Moreover, the time complexity of  \autoref{alg:decoder-hyperplane-2D} is at most $\poly(t,d, |\F|)$. 
\end{theorem}

\begin{proof}
The bound on the time complexity follows immediately from the observation that the construction of the word $\tilde{\vecr}$ from the received word $\vecr$ can be done in $O(t^2)$ time and the bound on the time complexity of the unique decoding algorithm for code $\mathcal{I}_{2}^\mathcal{H}$ in \autoref{cor:unique-decode-intermediate-2D}.

We now argue the correctness of the algorithm. Given the received word $\vecr$, the algorithm first constructs the received word $\tilde{\vecr}$ for the intermediate code $\mathcal{I}_{2}^\mathcal{H}$ and then invokes the unique decoding algorithm for the code $\mathcal{I}_{2}^\mathcal{H}$ given in \autoref{cor:unique-decode-intermediate-2D} on the received word $\tilde{\vecr}$. Thus, the following claim, together with the correctness of the unique decoding algorithm for $\mathcal{I}_{2}^\mathcal{H}$ in  \autoref{cor:unique-decode-intermediate-2D} implies the correctness of \autoref{alg:decoder-hyperplane-2D}.

\begin{claim}
The received word $\vecr$ is within half the minimum distance of the code $\GAP{2, d, t}^{\mathcal{H}}$ if and only if the word $\tilde{\vecr}$ is within half the minimum distance of the code $\mathcal{I}_{2}^\mathcal{H}$.  
\end{claim}
\begin{proof}[Proof of claim]
Let $\vecc$ be the codeword of the code $\GAP{2, d, t}^{\mathcal{H}}$ closest to the received word $\vecr$ and let $\tilde{\vecc} \in \F^{t(t-1)}$ be defined as 
\[
\forall i\neq j \in \{1, \ldots, t\}, \quad \tilde{\vecc}(i,j) = \tilde{\vecc}(j,i) = \vecc(\{i,j\}) . 
\]
From the definition of the code $\mathcal{I}_{2}^\mathcal{H}$, it follows that $\tilde{\vecc}$ is a codeword in $\mathcal{I}_{2}^\mathcal{H}$. In fact, $\vecc$ and $\tilde{\vecc}$ are encodings of the same message polynomial. Moreover, from the definitions of $\tilde{\vecr}$ and $\tilde{\vecc}$, it follows that $\Delta(\tilde{\vecc}, \tilde{\vecr}) = 2\cdot \Delta(\vecc, \vecr)$. Using this bound and the fact that the block length of $\mathcal{I}_{2}^\mathcal{H}$ is twice that of $\GAP{2, d, t}^{\mathcal{H}}$, we get that the fractional Hamming distances, $\delta(\tilde{\vecc}, \tilde{\vecr})$ and  $\delta(\vecc, \vecr)$ must be equal to each other. The claim now follows immediately from \autoref{lem:props-intermediate-code-2D}, which shows that the fractional minimum distances of these codes are equal.
\end{proof}
\end{proof}

\subsection{Decoding in higher dimensions}
In this section, we extend the ideas in the previous section from the bivariate to the multivariate case. Our technical argument is based on an induction on the number of variables. We start by setting up the notation. 
We have a set $\mathcal{H}$ of hyperplanes $\mathcal{H} = \{H_1, H_2, \ldots, H_t \}$ in $\F^m$ that are in general position, i.e., every subset of size $m$ of these hyperplanes intersect at a point, and no $(m+1)$ of them intersect anywhere. 
Without loss of generality (up to an invertible change of basis), we can assume that the hyperplanes can be expressed  as 
\[
H_i = \left\{(\veca, L_i(\veca)) : \veca \in \F^{m-1} \right\} ,
\]
where $L_i(\vecX) \in \F[\vecX]$ is an affine linear form on variables $X_1, X_2, \ldots, X_{m-1}$. 

The set of evaluation points is the set of $m$-wise intersection of these hyperplanes, and hence the corresponding polynomial evaluation code $\GAP{m, d, t}^{\mathcal{H}}$ has block length $\binom{t}{m}$, and from \autoref{thm:code-construction-geometric}, we have that its distance is $\binom{t-d}{m}$. This section aims to efficiently decode these codes from less than $\binom{t-d}{m}/2$ errors. As in the case of two dimensions, once again, we relate this code to an intermediate code $\mathcal{I}_{m}^\mathcal{H}$ of equal fractional distance and observe that 
$\mathcal{I}_{m}^\mathcal{H}$ is obtained by concatenating a Reed-Solomon code with inner codes that are all instances of $\GAP{m-1, d, t-1}$. These inner codes can be efficiently decoded from half their minimum distance via induction, and we can invoke \autoref{thm:BW} and \autoref{thm:GMD-decoding} to get our decoding algorithm for $\GAP{m, d, t}^{\mathcal{H}}$.

\paragraph*{An intermediate concatenated code: }We start by defining the intermediate code. Throughout this section, $\vecX$ denotes the $(m-1)$-tuple $(X_1, X_2, \ldots, X_{m-1})$ of variables. 

\begin{definition}\label{defn:intermediate-code-mD}
The message space of $\mathcal{I}_{m}^\mathcal{H}$ is the space of $m$-variate polynomials of total degree at most $d$ in $\F[\vecX,Y]$. The block length of the code is $m\cdot \binom{t}{m}$ and the alphabet is $\F$. We index the coordinates of a codeword of this code by pairs $(i,\vecj)$, where $i \in \{1, 2, \ldots, t\}$ and $\vecj$ is a subset of size $(m-1)$ of $\{1, 2, \ldots, t\}\setminus \{i\}$. The $(i,\vecj)$th coordinate of the codeword corresponding to a message polynomial $f$ is the evaluation of $f$ at the intersection of $m$ hyperplanes $\{H_i\} \cup \{H_{\ell} : \ell \in \vecj\}$. 
\end{definition}
Once again, it follows that code words of $\mathcal{I}_{m}^\mathcal{H}$ and $\GAP{m, d, t}^{\mathcal{H}}$ are in a natural bijection, and one can go from a codeword of the latter to the former by repeating every coordinate exactly $m$ times. The following analog of \autoref{lem:props-intermediate-code-2D} follows almost immediately from the discussion so far.  
 
\begin{lemma}\label{lem:props-intermediate-code-mD}
Let $C_{out}$ be the code $\mathrm{RS}(m,t,d,S)$ from \autoref{defn:RS-over-poly-rings} where $S$ is the set $\{L_1(\vecX), \ldots, L_t(\vecX)\}$ of $t$-distinct affine forms and $H_1, H_2, \ldots, H_t$ are hyperplanes in general position such that for every $i$,  $H_i = \{(\veca, L_i(\veca)) : \veca \in \F^{m-1}\}$.

For $i \in \{1, 2, \ldots, t\}$, let $C_{i} = \GAP{m-1, d, t-1}^{\mathcal{H}_i}$, where ${\mathcal{H}_i} = \{H_{j} \cap H_{i} : j \in \{1, 2, \ldots, t\}\setminus \{i\}\}$.

Then, the following are true. 
\begin{itemize}
\item \label{item:props-intermediate-cdoe-mD-concat} $\mathcal{I}_{m}^\mathcal{H}$ is the code obtained by the concatenation of  $C_{out}$ and $\mathcal{C}_{in} = (C_{1}, \ldots, C_{t})$.
\item \label{item:props-intermediate-cdoe-mD-dist} $\delta(\mathcal{I}_{m}^\mathcal{H}) = \delta(\GAP{m, d, t}^{\mathcal{H}}) = \delta(C_{out}) \cdot \delta(C_{i})$, for every $i \in \{1, 2, \ldots, t\}$.  

\end{itemize}
\end{lemma}
From \autoref{lem:props-intermediate-code-mD}, we note that to invoke \autoref{thm:GMD-decoding} to decode $\mathcal{I}_{m}^\mathcal{H}$, we need the algorithm from \autoref{thm:BW} and a unique decoding algorithm for $\GAP{m-1, d, t-1}$.  
For the latter, we note that if $\mathcal{H} = \{H_1, H_2, \ldots, H_t\}$ is a set of hyperplanes in $\F^m$ in general position, then for every $i \in \{1, 2, \ldots, t\}$, $\mathcal{H}_i = \{H_j \cap H_i : j \neq i\}$ are hyperplanes in $\F^{m-1}$ in general position. 
Thus, the code $C_{i}$ is indeed a GAP code for polynomial in one fewer variable and by the induction hypothesis, such GAP codes are efficiently decodable up to half their minimum distance. 
The above discussion, together with \autoref{thm:GMD-decoding}, immediately implies the following. 

\begin{lemma}\label{lem:unique-decode-intermediate-mD} 
   There is an algorithm, denoted by $\mathrm{DecodeIntermediateMultivariate}$, that decodes $\mathcal{I}_m^\mathcal{H}$ from error patterns of weight less than the minimum distance. 
    I.e., when the number of errors is less than half the minimum distance.
    Moreover, the algorithm is deterministic and on every input, makes at most $t$ function calls to a unique decoder for the polynomial evaluation code $\GAP{m-1, d, t-1}$ (\autoref{alg:decoder-hyperplane-mD}), at most $t$ function calls to the Berlekamp-Welch decoder for Reed-Solomon codes in  \autoref{thm:BW}, and performs at most $\poly(t^m,\log |\F|)$ additional computation.
\end{lemma}

\paragraph*{The decoding algorithm : }We now describe our decoding algorithm for $\GAP{m, d, t}^{\mathcal{H}}$ . 

\begin{minipage}{\algwidth}
\begin{algorithm}[H]
    \caption{$\mathrm{DecodeMultivariateGeometric}\left(m, t, d, \mathcal{H}, \vecr\right)$}
    \label{alg:decoder-hyperplane-mD}
    \KwIn{$m, t, d \in \N$ with $t \geq d$, a set $\mathcal{H} = \{H_1, \ldots, H_t\}$ of hyperplanes in $\F^m$ in general position and received word $\vecr \in \F^{{\binom{t}{m}}}$}
    \KwOut{The unique $m$-variate polynomial  $f \in \F[\vecX,Y]$ of degree at most $d$ such that $\Delta\left(\GAP{m, d, t}^{\mathcal{H}}(f),\vecr\right) < \binom{t-d}{m}/2$ or, $\bot$ if  no such polynomial exists.}
    \If{$m = 2$}{
    Output $\mathrm{DecodeBivariateGeometric}\left(t, d, \mathcal{H}, \vecr\right)$ (\autoref{alg:decoder-hyperplane-2D}). 
    }
    \Else{
    \For{$i\in \{1, \ldots, t\}$}{
        \For{$\vecj \subseteq \{1, \ldots, t\}\setminus \{i\}, |\vecj| = m-1$}{
        	$\tilde{\vecr}(i,\vecj) \gets \vecr\left(\{i\} \cup \vecj\right)$ \\
        }
        }
        Output $\mathrm{DecodeIntermediateMultivariate}\left(m,d,t,\mathcal{H},\tilde{\vecr}\right)$ (\autoref{lem:unique-decode-intermediate-mD}).
        } 
\end{algorithm}
 \end{minipage}
 \newline
  Note that the unique decoding algorithm for $\mathcal{I}_{m}^\mathcal{H}$ in \autoref{lem:unique-decode-intermediate-mD} makes function calls to  \autoref{alg:decoder-hyperplane-mD}, but for the number of variables being $(m-1)$, which is handled recursively. The following theorem sums up the analysis of \autoref{alg:decoder-hyperplane-mD}. 

\begin{theorem}\label{thm:algo-mD-hperplane-decoder-analysis}
    Let $m, t, d \in \N$ with $t \geq d$, and $\mathcal{H} = \{H_1, \ldots, H_t\}$ be a set of $t$ hyperplanes in $\F^m$ in general position. 
    Then $\mathrm{DecodeMultivariateGeometric}$ decodes $\GAP{m, d, t}^{\mathcal{H}}$ from error patterns of weight less than the minimum distance.
    I.e., when the number of errors is less than half the minimum distance.
    Moreover, the time complexity of  \autoref{alg:decoder-hyperplane-mD} is at most $\poly(t^m, \log |\F|)$. 
\end{theorem}

\begin{proof}
The time complexity of the algorithm can be upper bounded by the time taken to generate the word $\tilde{\vecr}$ and then the time taken in the function call to $\mathrm{DecodeIntermediateMultivariate}(m,d,t,\mathcal{H}, \tilde{\vecr})$. By \autoref{lem:unique-decode-intermediate-mD}, on each such function call, the algorithm $\mathrm{DecodeIntermediateMultivariate}$ makes $t$ calls to \autoref{alg:decoder-hyperplane-mD}, but now for the number of variables being $(m-1)$ and the number of hyperplanes being $(t-1)$ and $t$ calls to the Reed-Solomon decoder given in \autoref{thm:BW} which takes time $\poly(t^m, d^m, \log |\F|) \leq \poly(t^m, \log |\F|)$, in addition to some $\poly(t^m, \log |\F|)$ additional computation. Thus, the time complexity $T(m,t,d)$ of \autoref{alg:decoder-hyperplane-mD} can be upper bounded by the following recursion. 
\begin{align*}
T(2,d,t) &\leq \poly(t,d,\log |\F|)\\
\forall m > 2, T(m,d,t) &\leq t\cdot T(m-1,d,t-1) + \poly(t^m,\log |\F|) \, , 
\end{align*}
which implies that $T(m,d,t)$ is at most $\poly(t^m,\log |\F|)$. 

We now argue the correctness of the algorithm. This follows the outline of the proof of correctness of \autoref{alg:decoder-hyperplane-2D} in \autoref{thm:algo-2D-hperplane-decoder-analysis}. Given the received word $\vecr$, the algorithm first constructs the received word $\tilde{\vecr}$ for the intermediate code $\mathcal{I}_{m}^\mathcal{H}$ and then invokes the unique decoding algorithm for the code $\mathcal{I}_{m}^\mathcal{H}$ given in \autoref{cor:unique-decode-intermediate-2D} on the received word $\tilde{\vecr}$. Therefore, together with the correctness of the unique decoding algorithm for $\mathcal{I}_{m}^\mathcal{H}$ in  \autoref{lem:unique-decode-intermediate-mD}, the following claim implies the correctness of \autoref{alg:decoder-hyperplane-mD}.

\begin{claim}
The received word $\vecr$ is within half the minimum distance of the code $\GAP{m, d, t}^{\mathcal{H}}$ if and only if the word $\tilde{\vecr}$ is within half the minimum distance of the code $\mathcal{I}_{m}^\mathcal{H}$.  
\end{claim}
\begin{proof}[Proof of claim]
Let $\vecc$ be the codeword of the code $\GAP{m, d, t}^{\mathcal{H}}$ closest to the received word $\vecr$ and let $\tilde{\vecc}$ be defined as 
\[
\forall i\in \{1, \ldots, t\}, \vecj \subseteq \{1, \ldots, t\}\setminus \{i\}, |\vecj| = m-1, \quad \tilde{\vecc}(i,\vecj) = \vecc(\{i\} \cup \vecj) . 
\]
From the definition of the code $\mathcal{I}_{m}^\mathcal{H}$, it follows that $\tilde{\vecc}$ is a codeword in $\mathcal{I}_{m}^\mathcal{H}$. In fact, $\vecc$ and $\tilde{\vecc}$ are encodings of the same message polynomial. Moreover, from the definitions of $\tilde{\vecr}$ and $\tilde{\vecc}$, it follows that $\Delta(\tilde{\vecc}, \tilde{\vecr}) = m\cdot \Delta(\vecc, \vecr)$. The claim now follows from this bound, the fact that the block length of $\mathcal{I}_{m}^\mathcal{H}$ is also $m$ times the block length of $\GAP{m, d, t}^{\mathcal{H}}$, and the fact that the fractional minimum distances of these codes are equal to each other as shown in \autoref{lem:props-intermediate-code-mD}.
\end{proof}
\end{proof}

\begin{remark}
    $\GAP{m, d, t}$ can also be efficiently decoded from error \textbf{and erasure} patterns of weight at most the minimum distance.
\end{remark}

This follows from the fact that the GMD algorithm can handle erasures as long as each inner decoder can handle erasures. We will elaborate more on this in the next section.

\section{Local testability of GAP codes}\label{sec:testing}
Next, we discuss the local testability of GAP codes. 
We analyze two very natural local tests for GAP codes, the line-point test, and the plane-point test, and show their local testability via these tests. 

First, we fix some notations and define our tests. 
Throughout this section, let $\F$ be a field, $m, d, t \in \N$ be such that $t \geq d + m$, and let $\mathcal{H} = \{ H_1, \ldots, H_t\}$ be a set of hyperplanes in general position in $\F^m$. 
Let $\mathcal{S, L, P}$ denote the set of all $m$, $(m-1)$, and $(m-2)$-wise intersections of the hyperplanes in $\mathcal{H}$ respectively. 
Note that $\mathcal{S, L, P}$ correspond to points, lines, and planes in $\F^m$ respectively, and furthermore, $\mathcal{S}$ is the set of evaluation points of $\GAP{m, d, t}^{\mathcal{H}}$.

\begin{definition}[The line-point test]\label{def:lines-test}
Given query access to a received word $f: \mathcal{S}\to \F$, the line-point low-degree test is as follows.  
\begin{itemize}
\item pick a random line $\ell$ from the set $\mathcal{L}$ and a random point $x$ on $\ell$
\item let $g_{\ell}$ be a degree $d$ univariate polynomial that is closest to the restriction of $f$ on $\ell \cap \mathcal{S}$
\item accept if $g_{\ell}(x) = f(x)$, reject otherwise
\end{itemize}  
\end{definition}
The plane-point test is similar to the definition above, except that instead of picking a random line, we pick a random plane $P$ from $\mathcal{P}$ and a random point on this plane. We accept if and only if the degree $d$ bivariate polynomial closest to the restriction of $f$ on $P$ agrees with $f$ on the chosen point. 

Both of these tests have perfect completeness. 
Namely, if $f = P|_S$ for some polynomial $P(X_1, \ldots, X_m)$ of degree at most $d$, then the tests accept with probability 1.
Thus, we are interested in the soundness of the tests. 

The main result of this section is an analysis of the soundness of these tests. 
Specifically, we show that the plane-point test is a good local test for GAP codes with constant $m$ and constant rate $R < 1$. 

\begin{theorem}
\label{thm:local-test-main-theorem-plane-point}
For all $m \in \mathbb N$, $R < 1$, there exists a constant 
$K_{m,R}$ such that for any $m$-variate GAP code $C$ of rate $\leq R$,
and any received word $f$, if the plane-point test on $f$ has rejection probability $p$, then $$\delta_C(f) \leq K_{m,R}  \cdot p.$$
\end{theorem}

For the line-point test, we are only able to show soundness at rates below some $R_m = \exp(-m)$. 

\begin{theorem}
\label{thm:local-test-main-theorem}
For all $m \in \mathbb N$, there exists constants $R_m, K_{m}$
such that for any $m$-variate GAP code $C$ of rate $\leq R_m$,
and any received word $f$, if the line-point test on $f$ has rejection probability $p$, then $$\delta_C(f) \leq K_{m}  \cdot p.$$
\end{theorem}

For fixed $m$, if the GAP code has block length $N$, then the query complexity of the line-point test is $O(N^{1/m})$, and the query complexity of the plane-point test is $O(N^{2/m})$.
Then, a simple corollary of \autoref{thm:local-test-main-theorem-plane-point} is the local testability, with polynomially small query complexity, of high rate GAP codes.
The first time such parameters were achieved was by Viderman~\cite{V15}, although today constant rate locally testable codes with constant query complexity are known through the celebrated results of~\cite{DELLM,PK}.

\begin{theorem}
 \label{thm:local-test-main-actual}
For any $\lambda, R \in (0,1)$, there are GAP codes of arbitrarily large codeword length $N$, with rate $R$ and distance $\delta_{R,\lambda}$, that can be locally tested with $O(N^{\lambda})$ queries.
\end{theorem}
This follows immediately from \autoref{thm:local-test-main-theorem-plane-point} by taking $m$ big enough so that $\frac{2}{m} < \lambda$.

\subsection*{Additional notation }
Before proceeding further, we set up some additional notation and conventions for this section that we repeatedly appeal to. 
In many of the technical statements in this section, we work with an appropriate collection of hyperplanes $\mathcal{H}$ in $\F^m$ in general position and study properties of functions on the point set $\mathcal{S}$ obtained by taking $m$-wise intersection of these hyperplanes. 
In this setting, unless otherwise stated, all functions that we study should be thought of as functions on the set $\mathcal{S}$ or its intersection with the appropriate domain at hand. 
For instance, when we say that for an $H \in \mathcal{H}$, let $f_H$ be the restriction of $f$ to the hyperplane $H$, we mean that $f_H$ is a function of the form $f_{H}:\mathcal{S} \cap H \to \F$. 
We also follow this convention of denoting by $f_H$ the restriction of a global function $f$ to a hyperplane (or, more generally, an affine subspace) $H$. 
We also say that two functions $f, g$ are consistent with each other if they take the same values on all points at the intersection of their domains. 
For example, given a function $f$ on a hyperplane $H$ and a function $g$ on a hyperplane $H'$,  we say that $f$ and $g$ are consistent if they agree everywhere on $H \cap H' \cap \mathcal{S}$. 
Otherwise, they are said to be inconsistent.  

Throughout this section, the field $\F$ is assumed to be large enough, essentially so that $\F^m$ contains $t$ hyperplanes in general position. 
From \autoref{obs:example-hype-general-position}, we know that $|\F| \geq t$ suffices for this. 

In terms of the asymptotic behavior of the parameters involved in the results in this section, we should think of the number of variables $m$ as an arbitrary constant, and parameters $d, t$ and the field size $|\F|$ to be growing asymptotically. We also note that in our proofs, we haven't tried to achieve the best possible quantitative bounds on various parameters involved. We anticipate that these bounds can be tightened further using a more careful analysis.  

\subsection*{Organization}

In \autoref{sec:local-char}, we prove a local characterization lemma which states that the property of being degree $d$ on restrictions to each hyperplane is a characterization of degree $d$ polynomials. 
More precisely, if there are $(m-1)$-variate degree $d$ polynomials $g_H$ for each $H \in \mathcal{H}$, such that each pair $(g_H, g_{H'})$ are consistent on their intersection, then there is a $m$-variate degree $d$ polynomial $f$ such that $f$ is consistent with $g_H$ for all $H \in \mathcal{H}$.

We then prove the soundness of the plane-point test in \autoref{sec:plane-point} and that of the line-point test in \autoref{sec:line-point}, \autoref{sec:robust-local-characterization} and \autoref{sec:divisibility}. \autoref{sec:divisibility} contains a general divisibility lemma (\autoref{lem:ps-style-lemma-md}) relating divisibility on hyperplane restrictions with global divisibility, a statement of independent interest.

The proofs of soundness of both tests use a robust version of the local characterization lemma, which states that if almost all (though not necessarily all) of the hyperplane polynomials are consistent with each other, then they can be stitched together into a global degree $d$ polynomial that is consistent with many of them. We have two alternate versions of the robust local characterization, 
one with better parameters that works for $m \geq 3$ (\autoref{sec:plane-point}). 
one with worse parameters that works for $m \geq 2$ (\autoref{sec:robust-local-characterization}).

The first robust local characterization lemma is based on the Raz-Safra~\cite{razSubconstantErrorprobabilityLowdegree1997} and Ben-Sasson-Sudan~\cite{ben-sassonRobustLocallyTestable2006} hyperplane consistency graph method. The second one is more
algebraic, and uses a refined version of the Polischuk-Spielman method~\cite{PS}
based on Bezout's theorem, paying careful attention to intersection multiplicities.



\subsection{Local characterization}\label{sec:local-char}

If $f$ is a $m$-variate degree $d$ polynomial, the restrictions of $f$ to lines are univariate polynomials of degree at most $d$. 
In this section, we prove something of a converse to that statement. 
Formally, we show the following.

\begin{theorem}\label{thm:local-characterization-mD}
Let  $m, d, t \in \N$ be such that $t \geq  d + m$, $\F$ be a large enough field, and let $\mathcal{H} = \{ H_1, \ldots, H_t\}$ be a set of hyperplanes in general position in $\F^m$. 
Let $\mathcal{S} \subseteq \F^m$ denote the set of all $m$-wise intersection points of the hyperplanes in $\mathcal{H}$  
and let $\mathcal{L}$ be the set of lines determined by the $(m-1)$-wise intersection of hyperplanes in $\mathcal{H}$. 
For every line $\ell$ in $\mathcal{L}$, let $g_{\ell}$ be a univariate polynomial of degree at most $d$ defined on $\ell$.

If for all distinct pairs of lines $\ell_1, \ell_2 \in \mathcal{L}$, the line polynomials $g_{\ell_1}$ and $g_{\ell_2}$ agree on the point $\ell_1 \cap \ell_2 \in \mathcal{S}$, then there is an $m$-variate degree $d$ polynomial $f$ such that for every line $\ell \in \mathcal{L}$, the restriction of $f$ to the line $\ell$ equals $g_{\ell}$. 
\end{theorem}

This statement talks about deducing the $m$-dimensional structure from $1$-dimensional restrictions. The proof will recursively use the following lemma, which
deduces $m$-dimensional structure from $(m-1)$-dimensional restrictions.

\begin{lemma}\label{lem:local-consistency-to-global-consistency}
Let  $m, d, t \in \N$ be such that $t \geq d + m$ and $\F$ be a large enough field. 
Let $\mathcal{H} = \{ H_1, \ldots, H_t\}$ be a set of hyperplanes in general position in $\F^m$ and let $\mathcal{S} \subseteq \F^m$ denote the set of all $m$-wise intersection points of the hyperplanes in $\mathcal{H}$. 
For every hyperplane $H$ in $\mathcal{H}$, let $g_{H} : H \to \F$ be a degree $d$ polynomial on $(m-1)$ variables. 

If for all distinct pairs of hyperplanes $H, H' \in \mathcal{H}$, the polynomials $g_{H}$ and $g_{H'}$ agree on $H \cap H' \subseteq \mathcal{S}$, then there is an $m$-variate degree $d$ polynomial $f : \mathcal{S} \to \F$ such that for every hyperplane $H \in \mathcal{H}$, the restriction of $f$ to $H$ equals $g_{H}$. 
\end{lemma}

\begin{proof}
For every $i \in \{1, 2, \ldots, t\}$, let $\mathcal{H}_i$ be the set $\{ H_j \cap H_i : j \neq i\}$ of subspaces of $\F^m$ of co-dimension $2$. 
However, we will instead think of $\mathcal{H}_i$ as a set of $(t-1)$ hyperplanes in the $(m-1)$ dimensional ambient space $\F^{m-1}$ (identified with $H_i$). 
Moreover, since $\mathcal{H}$ is a set of hyperplanes in $\F^m$ in general position, we have that for every $i$, $\mathcal{H}_i$ is a set of hyperplanes in general position in $\F^{m-1}$. 
Let $\mathcal{S}_i$ be the points given by $(m-1)$-wise intersection of hyperplanes in $\mathcal{H}_i$. 
Note that  $\mathcal{S}_i$ is precisely the subset of points of $\mathcal{S}$ that are contained in the hyperplane $H_i$. 

Let $v: \mathcal{S} \to \F$ be a function defined as follows. 
For any $\veca \in \mathcal{S}$, let  $\{H_{i_1}, H_{i_2}, \ldots, H_{i_m} \}\subseteq \mathcal{H}$ be the hyperplanes that intersect at $\veca$. 
Then, $v(\veca)$ is defined to be equal to $g_{H_{i_j}}(\veca)$ for $j \in \{1, 2, ,\ldots, m\}$. 
Since the hyperplane polynomials are mutually consistent with each other, the function $v$ is well-defined.  
We now define the global degree $d$ polynomial $f$ using the function  $v$. \\

\noindent
\textbf{Defining the global function: }
Let $\mathcal{H}'$ be an arbitrary subset of $\mathcal{H}$ of size exactly $(d + m)$, and let $\mathcal{S}'$ be the set of points obtained by taking $m$-wise intersection of hyperplanes in $\mathcal{H}'$. 
Clearly, $\mathcal{S}'$ is a subset of $\mathcal{S}$. 
Let $f$ be the \emph{unique} degree $d$ polynomial on $m$-variables such that $f(\veca) = v(\veca)$ for every $\veca \in \mathcal{S}'$. 
The existence and uniqueness of $f$ again follows from the fact that the set $\mathcal{S}'$ is an interpolating set for degree $d$ polynomials on $m$ variables (\autoref{thm:geometric-hitting-set}).

We now argue that for every $i$, the restriction of the global function $f$ to any hyperplane $H_i \in \mathcal{H}$ equals the function $g_{H_i}$. 
We do this in two steps. \\

\noindent
\textbf{Case I - $H_i \in \mathcal{H}'$ : } Let $\mathcal{S}_i'$ of points obtained by taking the intersection of any $(m-1)$ hyperplanes in $\mathcal{H}'\setminus \{H_i\}$ together with $H_i$, and let $f_{H_i}$ be the degree $d$ polynomial (on $(m-1)$-variables) obtained by restricting $f$ to $H_i$. 
Note that $\mathcal{S}_i'$ is a subset of $\mathcal{S}_i \cap \mathcal{S}'$ and is an interpolating set for degree $d$ polynomials on the space $H_i$ since it is the set of $(m-1)$-wise intersections of $(d + m-1)$ hyperplanes in general position in this $(m-1)$-dimensional ambient space (\autoref{thm:geometric-hitting-set}). 
From the definition of $f$, and the properties of $f_{H_i}$ discussed earlier, we know that for every point $\veca \in \mathcal{S}_i'$, $f(\veca) = f_{H_i}(\veca) = g_{H_i}(\veca)  = v(\veca)$. 
Note that we use the fact that $\mathcal{S}_i'$ is a subset of $\mathcal{S}'$ for this. 
But, $f_{H_i}$ and $g_{H_i}$ are both degree $d$ functions on the space $H_i$ that agree everywhere on the set $\mathcal{S}_i'$ which is an interpolating set for such functions. 
Thus, they must be equal to each other. \\

\noindent
\textbf{Case II - $H_i \notin \mathcal{H}'$ : } Now, the set $\mathcal{S}_i'$ of points obtained by taking the intersection any $(m-1)$ hyperplanes in $\mathcal{H}'$ together with $H_i$. 
$f_{H_i}$ is the restriction of $f$ to $H_i$  as defined for Case I. 
Let $\veca \in \mathcal{S}_i'$ be any point and let $H_j \in \mathcal{H}'$ be such that $\veca \in H_j$. 
So, we have $g_{H_i}(\veca) = g_{H_j}(\veca) = f_{H_j}(\veca)$, since $g_{H_i}$ and $g_{H_j}$ are consistent with each other and $g_{H_j} = f_{H_j}$ from the case above. 
But then, $f_{H_j}$ and $f_{H_i}$ are both restrictions of the same global function $f$, and $\veca$ is a common point in the domain of these functions. 
So,  $f_{H_i}(\veca) = f_{H_j}(\veca)$. 
Therefore, we get that  $g_{H_i}(\veca) = g_{H_j}(\veca) = f_{H_i}(\veca)$. 
Hence, we have two $(m-1)$-variate degree $d$ polynomials on  $g_{H_i}$ and $f_{H_i}$ on $H_i$ that are equal to each other on the set $\mathcal{S}_i'$, which is again an interpolating set for such polynomials. 
Thus, they must be equal to each other. 
\end{proof}


We can now complete the proof of the local characterization from $1$-dimensional restrictions.

\begin{proof}[Proof of \autoref{thm:local-characterization-mD}]
The proof of the theorem is via induction on $m$, with the $m= 2$ case of \autoref{lem:local-consistency-to-global-consistency} being the base case. 
So, we now assume that $m> 2$ and the theorem is true in $(m-1)$ dimensions and prove it for $m$ dimensions. 
The overall structure of the proof follows that of the proof of \autoref{lem:local-consistency-to-global-consistency}.\\

For every $i \in \{1, 2, \ldots, t\}$, let $\mathcal{H}_i$ be the set $\{ H_j \cap H_i : j \neq i\}$ of subspaces of $F^m$ of co-dimension $2$, which as before, we think of $\mathcal{H}_i$ as a set of $(t-1)$ hyperplanes in the $(m-1)$ dimensional ambient space $\F^{m-1}$ (identified with $H_i$). 
Moreover, since $\mathcal{H}$ is a set of hyperplanes in $\F^m$ in general position, we have that for every $i$, $\mathcal{H}_i$ is a set of hyperplanes in general position in $\F^{m-1}$. 
Let $\mathcal{S}_i$ be the points given by $(m-1)$-wise intersection of hyperplanes in $\mathcal{H}_i$, $\mathcal{L}_i$ be the lines given by $(m-2)$-wise intersection of hyperplanes in $\mathcal{H}_i$. 
Note that  $\mathcal{S}_i$ is precisely the subset of points of $\mathcal{S}$ that are contained in the hyperplane $H_i$ and $\mathcal{L}_i$ is precisely the subset of lines of $\mathcal{L}$ that are contained in the hyperplane $H_i$.   
Now, clearly, the set of line polynomials for lines in $\mathcal{L}_i$ also satisfies the hypothesis of the theorem, namely that any of two of these polynomials are consistent. 
But now, we are in $(m-1)$ dimensions. Hence, by the induction hypothesis, we get that there is a degree $d$ polynomial $f_{H_i}$ such that for every line $\ell \in \mathcal{L}_i$, the restriction of $f_{H_i}$ on $\ell$ equals the polynomial $g_{\ell}$.

Moreover, since all the line polynomials are consistent with each other and the hyperplane polynomials $f_{H_i}$ are consistent with all line polynomials on the respective hyperplane, we get that the hyperplane polynomials $f_{H_i}$ obtained above must also be consistent with each other. 
Thus, we are in the setting of \autoref{lem:local-consistency-to-global-consistency}, and get that there is an $m$-variate degree $d$ polynomial $f$ that is consistent with all the hyperplane polynomials $f_{H_i}$. 
Since these hyperplane polynomials are consistent with the line polynomials, we get that $f$ is consistent with all the line polynomials, thereby completing the proof of the theorem.

\end{proof}

\subsection{Soundness of the plane-point test}\label{sec:plane-point}

In this subsection, we prove \autoref{thm:local-test-main-theorem-plane-point}.

We begin by proving a robust analogue of the local characterization lemma, \autoref{lem:local-consistency-to-global-consistency}, for $m \geq 3$.
Here, in the hypothesis, we are only guaranteed that the low-degree functions on \emph{most} (as opposed to all) pairs of hyperplanes are consistent with each other. 
The conclusion again is in the spirit of that in \autoref{lem:local-consistency-to-global-consistency} and guarantees a single low degree polynomial \emph{explaining} most of these hyperplane polynomials.


\subsubsection{Robust local characterization}
\begin{lemma}[Robust Local Characterization]\label{thm:rlc-planes}
    Let $m \in \mathbb{N}_{\geq 3}$, and $\rho, \beta, c \in (0,1)$ be constants.
    Let $t$ be a growing parameter and $d = \rho t$.
    Let $H_1,...,H_t \subset \mathbb{F}_q^m$ be hyperplanes in general position, and let $g_1,...,g_t$ be corresponding $(m-1)$-variate polynomials of degree at most $d$. 

    Suppose at least $1 - c$ fraction of pairs of the $g_i$ are consistent, and 
    \begin{equation}
        1 - \rho > \sqrt{2c} + \beta.
    \label{eq:rlc-cond}
    \end{equation}

    Then, there exists some $m$-variate polynomial $h$ such that $h|_{H_i} \equiv g_i$ for at least a $1 - c/\beta$ fraction choices of $i \in [t]$.
\end{lemma}

\begin{proof} 
    The proof closely follows \cite{ben-sassonRobustLocallyTestable2006}.
    Define an inconsistency graph $G$ as follows. 
    The vertices are $[t]$, and there is an edge between $i$ and $j$ if and only if $g_i$ and $g_j$ are inconsistent (they are not equal to each other on $H_i \cap H_j$). 
    The goal is to find a large independent set in $G$ (corresponding to a large set of pairwise consistent polynomials) and apply \autoref{lem:local-consistency-to-global-consistency}.
    We first show that $G$ has the following property. 
    \begin{claim}
        If there is an edge between $i$ and $j$, then either the degree of $i$ or the degree of $j$ is at least $\frac{1}{2}(1 - \rho)t(1 - o(1))$.
    \end{claim}

    Suppose $i$ and $j$ are inconsistent, then $g_i$ and $g_j$ restricted to $H_i \cap H_j$ are distinct codewords of $\GAP{m-2, d, t-2}$, and hence disagree on at least $\Delta$ points, where $\Delta$ is the distance of $\GAP{m-2, d, t-2}$.
    Let $S$ be the subset of points where $g_i$ and $g_j$ differ. 
    Recall $m$-wise intersections of hyperplanes specify the points, and thus, we can think of $S$ as being a subset of $\binom{[t] \setminus \{i, j\}}{m-2}$ (since points in $S$ must lie on $H_i$ and $H_j$). 
    Let $T$ be the set of indices of the hyperplanes that contain some disagreement between $g_i$ and $g_j$. 
    For any $k \in T$, $g_k$ must be inconsistent with at least one of $g_i$ and $g_j$. 
    We now bound the size of $T$.
    \begin{align*}
        & |S| \leq \binom{|T|}{m-2}\\
        & \implies \Delta \leq \binom{|T|}{m-2}\\
        & \implies (1 - \rho)^{m-2}\binom{t-2}{m-2} \leq \binom{|T|}{m-2}\\
        & \implies (1 - \rho)^{m-2}\leq \frac{|T|!}{(|T| - m-2)!}\cdot\frac{(t-2-m-2)!}{(t - 2)!}\\
        & \implies (1 - \rho)^{m-2}\leq \prod_{i=0}^{m-3}\frac{|T| - i}{t - 2 - i}\\
        & \implies (1 - \rho)^{m-2}\leq \left( \frac{|T|}{t-2} \right)^{m-2}\\
        & \implies (1 - \rho)(t - 2)\leq |T|.
    \end{align*}
    The claim then follows from applying the Pigeonhole Principle.

    Thus, the vertices with degree $< \frac{1}{2}(1 - \rho)t(1 - o(1))$ form an independent set, $I$. 
    Let $z$ be the fraction of vertices of degree at least $\frac{1}{2}(1 - \rho)t(1 - o(1))$. 
    Counting degrees, we have that 
    $\frac{zt^2}{2}(1 - \rho)(1 - o(1)) \leq 2 c \binom{t}{2}$, which implies $z \leq \frac{2c}{1 - \rho} + o(1)$.

    To apply the local characterization lemma (\autoref{lem:local-consistency-to-global-consistency}), we need $(1 - z)t \geq d + m$, which is true when 
    \begin{align*}
        1 - z \geq \rho + m/t \\
        \impliedby 1 - \rho \geq \frac{2c}{1-\rho} + o(1)\\
        \impliedby (1 - \rho)^2 \geq 2c + o(1)\\
        \impliedby 1 - \rho \geq \sqrt{2c} + o(1)
    \end{align*}
    which is satisfied by the assumption of the theorem.

    Applying \autoref{lem:local-consistency-to-global-consistency}, we have that there exists a $m$-variate polynomial $h$ consistent with each of the polynomials $g_i$ for $i \in I$, which is a set of fractional size at least $1 - z$, we'll show that it must be consistent with an even larger set of polynomials.

    Let $i \in [t]$, and suppose $g_i$ is consistent with at least $d + 1$ polynomials in $I$. 
    This implies that $g_i|_{H_j} \equiv h|_{H_i \cap H_j}$ for at least $d + 1$ values of $j$, which implies $h|_{H_i} \equiv g_i$ since both of these polynomials have degree at most $d$. 
    Hence, $h$ is consistent with $g_i$ as well. 
    For any $i$, let $X_i$ be the number of polynomials that are inconsistent with $i$, we have $\E_{i \in [t]}[X_i] = ct$. 
    Note that $t - X_i$ is the number of polynomials consistent with $i$. 
    If $t - X_i \geq zt + d + 1$, then $g_i$ is consistent with at least $d + 1$ polynomials in $I$, hence also consistent with $h$. 
    We'll show that this is true for many $i$. 
    By Markov's inequality,
    \begin{align*}
        \Pr(X_i > (1 - z)t -d -1) &\leq \frac{ct}{(1 - z)t -d -1}\\
    & \leq \frac{c}{1 - z -\rho -1/t}\\
    & = \frac{c}{1 - z -\rho -1/t}\\
    & \leq \frac{c}{\beta}.
    \end{align*}
    Note that we get the last inequality since $z \leq \frac{2c}{1 - \rho} \leq \frac{2c}{\sqrt{2c}} = \sqrt{2c}$, and $1 - \rho > \sqrt{2c} + \beta$.
    Thus, at least a $1 - c/\beta$ fraction of the $g_i$s is consistent with $h$ as required.
\end{proof}

\subsubsection{Analysis of the plane-point test}

We now analyze the plane-point test and prove the following lemma.

\begin{lemma}\label{lem:pl-v-po}
    Let $m \in \mathbb{N}_{\geq 2}$, and $\rho, p, \beta \in (0,1)$ be constants.
    Let $t$ be a sufficiently large growing parameter, and let $d/t \leq \rho$.
    Define the sequences $(Q_{m, \rho})_{m \geq 3}$ and $(P_{m, \rho})_{m \geq 3}$ as follows. 
    $$
    P_{m, \rho} = \begin{cases}
        1 & m = 2\\
        \frac{(1 - \rho - \beta)^2(1 -\rho)}{4} \cdot P_{m-1, \rho}& m \geq 3\\
    \end{cases}
    $$
    $$
    Q_{m, \rho} = \begin{cases}
        1 & m = 2\\
        \frac{3}{P_{m-1,\rho}(1 - \rho)^{m-2}\beta} & m \geq 3
    \end{cases}
    $$
    Let $C = \GAP{m, d, t}$, and let $f$ be a received word. 
    Suppose the plane-point test accepts $f$ with at least probability at least $1 - p$. 
    Then, if $p \leq P_{m, \rho}$, then 
    $$
    \delta_C(f) \leq Q_{m, \rho} p 
    $$
\end{lemma}

\begin{proof}

The proof of the lemma is by induction on $m$. 

For the base case, $m = 2$. 
We query all the points since the entire code lies on a two-dimensional plane. 
Let $g$ be the closest bivariate polynomial to $f$, then the acceptance probability is $\E_{x}[\1_{g(x) = f(x)}] = \alpha(g, f)$. Thus, we have $1 - p \geq \alpha(g, f)$, so $\delta_C(f) \leq p$ as required.

Let $m \geq 3$ for the inductive step, and suppose the planes test works for the $(m-1)$-variate GAP code. 
Let $f_i = f|_{H_i}$.
We give a brief proof overview below.
\begin{enumerate}
    \item Using the inductive hypothesis, find $(m-1)$-variate polynomials $g_1, g_2,..., g_t$ such that $g_i$ is close to $f_i$. 
    \item Then, we will use \Cref{thm:rlc-planes} (Robust Characterization Lemma) on $g_1,...,g_t$ to extract an $m$-variate polynomial $h$ consistent with most of the $g_i$. To do so, we must show that many pairs $g_i$ and $g_j$ are consistent.
    \item Finally, we will bound the distance between $h$ and $f$.
\end{enumerate}

\paragraph{Finding the $g_i$.}
Let $p_i$ be the rejection probability given that the selected plane lies on $H_i$. 
Call $i$ \emph{good} if $p_i \leq P_{m-1,\rho}$, that is $p_i$ satisfies the conditions of the $(m-1)$-variate version of the theorem.
A simple calculation shows that the average rejection probability on each hyperplane is the overall rejection probability, i.e. 
$\E[p_i] = p$.
Applying Markov's Inequality, we have 
\begin{align*}
    \Pr[i \text{ is bad}]\leq \Pr[p_i > P_{m-1,\rho}] \leq \frac{p}{P_{m-1,\rho}}.
\end{align*}
Let $\gamma = \frac{p}{P_{m-1,\rho}}$. 
Then, at least a $1 - \gamma$ fraction of the $i$s are good.
For each $i \in [t]$, let $g_i$ be the $(m-1)$-variate polynomial obtained by applying the $(m-1)$-variate version of the theorem if $i$ is good and the $0$ polynomial otherwise. 

Note that if $i$ is bad, $g_i$ might be far from $f_i$. 
However, since most of the indices $i$ are good, we will show that the \emph{average} agreement between $g_i$ and $f_i$ is large.
To simplify the notation a bit, let $\alpha_i = \alpha(g_i , f_i)$, and $\eta_i = 1 - Q_{m-1,\rho}p_i$. 
\begin{align*}
    \E_{i}[\alpha(g_i, f_i)] &= \E_{i}[\alpha_i]\\
    &= \E_{i}[\alpha_i - \eta_i + \eta_i]\\
    &= \E_{i}[\eta_i] + \E[\alpha_i - \eta_i]\\
    &= 1 - Q_{m-1,\rho}p + \E[\alpha_i - \eta_i].
\end{align*}
Focusing on the $\E[\alpha_i - \eta_i]$, we have that for each good $i$, $\alpha_i \geq \eta_i$ by the distance guarantee of the $(m-1)$-variate version of the theorem, and for each $i$ that is not good (of which there are most a $\gamma$ fraction), $\alpha_i - \eta_i \geq Q_{m-1,\rho}p_i - 1 \geq Q_{m-1,\rho}P_{m-1,\rho} - 1$, where we get the first inequality since the agreement is always at least zero, and the second inequality because $i$ is not good. 
Thus, we have 
$$\E_{i}[\alpha(g_i, f_i)] \geq 1 - Q_{m-1,\rho}p + \gamma(Q_{m-1, \rho}P_{m-1,\rho} - 1) = 1 - \gamma.$$

\paragraph{Many pairs are consistent.}
Now that we know the average agreement between the $g_i$ and $f_i$ is large, we will show that many pairs, $(g_i, g_j)$, are consistent.
Towards this goal, we first show that the average agreement of $g_i$ and $g_j$ is high.
\begin{align*}
    \E_{i, j}[\alpha(g_i, g_{j})] &=\E_x\E_{H_i, H_j \ni x}[\1_{g_i(x) = g_{j}(x)}]\\
                                    &\geq\E_x\E_{H_i, H_j \ni x}[\1_{g_i(x) = f(x)}\1_{g_{j}(x) = f(x)}]\\
                                    &=\E_x[\E_{i : x \in H_i}[\1_{g_i(x) = f(x)}]]^2\\
                                    &\geq [\E_x\E_{i : x \in H_i}[\1_{g_i(x) = f(x)}]]^2\\
                                    &=[\E_{i}[\alpha(g_i, f_i)]]^2\\
                                    &\geq (1 - \gamma)^2\\
                                    &> 1 - 2\gamma.
\end{align*}

Let $c$ be the fraction of inconsistent pairs. 
If a pair is consistent, they have agreement 1. 
On the other hand, if a pair is not consistent, they have agreement at most $1 - \delta_{m-2}$, which is $1 - (1 -\rho)^{m-2}$.
Thus, 
\begin{align*}
    1 - 2\gamma < 1 - c + c(1 - (1 - \rho)^{m-2})\\
    \implies c < \frac{2\gamma}{(1 - \rho)^{m-2}}.
\end{align*}

\paragraph{Applying \Cref{thm:rlc-planes}.}
To apply \Cref{thm:rlc-planes}, we need to verify condition \ref{eq:rlc-cond} holds.

In particular, we need 
\begin{align*}
    & (1 - \rho - \beta)^2 \geq 2c\\
    & \impliedby \frac{(1 - \rho - \beta)^2}{2} \geq \frac{2p}{P_{m-1,\rho}(1 - \rho)^{m-2}}\\
    & \impliedby \frac{(1 - \rho - \beta)^2(1 - \rho)^{m-2}}{4} \cdot P_{m-1,\rho} \geq p\\
    & \impliedby P_{m, \rho} \geq p.
\end{align*}

Applying \Cref{thm:rlc-planes}, we obtain a $m$-variate polynomial $h$ consistent with at least a $1 - c / \beta$ fraction of the $g_i$. Finally, we show that $h$ is not too far from $f$.

\paragraph{Agreement between $h$ and $f$.} 
Let $h_i$ be the restriction of $h$ to $H_i$.
Let $X_i = \alpha(h_i, f_i)$, and $Y_i = \alpha(f_i, g_i)$.
$$
\alpha(h, f) = \E[X_i] = \E[X_i + Y_i - Y_i] > 1 - \gamma + \E[X_i - Y_i] \geq 1 - \gamma - c/\beta.
$$
where we get the last inequality since $g_i \neq h_i$ on at most a $c/\beta$ fraction of $i \in [t]$, and the different of two number each in $(0,1)$ is at least -1.

Substituting bound on $c$, 
\begin{align*}
\alpha(h, f) &\geq 1 - \gamma - \frac{2\gamma}{(1 - \rho)^{m-2}\beta} \\
             &= 1 - p \cdot \frac{1}{P_{m-1, \rho}}(1 + \frac{2}{(1 - \rho)^{m-2}\beta})\\
&> 1 - p(\frac{3}{P_{m-1, \rho}(1 - \rho)^{m-2}\beta})\\
&= 1 - pQ_{m, \rho}
\end{align*}
Therefore, $\delta_C(f) \leq pQ_{m, \rho}$, as required.

\paragraph{Bounds on $P_{m, \rho}, Q_{m, \rho}$.} 
\begin{align*}
    P_{m, \rho} = \prod_{i=0}^{m-3}\left(\frac{(1 - \rho - \beta)^2(1 -\rho)^{m-2-i}}{4}\right) \geq \frac{(1 - \rho - \beta)^{2m- 2}(1 -\rho)^{m^2}}{4^{m-2}}
\end{align*}
Then,
$$
Q_{m, \rho} = \frac{3}{P_{m-1, \rho}(1 - \rho)^{m-2}\beta} \leq \frac{3 \cdot 4^{m-3}}{\beta(1 - \rho - \beta)^{2m- 4}(1 -\rho)^{(m-1)^2 + m-2}}
$$

Note that the bounds on $P_{m, \rho}$ and $Q_{m, \rho}$ are exponential in $m$; however, since $m, \rho$ and $\beta$ and constants, $P_{m, \rho}$, and $Q_{m, \rho}$ are also constants.

\end{proof}

To prove \Cref{thm:local-test-main-theorem-plane-point}, apply \Cref{lem:pl-v-po} with $\rho = R^{1/m}$, $\beta = \frac{1}{2}(1 - \rho)$, and $C_{m,R} = \max\{1/P_{m,\rho}, Q_{m,\rho}\}$.

\subsection{Soundness of the line-point test}\label{sec:line-point}

In this subsection, we prove~\autoref{thm:local-test-main-theorem} on
the soundness of the line-point test. 
The proof will use the soundness of the plane-point test, along with a
robust local characterization of bivariate polynomials based on restrictions to
lines. We now state this robust local characterization statement.

\begin{theorem}\label{thm:robust-local-consistency-to-global-consistency-2d}
There exists a constant $\epsilon_0 \in (0,1)$ such that for all $\epsilon \in [0,1)$, with $\epsilon < \epsilon_0$, $d, t \in \N$ with $t >  \max\{\frac{d+2}{1-\sqrt{\epsilon}}, \frac{2d}{1-3\sqrt{\epsilon}}\}$, and sufficiently large field $\F$, the following is true. 

Let $\mathcal{H} = \{ \ell_1, \ldots, \ell_t\}$ be a set of lines in general position in $\F^2$. 
Let $\mathcal{S} \subseteq \F^m$ denote the set of all pairwise intersection points of the lines in $\mathcal{H}$. 
For every line  $\ell$ in $\mathcal{H}$, let $g_{\ell} : \ell \to \F$ be a degree $d$ univariate polynomial.

If the polynomials $g_{\ell}$ and $g_{\ell'}$ agree with each other on $\ell \cap \ell'$ for at least $(1-\epsilon)\binom{t}{2}$ pairs $\{\ell, \ell'\} \subseteq \mathcal{H}$, then there is an bivariate degree $d$ polynomial $f : \mathcal{S} \to \F$ and a set $\tilde{\mathcal{H}} \subseteq \mathcal{H}$ of size at least $(1-2\epsilon)t$ such that for every $\ell \in \tilde{\mathcal{H}}$, the restriction of $f$ on $\ell$ equals $g_{\ell}$. 
\end{theorem}

This can be viewed as a robust analogue of \autoref{lem:local-consistency-to-global-consistency} for the case $m=2$. We already proved such a robust analogue for $m > 2$ in \autoref{thm:rlc-planes} during the analysis of the plane-point test. The proof appears in \autoref{sec:robust-local-characterization}.

The key ingredient of this robust local characterization is a general divisibility lemma, given below.

\begin{lemma}\label{lem:ps-style-lemma-md}
Let $\F$ be any field, $e, d, t, m$ be natural numbers, and $\mathcal{H}$ be a set of $t$ hyperplanes in $\F^m$ in general position. 
$E(X_1, \ldots, X_m)$ and $P(X_1, \ldots, X_m)$ are polynomials in $\F[X_1, \ldots, X_m]$ of degree equal to $e$ and $d$ respectively. 
For any hyperplane $H \subseteq \F^m$, let $E_H(Z_1, \ldots, Z_{m-1})$ and $P_{H}(Z_1, \ldots, Z_{m-1})$ respectively be the $m-1$-variate polynomials in $\F[Z_1, \ldots, Z_{m-1}]$ obtained by the restriction of $E$ and $P$ to the hyperplane $H$. 

If $t > 2d + e$ and for all hyperplanes $H \in \mathcal{H}$, there exists a polynomial $Q_{H}(Z_1, \ldots, Z_{m-1})$ such that 
\[
P_{H}(Z_1, \ldots, Z_{m-1})  = E_{H}(Z_1, \ldots, Z_{m-1}) \cdot Q_{H}(Z_1, \ldots, Z_{m-1}) \, ,
\]
then, there exists a polynomial $Q(X_1, \ldots, X_m) \in \F[X_1, \ldots, X_m]$ such that 
\[
P(X_1, \ldots, X_m) = E(X_1, \ldots, X_m) \cdot Q(X_1, \ldots, X_m) \, .
\]
\end{lemma}

The proof appears in \autoref{sec:divisibility}. For the robust local characterization of bivariate polynomials with lines, \autoref{thm:robust-local-consistency-to-global-consistency-2d}, we only need the $m = 2$ case of this lemma. Nevertheless we state and prove the result for general $m$ because it seems to be a basic statement of independent interest.

In terms of the quantitative parameters, the following simple example shows that at least for $m=2$, the bounds on $t$ in the lemma are essentially tight. 

\paragraph*{An almost tight example: } Let $E(X_1, X_2)$ be the polynomial $X_1$ of degree $1$ and let $P(X_1, X_2)$ be the polynomial $\prod_{i = 1}^d (X_2 - i^2)$ of degree $d$. Let us now consider the lines $\ell_i$ defined by solutions to the linear equation $X_2 = i\cdot X_1 + i^2$ as $i$ varies in the set $\{-1, 1, -2, 2, \ldots, -d, d\}$. Clearly, for each of these lines, the restriction of $P$ on the line is divisible by the restriction of $E$ on the line. However, $P$ is not divisible by $E$ in the ring $\F[X_1, X_2]$. 

\autoref{lem:ps-style-lemma-md} says that if the number of such lines exceeds $2d+1$, then divisibility of restrictions guarantees divisibility as bivariates and this example shows that the parameter $t$ in the statement of \autoref{lem:ps-style-lemma-md} must be at least $2d$ for this to happen. 

\subsubsection{Proof of \Cref{thm:local-test-main-theorem} in two dimensions}\label{sec:proof-of-testing-theorems-2D}

We recall the $m=2$ case of \autoref{thm:local-test-main-theorem}. 
Throughout this section, we denote the fractional agreement between two functions $f, g$ by $\alpha(f,g)$.
\begin{theorem}\label{thm:local-test-main-theorem-2D}
There exists a constant $\epsilon_0 \in (0,1)$ such that for all $\epsilon \in [0,1)$ with $\epsilon < \epsilon_0$, natural numbers $t,d$ with $t > \max\{\frac{d+2}{1-\sqrt{\epsilon}}, \frac{2d}{1-3\sqrt{\epsilon}}\}$, and sufficiently large field $\F$, the following is true.

Let $\mathcal{H} = \{ \ell_1, \ldots, \ell_t\}$ be a set of lines in general position in $\F^2$. 
Let $\mathcal{S} \subseteq \F^2$ denote the set of all pairwise intersection points of the lines in $\mathcal{H}$. 
If $f: \mathcal{S} \to \F$ is a function for which the line-point test on $f$ accepts with probability at least $(1-\epsilon)$, then there exists a degree $d$ polynomial $h \in \F[X, Y]$ such that $f$ and $h$ agree on at least $(1-5\epsilon-o(1))$ fraction of points in $\mathcal{S}$. 

\end{theorem}

\begin{proof}
For every line $\ell \in \mathcal{H}$, let $g_{\ell}$ be the degree $d$ polynomial that is closest to the restriction of $f$ on $\ell$ (henceforth denoted by $f_{\ell}$ for brevity). Let $\alpha(f_{\ell}, g_{\ell})$ denote the fractional agreement between $f_{\ell}$ and $g_{\ell}$. 
Since the line-point passes with probability $(1-\epsilon)$, we have that 
\[
\E_{\ell}[\alpha(f_{\ell}, g_{\ell})] = \E_{\ell, x \in \ell} [\mathbb{1}_{g_{\ell}(x) = f(x)}] \geq (1-\epsilon) \, .
\]
From the above statement, we would like to conclude that for most pairs of lines $\ell, \ell'$, the polynomials $g_{\ell}$ and $g_{\ell'}$ are consistent with each other, i.e. 
they agree on $\ell \cap \ell'$. 
To this end, we have the following sequence of inequalities. 
\begin{align*}
\E_{\ell, \ell'}[\mathbb{1}_{g_{\ell}(\ell \cap \ell') = g_{\ell'}(\ell \cap \ell')}] =& \E_{x} \E_{\ell, \ell' \ni x} [\mathbb{1}_{g_{\ell}(x) = g_{\ell'}(x)}] \\
\geq &  \E_{x} \E_{\ell, \ell' \ni x} [\mathbb{1}_{g_{\ell}(x) = f(x)} \cdot \mathbb{1}_{g_{\ell'}(x) = f(x)}] \\
= &  \E_{x} \left(\E_{\ell\ni x} [\mathbb{1}_{g_{\ell}(x) = f(x)}]\right)^2 \\
\geq &  \left( \E_{x} \E_{\ell\ni x} [\mathbb{1}_{g_{\ell}(x) = f(x)}]\right)^2 \\
\geq & (1-\epsilon)^2 \\
\geq & 1-2\epsilon \, .
\end{align*}
Here, $x$ ranges over the points in $\mathcal{S}$ and the lines $\ell, \ell'$ range over lines in the set $\mathcal{L}$, and the third inequality relies on an application of the Cauchy-Schwartz inequality.

Thus, we have that the for at least $(1-2\epsilon - 1/t)$ fraction of all pairs of distinct lines $\ell, \ell'$  in $\mathcal{L}$ are consistent with each other. 
Therefore, from the robust local characterization theorem (\autoref{thm:robust-local-consistency-to-global-consistency-2d}) to this setting ( we note that the inequalities needed in the hypothesis are satisfied for our choice of parameters here), we get that there is bivariate polynomial $h$ of degree at most $d$ such that for at least $(1-4\epsilon - 2/t)t$ lines $\ell$ in $\mathcal{H}$, we have that 
\[
h_{\ell} = g_{\ell} \, .
\]
Thus, on lines $\ell$ where $h_{\ell} = g_{\ell}$, we have that $\alpha(h_{\ell}, f_{\ell}) = \alpha(g_{\ell}, f_{\ell})$ and on the remaining $(4\epsilon + 2/t)$ fraction of lines, we have that $\alpha(h_{\ell}, f_{\ell}) \geq 0$ (and $\alpha(g_{\ell}, f_{\ell})$ could be as large as $1$). 
Thus, the overall fractional agreement between $f$ and $h$ is at least $1-\epsilon - 4\epsilon - 2/t = (1-5\epsilon - o(1))$. 

\end{proof}
An easy corollary of the above theorem, whose form will be useful for us in the higher dimensional analysis of the test is stated below. 
The proof follows from the fact that if the parameter $\epsilon$  is chosen to be sufficiently compared to the parameter $\epsilon_0$ in \autoref{thm:local-test-main-theorem-2D}, then the corollary below does not guarantee any non-trivial agreement between $f$ and $h$, else, it guarantees an agreement that is only weaker than that in \autoref{thm:local-test-main-theorem-2D}.  

\begin{corollary}\label{cor:smooth-2D-test}
    There is a constant $\epsilon_0 \in (0,1)$ such that for every $\epsilon \in [0,1)$, natural numbers $t,d$ with $t > \max\{\frac{d+2}{1-\sqrt{\epsilon}}, \frac{2d}{1-3\sqrt{\epsilon}}\}$, and sufficiently large field $\F$, the following is true. 

Let $\mathcal{H} = \{ \ell_1, \ldots, \ell_t\}$ be a set of lines in general position in $\F^2$. 
Let $\mathcal{S} \subseteq \F^2$ denote the set of all pairwise intersection points of the lines in $\mathcal{H}$. 
If $f : \mathcal{S} \to \F$ is a function for which  the line-point test  on $f$ accepts with probability at least $(1-\epsilon)$, then there exists a degree $d$ polynomial $h \in \F[X,Y]$ such that $f$ and $h$ agree on at least $(1-\frac{5\epsilon+o(1)}{\epsilon_0})$ fraction of points in $\mathcal{S}$. 
\end{corollary}


\subsubsection{Proof of \Cref{thm:local-test-main-theorem} in higher dimensions}
\label{sec:line-point-mD}

We now give an analysis of the line-point test for $m \geq 3$. 
We do this in two steps: we use \autoref{lem:line_point_to_plane_point} to show that if the line-point test passes for a function $f$ with high probability, the plane-point test also passes with high probability; then we use the soundness of the plane-point test, \autoref{thm:local-test-main-theorem-plane-point}, to deduce that $f$ is close to low degree. 

\begin{lemma}\label{lem:line_point_to_plane_point}
Let $\mathcal{H} = \{ H_1, \ldots, H_t\}$ be a set of hyperplanes in general position in $\F^m$. Let $\mathcal{S} \subseteq \F^m$ denote the set of all $m$-wise intersection points of the hyperplanes in $\mathcal{H}$ and $f:\mathcal{S}\to \F$ be a function. If the line-point test on function $f$ accepts with probability at least $(1-\epsilon)$, then the plane-point test on $f$ accepts with probability at least $(1 - \epsilon/\epsilon_0 - o(1))$. 
\end{lemma}
\begin{proof}
Since the hyperplanes in $\mathcal{H}$ are in general position, we have that the number of lines on every plane is the same, and every line is on the same number of planes. 
Thus, the distribution on $\mathcal{L}$ obtained by sampling a uniformly random plane $P$ from $\mathcal{P}$ and then sampling a uniformly random line $\ell$ from the set of lines contained in $P$ is uniform. 
Following the notation in \autoref{thm:local-test-main-theorem-2D}, we denote $g_{\ell}$ to be the degree $d$ univariate that is closest to the restriction $f_{\ell}$ of $f$ on the line $\ell$, and $\alpha()$ denotes the fractional agreement of two functions. 
Therefore, we have
\begin{align*}
\E_{\ell \in \mathcal{L}}\left[\alpha(f_{\ell}, g_{\ell}) \right] =& \E_{P \in \mathcal{P}} \left(\E_{\ell \in \mathcal{L}, \ell \subseteq P} \left[ \alpha(f_{\ell}, g_{\ell}) \right]\right) \, ,
\end{align*}
and these quantities are at least $(1-\epsilon)$ since the line-point test passes with probability at least $(1-\epsilon)$. 

We now note that for any plane $P$, the quantity $\left(\E_{\ell \in \mathcal{L}, \ell \subseteq P} \left[ \alpha(f_{\ell}, g_{\ell}) \right]\right)$
is precisely the probability that the line-point test passes when invoked on the restriction $f_P$ of $f$ on the plane $P$. 
Let this probability be denoted by $\epsilon_P$. 
From \autoref{cor:smooth-2D-test}, we have that for a bivariate function (namely, $f_P$ here), if the line-point test passes with probability $\epsilon_P$, then there is a degree $d$ polynomial $h_P$ such that $\alpha(h_P, f_P)$ is at least $1 - 5\epsilon_P/\epsilon_0 - o(1)$ , for some fixed constant $\epsilon_0 \in (0,1)$. 
Thus, we have that
\[
\E_{P \in \mathcal{P}} \alpha(h_P, f_P) \geq \E_{P \in \mathcal{P}} (1 - 5\epsilon_P/\epsilon_0 - o(1)) \, .
\]
Now, using the fact that $\E_P(1-\epsilon_P) = \E_{\ell \in \mathcal{L}}\left[\alpha(f_{\ell}, g_{\ell}) \right] \geq (1-\epsilon)$, we get that  
\[
\E_{P \in \mathcal{P}} \alpha(h_P, f_P) \geq  1 - 5\epsilon/\epsilon_0 - o(1)\, .
\]
Now, for every plane $P$, if $h_P$ is replaced by the degree $d$ bivariate closest to $f_P$, the quantity on the left-hand side can only increase. 
Thus, we have that the plane-point test on the function $f$ passes with probability at least $(1 - 5\epsilon/\epsilon_0 - o(1))$, as we wanted to show. 
\end{proof}

\subsection{Robust local characterization of bivariate polynomials using restrictions to lines}\label{sec:robust-local-characterization}
In this section, we prove \autoref{thm:robust-local-consistency-to-global-consistency-2d}.


\begin{proof}[Proof of \autoref{thm:robust-local-consistency-to-global-consistency-2d}]
Let $\mathcal{S}'$ be the set of points $\veca \in \mathcal{S}$ such that the two lines $\ell, \ell'$ in $\mathcal{H}$ that pass through $\veca$ agree with each other on $\veca$, i.e., $g_{\ell}(\veca) = g_{\ell'}(\veca)$. 
From the hypothesis of the theorem, we have that $|\mathcal{S}'| \geq (1-\epsilon) \cdot |\mathcal{S}| = (1-\epsilon) \cdot \binom{t}{2}$. \\

\noindent
\textbf{An error locator polynomial: } Our first observation is that there is a non-zero bivariate polynomial $E(x,y)$ of degree at most $\sqrt{\epsilon} t$ that vanishes on all points of the set $\mathcal{S}\setminus\mathcal{S}'$. 
To see this, we think of the coefficients of $E$ as formal variables of a polynomial of degree $\sqrt{\epsilon} t$ and set up a system of homogeneous linear constraints asserting that $E$ vanishes on all points in  $\mathcal{S}\setminus\mathcal{S}'$. 
The number of constraints is at most $\epsilon \binom{t}{2}$, whereas the number of variables is $\binom{\sqrt{\epsilon} t+2}{2}$, which exceeds the number of constraints. 
Thus, there is a non-zero solution. 
Such a polynomial $E$ plays the role of an \emph{error locator} for the rest of the proof. 
Let $e \leq \sqrt{\epsilon} t$ denote the degree of $E$. \\

\noindent
\textbf{A global low degree polynomial: }For every $\ell \in \mathcal{H}$, let $P_{\ell}(z)$ be the polynomial 
\[
P_{\ell}(Z) := g_{\ell}(Z) \cdot E_{\ell}(Z) \, ,
\]
where $E_{\ell}$ is the restriction of $E(X,Y)$ to the line $\ell$. Clearly, 
\[
\deg(P_{\ell}) = \deg(g_{\ell}) + \deg(E_{\ell}) \leq d + \deg(E_{\ell}) \, .
\]
We claim that for every pair of distinct lines $\ell, \ell' \in \mathcal{H}$, the polynomials $P_{\ell}$ and $P_{\ell'}$ are consistent with each other, i.e. 
they agree on the point $\veca$ where $\{\veca\} = \ell \cap \ell'$. 
To argue this, we consider two cases based on whether or not $\veca$ is in $\mathcal{S}’$. 
If $\veca \in \mathcal{S}'$, then we have 
\[
P_{\ell}(\veca) =  g_{\ell}(\veca) \cdot E_{\ell}(\veca) =  g_{\ell'}(\veca) \cdot E_{\ell'}(\veca)  = P_{\ell'}(\veca)\, .
\]
Here the second equality follows from the fact that since $\veca$ is in $\mathcal{S}'$, and $\ell$ and $\ell'$ are incident to it, the corresponding line polynomials $g_{\ell}$ and $g_{\ell'}$ must agree at $\veca$, and the third equality follows from the fact that by definition, $E_{\ell}$ and $E_{\ell'}$ are restrictions of $E$ and hence must agree at $\ell \cap \ell'$. 

If $\veca \notin \mathcal{S}'$, then by the definition of $E$, we get that 
\[
P_{\ell}(\veca) = P_{\ell'}(\veca)=  g_{\ell}(\veca) \cdot E_{\ell}(\veca) =  g_{\ell'}(\veca) \cdot E_{\ell'}(\veca)  = 0 \, .
\]

Applying the local characterization lemma (\autoref{lem:local-consistency-to-global-consistency}) to this collection $\{P_{\ell}: \ell \in \mathcal{H}\}$ of line polynomials of degree $d + e$, we get that if $t \geq d + e + 2$ (which holds since $t > \frac{d+2}{1-2\sqrt{\epsilon}}$ from the hypothesis of the theorem), then there is a global polynomial $P(X,Y)$ of degree at most $(d+e)$ such that for every line $\ell \in \mathcal{H}$, the restriction of $P$ on $\ell$ equals $P_{\ell}$ (hence justifying the notation). 

Now, from the above conclusion and the definition of $P_{\ell}$, we get that for every $\ell \in \mathcal{H}$, the restriction $E_{\ell}$ of the error locator polynomial $E$ on $\ell$ divides the restriction $P_{\ell}$ of $P$ on the line $\ell$. 
Since $t > \frac{2d}{1-3\sqrt{\epsilon}}$ and $e \leq \sqrt{\epsilon}t$, we get that $t > 2d + 3e = 2(d + e) + e \geq 2\max_{\ell}\{\deg(P)\} + \deg(E)$. 
Thus, from the divisibility lemma (\autoref{lem:ps-style-lemma-md}), we have that the bivariate polynomial $E(X,Y)$ divides the bivariate polynomial $P(X,Y)$.

Let $f(X,Y)$ be the resulting quotient $P(X,Y)/E(X,Y)$. 
Clearly, $\deg(f)$ is equal to $\deg(P) - \deg(E)$, which is at most $(d + e) - e = d$. 
To prove the theorem, it now suffices to show that the restriction $f_{\ell}$ of $f$ on a line $\ell$ equals $g_{\ell}$ for most lines $\ell$. \\

\noindent
\textbf{Agreement with many lines polynomials : }From the definition of $f$, we get that for every $\ell \in \mathcal{H}$, 
\[
P_{\ell}(Z) = f_{\ell}(Z) \cdot E_{\ell}(Z) \, ,
\]
but from the definition of $P_{\ell}$, we have that 
\[
P_{\ell}(Z) = g_{\ell}(Z) \cdot E_{\ell}(Z) \, .
\]
Thus, if $E_{\ell}(Z)$ is not identically zero, it must be the case that $f_{\ell}(Z)$ equals $g_{\ell}(Z)$. 
We now recall \autoref{clm:no-of-bad-lines} from the proof of \autoref{lem:ps-style-lemma-2d} which shows that the number of lines $\ell$ where $E(X,Y)$ can vanish identically is at most $\deg(E) = e \leq \sqrt{\epsilon} t$. 
Let $\mathcal{B}$ denote the set of all such lines. 
Thus, on all lines $\ell \in \mathcal{H}\setminus \mathcal{B}$, we get that $f_{\ell}$ must equal $g_{\ell}$. 
Thus, the number of such \emph{good} lines is at least $(1-\sqrt{\epsilon})t$. 
To complete the proof, we now argue that the fraction of such good lines is even higher. \\

\noindent
\textbf{An improved bound on good lines: } For a line $\ell \in \mathcal{H}$, let $\mathcal{I}_{\ell} \subseteq \mathcal{H}$ be the set of lines $\ell'$ such that $g_{\ell}$ and $g_{\ell'}$ do not agree on $\ell \cap \ell'$. 
Thus, $\mathcal{I}_{\ell}$ is the set of lines that are \emph{inconsistent} with $\ell$. 
We say that the lines outside $\mathcal{I}_{\ell}$ are \emph{consistent} with $\ell$. 
From the hypothesis of the theorem 
\[
\E_{\ell \in \mathcal{H}} \left[ |\mathcal{I}_{\ell}| \right] \leq \epsilon t \, .
\] 
Thus, by Markov's inequality, we get that 
\[
\Pr_{\ell \in \mathcal{H}} \left[ |\mathcal{I}_{\ell}| \geq ((1-\sqrt{\epsilon})t-d) \right] \leq \frac{\epsilon t}{(1-\sqrt{\epsilon})t-d} \leq 2\epsilon \, ,
\]
where the last inequality follows from the fact that $t > \frac{2d}{1-3\sqrt{\epsilon}} \geq \frac{2d}{1-2\sqrt{\epsilon}}$.
Thus, we have that for $(1-2\epsilon)t$ of the lines $\ell$, the set $\mathcal{I}_{\ell}$ of lines that are inconsistent with $\ell$ is of size less than $((1-\sqrt{\epsilon})t-d)$. 
Let $\ell$ be one such line. 
Then, $\ell$ is consistent with at least $(\epsilon \sqrt{t} + d + 1)$ lines in $\mathcal{H}$. 
Among these consistent lines, at most $|\mathcal{B}| \leq \sqrt{\epsilon}t$ of these can be in the set $\mathcal{B}$. 
Thus, there are at least $(d+1)$ lines $\ell'$ such that $\ell$ and $\ell'$ are consistent with each other, i.e. 
$g_{\ell}$ and $g_{\ell'}$ agree on $\ell \cap \ell'$, and $f_{\ell'} = g_{\ell'}$. 
So, there are at least $(d+1)$ distinct points on $\ell$, namely its distinct intersection points $\veca$ with $(d+1)$ lines $\ell' \notin \mathcal{B}$ that $\ell$ is consistent with, such that 
\[
f_{\ell}(\veca) = f_{\ell'}(\veca) = g_{\ell}(\veca) = g_{\ell'}(\veca) = f(\veca) \, . 
\]
Thus, $f_{\ell}$ and $g_{\ell}$ must be equal to each other. 

Therefore, the set of lines $\ell$ on which the restriction of $f$ equals $g_{\ell}$ is of size at least $(1-2\epsilon)t$. 
This completes the proof of the theorem. 
\end{proof}

\subsection{The divisibility lemma}
\label{sec:divisibility}
In this subsection, we prove the divisibility lemma, \autoref{lem:ps-style-lemma-md}. 


\subsubsection{Proof of Lemma \ref{lem:ps-style-lemma-md} in 2 dimensions}\label{sec:ps-2-dim}

We now prove the technically easier $m=2$ case of \autoref{lem:ps-style-lemma-md},
which we state below as a lemma.
\begin{lemma}\label{lem:ps-style-lemma-2d}
Let $\F$ be any field, $e, d, t$ be natural numbers, and $\mathcal{H}$ be a set of $t$ lines in $\F^2$ in general position. 
$E(X,Y)$ and $P(X,Y)$ are polynomials in $\F[X,Y]$ of degree equal to $e$ and $d$ respectively. 
For any line $\ell \subseteq \F^2$, let $E_\ell(Z)$ and $P_{\ell}(Z)$ respectively be the univariate polynomials in $\F[z]$ obtained by the restriction of $E$ and $P$ to the line $\ell$. 

If $t > 2d + e$ and for all lines $\ell \in \mathcal{H}$, there exists a polynomial $Q_{\ell}(Z)$ such that 
\[
P_{\ell}(Z)  = E_{\ell}(Z) \cdot Q_{\ell}(Z) \, ,
\]
then, there exists a polynomial $Q(X,Y) \in \F[X,Y]$ such that 
\[
P(X,Y) = E(X,Y) \cdot Q(X,Y) \, .
\]
\end{lemma}

\subsection*{Bezout's theorem related preliminaries}
To prove the lemma, we first take a detour and discuss Bezout's theorem and some of its applications, which turn out to be crucial for the rest of our analysis.  
We start with a discussion of the bivariate case first before moving on to high dimensions. 

For the rest of this section, we assume $\F$ is an algebraically closed field.
We begin by defining the intersection multiplicity of two bivariate polynomials at a point $(\alpha, \beta) \in \F^2$. 
The definition is in terms of the local ring of $\F^2$ at that point, which we now define.

\begin{definition}[Local ring at a point in $\F^2$]
Let $p = (\alpha, \beta) \in \F^2$. 
The local ring of $p$, denoted by $\OO_p$ is defined as:
$$ \OO_p = \left\{ \frac{A(x, y)}{S(X,Y)} \mid A(X,Y), S(X,Y) \in \F[X,Y], S(\alpha, \beta) \neq 0 \right\},$$
with ring operations being the usual operations on rational functions in $X, Y$.
\end{definition}


We now define the notion of intersection multiplicity of curves at a point. 
\begin{definition} [Intersection Multiplicity of curves in $\F^2$~{\cite[Section 3.3]{fultoncurves}}]

 Let $A(X,Y), B(X,Y) \in \F[X,Y]$ be relatively prime polynomials.
 Let $p \in \F^2$ be a point such that 
 $F(p) = G(p) = 0$.
 
 Then {\em the intersection multiplicity of $A$ and $B$ at $p$}, denoted $\imult(A,B;p)$ is defined by:
 $$\imult(A, B; p) = \dim_{\F} \left( \OO_{p} / I\right),$$
 where $\OO_p$ is the local ring of the point $p \in \F^2$,
 and $I$ is the ideal of $\OO_p$ generated
 by $A$ and $B$.
\end{definition}

To help get some intuition for this notion, we mention its relation with the more familiar notion of the vanishing multiplicity of a polynomial at a point. 
First, if $F$ vanishes with multiplicity $m_1$ at $p$ and $G$ vanishes with multiplicity $m_2$ at $p$, then the intersection multiplicity of $F$ and $G$ at $p$ is at least $m_1 \cdot m_2$. 
It could be more; it will be strictly greater than $m_1 \cdot m_2$ if (and only if) the curves cut out by $F$ and $G$ have a common tangent at $p$. 

We are now ready to state Bezout's theorem for curves in a two-dimensional space. 

\begin{theorem}[Bezout's theorem for curves in $2$-dimensional space~{\cite[Section 5.3]{fultoncurves}}]\label{thm:bezout-2d} 
 Let $A(X,Y)$, $B(X,Y)$ in $\F[X,Y]$ be relatively prime polynomials
 of degrees $d,e$ respectively.
 Let $p_1, \ldots, p_s \in \F^2$ be common zeroes of $F$ and $G$.
 
 Then
 $$ \sum_{i=1}^s  \imult(A, B; p_i) \leq d \cdot e.$$
\end{theorem}

We can now prove the following technical lemma, which we will crucially use later in proving the testability of 2-dimensional GAP codes.
 \begin{lemma}\label{lem:univariate-mult-intersection-mult}
Let $\F$ be an algebraically closed field and $U(X,Y), V(X,Y)$ be non-zero polynomials. 
Let $a, b, \alpha\in \F$ and $k \in \N$ be such that the univariate polynomials $U(Z,aZ+b), V(Z,aZ+b)$ vanish with multiplicity at least $k$ at $\alpha$. 

Then, the intersection multiplicity of the curves $U(X,Y)  = 0$ and $V(X,Y) = 0$ at the point $(\alpha, a\alpha + b)$ is at least $k$. 
\end{lemma}
\begin{proof}
 Let $p = (\alpha, a \alpha + b) \in \F^2$.

Now since $U(Z, aZ+b)$ vanishes with multiplicity at least $k$ at $\alpha$. 
Then $(Z-\alpha)^k$ divides $U(Z, aZ+b)$,
and this means that $U(X,Y)$ can be written in the form:
$$ U(X,Y) =   (Y-aX-b)\cdot A(X,Y) + (X-\alpha)^k \cdot B(X,Y)$$
for some $A(X,Y), B(X,Y) \in \F[X,Y]$.
Similarly $V(Z, uZ+b)$ vanishes with multiplicity at least
$k$ at $\alpha$, and thus $V(X,Y)$ is of the form:
$$ V(X,Y) =   (Y-aX-b)\cdot A'(X,Y) + (X-\alpha)^k \cdot B'(X,Y)$$
for some $A'(X,Y), B'(X,Y) \in \F[X,Y]$.

Thus, the ideal $I = \langle U, V \rangle$ of $\OO_p$ 
is contained in the ideal $J = \langle (Y-aX-b), (X-\alpha)^k\rangle$ of $\OO_p$. 
This means that
$$ \imult(U,V ; p) = \dim_\F(\OO_p/I) \geq \dim_\F(\OO_p/J).$$
Thus, to complete the proof, it suffices to show that $\dim_{\F}(\OO_p/J) \geq k$.
To see this\footnote{An alternative proof of this fact is as follows. 
$\dim_{\F}(\OO_p/J)$ equals the intersection multiplicity of the polynomials $Y - a X- b$ and $(X-\alpha)^k$ at the point $(\alpha,a\alpha + b)$, and this should be $k$ (by Bezout's theorem) because it is the intersection multiplicity at the unique point of intersection (in the full projective plane) of two relatively prime polynomials, 
one of degree $k$ and one of degree $1$.}, we will show that the elements
$1, (X-\alpha), \ldots, (X-\alpha)^{k-1}$ are $\F$-linearly independent in the $\F$-vector space $\OO_p/J$. 

We do this via contradiction. 
Suppose these elements are linearly dependent over $\F$ in $\OO_p/J$. 
Then there exist
$a_0, \ldots, a_{k-1} \in \F$, not all $0$, 
such that
$$ \sum_{i=0}^{k-1} a_i (X-\alpha)^i \in J$$
in the ring $\OO_z$.
Let $j$ be the smallest such that $a_j \neq 0$.
This means (recalling the definition of the local ring) that there exists some $S(X,Y) \in \F[X,Y]$
with $S(\alpha, a\alpha + b) \neq 0$, and some $R(X,Y), R'(X,Y) \in \F[X,Y]$ such that we have the following equation in $\F[X,Y]$.
$$ S(X,Y) \cdot (X-\alpha)^{j} \cdot \left( \sum_{i=j}^{k-1} a_i (X-\alpha)^{i-j} \right) = R(X,Y) \cdot (Y-aX-b) + R'(X,Y) \cdot (X-\alpha)^k.$$
Setting $Y = aX + b$ in this equation, we get
$$ S(X,aX+b) \cdot (X-\alpha)^{j} \cdot \left( \sum_{i=j}^{k-1} a_i (X-\alpha)^{i-j} \right) =  R'(X,aX+b) \cdot (X-\alpha)^k.$$
Since both sides are divisible by $(X-\alpha)^j$, we can cancel it from both sides, and we get
$$ S(X,aX+b) \cdot \left( a_j + \sum_{i=j+1}^{k-1} a_i (X-\alpha)^{i-j} \right) =  R'(X,aX+b) \cdot (X-\alpha)^{k-j}.$$
Finally, substituting $X = \alpha$ into this equation and using the fact that $S(\alpha, a\alpha + b) \neq 0$, we get that $a_j = 0$, a contradiction.

This gives us our desired lower bound on the intersection multiplicity of $U(X,Y)$ and $V(X,Y)$ at the point $(\alpha, a \alpha + b)$.
\end{proof}

\subsection*{Proof of \autoref{lem:ps-style-lemma-2d}}

We now proceed with the proof of \autoref{lem:ps-style-lemma-2d}. 
\begin{proof}[Proof of \autoref{lem:ps-style-lemma-2d}]
Let $E(X,Y) = G(X,Y)\cdot E'(X,Y)$ and $P(X,Y)  = G(X,Y) \cdot P'(X,Y)$, for polynomials $E', P', G$ where $G$ is the GCD of $E$ and $P$. 
Therefore, $ E’ $ and $ P’ $ are relatively prime. 
From the hypothesis of the lemma, we get that for every line $\ell \in \mathcal{H}$, we have that there exists a polynomial $Q_{\ell}(Z)$ such that 
\[
P_{\ell}'(Z)\cdot G_{\ell}(Z)  = E'_{\ell}(Z) \cdot G_{\ell}(Z) \cdot Q_{\ell}(Z) \, ,
\]
where $E_{\ell}', P_{\ell}', G_{\ell}$ denote the restrictions of $E', P'$ and $G$ respectively to the line $\ell$. 
If the degree of $E'$ is zero, i.e. 
it is a constant, then clearly, $E(X,Y)$ must divide $P(X,Y)$ with quotient being a constant multiple of $P'(X,Y)$ and we are done. 
So, for the rest of this proof, we assume that the degree of $E'$ is at least one. 
Let $e'$ denote the degree of $E'(X,Y)$ and $d'$ denote the degree of $P'(X,Y)$. 

We say that a line $\ell \in \mathcal{H}$ is \emph{bad} if either $G_{\ell}$ is identically zero, or the degree of $E_{\ell}'$ is strictly less than the degree of $E'(X,Y)$. 
Let $\mathcal{H}'$ be the subset of $\mathcal{H}$ are \emph{not} bad. 
This is the set of lines of interest to us for the rest of the proof, and the following claim, whose proof we defer to the end, gives a lower bound on the number of such lines. 

\begin{claim}\label{clm:no-of-bad-lines}
\[
|\mathcal{H}'| \geq |\mathcal{H}| - e \, .
\]
\end{claim}

For any such line $\ell \in \mathcal{H}'$, we have that the univariate polynomial $E_{\ell}'$ has degree equal to $e'$ and divides the univariate polynomial $P_{\ell}'$. 
In this case, each of the zeroes $\alpha$ of $E_{\ell}'(Z)$ is also a zero of $P_{\ell}'$, and in particular, this means that the point $(\alpha, m\alpha + b)$ is a common zero of the bivariates $E'(X,Y)$ and $P'(X,Y)$. 
Here $m, b \in \F$ are such that $\ell := \{(a,am + b): a \in \F\}$. 
Note that the zero $\alpha$ might not be in $\F$ but certainly lives the algebraic closure $\overline{\F}$ of $\F$. 

The idea for the rest of the proof is that if the zeroes of $E_{\ell}'$ happened to be all distinct, then, for each line $\ell$ in $\mathcal{H}'$, we get $e'$ distinct common zeroes of $E'(X, Y)$ and $P'(X, Y)$ in $\overline{\F}^2$. 
Moreover, we know that these polynomials do not have a common factor. 
Hence, from Bezout's theorem (\autoref{thm:bezout-2d}), we get that 
\[
\frac{1}{2}\cdot e'\cdot |\mathcal{H}'| \leq \deg(P') \cdot \deg(e') = d'\cdot e' , 
\]
where the factor $1/2$ is to account for the fact that  each common zero $(\alpha, m\alpha + b)$ can be a point of intersection of two lines in $\mathcal{H}'$, and be counted twice. 
Moreover, since $\mathcal{H}'$ is a set of lines in general position, no point $(\alpha, m\alpha + b)$ can lie on three or more lines in $\mathcal{H}'$. 
This gives an upper bound of $2d' \leq 2d$ on $|\mathcal{H}'|$ and of $(2d + e)$ on $|\mathcal{H}|$ using \autoref{clm:no-of-bad-lines}. 

The technical issue with this argument is that the polynomial $E_{\ell}'$ might not have distinct zeroes but have zeroes $\alpha$ with high multiplicity, and this weakens the counting argument above and only yields a bound of $d'e' + e \leq de + e$ on $|\mathcal{H}|$. 
To obtain a better bound, we note from \autoref{lem:univariate-mult-intersection-mult} that a zero of \emph{multiplicity} $k$ of $E_{\ell}'$ (and hence $P_{\ell}'$) also yields a common point on the curves of $E'(X,Y) = 0$ and $P'(X,Y) = 0$ of {intersection multiplicity} $k$. 
Now, using \autoref{thm:bezout-2d} and the counting argument above, we get the lemma.
\end{proof}

Next, we prove \autoref{clm:no-of-bad-lines}. 
\begin{proof}[Proof of \autoref{clm:no-of-bad-lines}]
Let a line $\ell$ be parameterized by its slope $m$ and intercept $b$ such that $\ell := \{(a, am + b): a \in \F\}$. 
We will show that if $\ell$ is \emph{bad}, then its slope $m$ must be a zero of a non-zero univariate polynomial of low degree.  
Thus, there are only a few choices of the slope $m$. 
Moreover, since the lines in $\mathcal{H}$ are in general position, there is at most one line of slope $m$ for any $m \in \F$. 
Thus, the number of bad lines must be small. 

Let $e' \geq 1$ be the degree of $E'(X,Y)$, and let $U_i(X,Y)$ denote the homogeneous component of $E'(X,Y)$ of degree equal to $i$. 
Thus, 
\[
E'(X,Y) = \sum_{i = 0}^{e'} U_i(X,Y) \, .
\]
Let $U_{e'}(X,Y) := \sum_{j = 0}^{e'} u_{e',j} Y^jX^{e'-j}$ for constants $u_{e',j}$, not all of which are zero. 
Thus, the restriction of $E'$ on $\ell := \{(a,am + b): a \in \F\}$ is given by 
\[
E_{\ell}'(Z) = E'(Z,mZ+b) = \sum_{i = 0}^{e'} U_i(Z,mZ+b) \, ,
\]
and the coefficient of $Z^{e'}$ in $E_{\ell}'(Z)$ equals $U_{e'}(1,m) = \sum_{j = 0}^{e'} u_{e',j} m^j$. 
In other words, the degree of  $E_{\ell}'(Z)$ is less than $e'$ if and only if the slope $m$ of $\ell'$ is a zero of the univariate polynomial $(\sum_{j = 0}^{e'} u_{e',j}\eta^j)$ ($\eta$ is a formal variable here). 
Now, $U_{e'}(X,Y)$ is non-zero, we have that  $(\sum_{j = 0}^{e'} u_{e',j}\eta^j)$ is a non-zero polynomial of degree $e'$, there are at most $e'$ such choices of $m$. 

A very similar argument implies also that the number of lines $\ell$ on which $G(X, Y)$ vanishes identically, i.e. 
$G_{\ell}$ is identically zero is also bounded by $\deg(G)$. 

Thus, the slope $m$ of any bad line must come from a set of size at most $\deg(E') + \deg(G) = \deg(E) = e$, and since there are no parallel lines in $\mathcal{H}$, the number of bad lines is at most $e$. 
\end{proof}

\subsubsection{Proof of Lemma \ref{lem:ps-style-lemma-md} for general $m$}\label{sec:line-point-mD-divisibility}

Our proof of this lemma follows the same high-level strategy as for the $2$-dimensional case. 
 However, because we need to deal with a higher-dimensional version of Bezout’s theorem and higher-dimensional intersection multiplicities, we end up needing some slightly heavier commutative algebra. 
 For algebra and algebraic geometry background, we refer the reader to~\cite{Eisenbud,Hartshorne}. 
 We only define the
 algebraic concepts about which we need to reason.
 
 \subsubsection*{Bezout's theorem related preliminaries}
 We now give a quick introduction to the statement of Bezout's theorem that we need.
 
 Let $A(X_1, \ldots, X_m)$ and $B(X_1, \ldots, X_m)$ be relatively prime polynomials; they define codimension $1$ varieties $V(A)$ and $V(B)$ in $\F^m$. 
 We want to study their intersection. 
 Since $A$ and $B$ are relatively prime, their intersection will be codimension $2$.
 
Informally,  Bezout's theorem for this situation is the following.
 Let $W_1, \ldots, W_s \subseteq \F_q^m$ be the irreducible components of the intersection of $V(A)$ and $V(B)$ (they are all codimension $2$ by
 ~\cite[Theorem I.7.1]{Hartshorne}). 
 For this setup, we can define the intersection multiplicity $\imult(A,B;W_i)$ of $A$ and $B$ at each variety $W_i$. 
 Bezout's theorem says that
 $$ \sum_{i} \imult(A,B;W_i) \cdot \deg(W_i) \leq \deg(A) \cdot \deg(B),$$
 where $\deg(W_i)$ is the degree of the variety $W_i$.  
 In fact, there is equality in this, provided that $A$ and $B$ do not have a dimension $m-2$ component contained in the ``hyperplane at infinity''.

 To state Bezout’s theorem formally and for our proof, we will need to understand something about intersection multiplicities. 
 We start with the definition.
 
 \begin{definition}[Intersection multiplicity in $m$ dimensions {\cite[Section I.7]{Hartshorne}}]\footnote{There are some minor differences between what we write here and the definition in~\cite{Hartshorne}. 
     \begin{enumerate}
            \item We only talk about affine varieties.
            \item We only talk about the intersection of codimension $1$ varieties, while in~\cite{Hartshorne}, one of the varieties is allowed to be of arbitrary codimension.
            \item We do not require $A$ and $B$ to be irreducible. Since we are in the codimension $1$ setting, $A$ and $B$ can be factored into irreducible factors, and everything generalizes since the definition of intersection multiplicity satisfies the bilinearity relations $\imult(A_1A_2, B; W) = \imult(A_1, B;W) + \imult(A_2,B;W)$ and $\imult(A, B_1B_2; W) = \imult(A, B_1;W) + \imult(A,B_2;W)$ whenever the $A$'s and $B$'s are relatively prime.
    \end{enumerate}
    }
    Let $A(X_1, \ldots, X_m)$, $B(X_1, \ldots, X_m) \in \F[X_1, \ldots, X_m]$ be relatively prime polynomials, and let $V(A), V(B) \subseteq \F^m$ be their zero sets.
    Let $W$ be an irreducible variety in $\F^m$ of dimension $2$.
    Let $\mathfrak p$ be the prime ideal
    of $W$ in
    $\F[X_1, \ldots, X_m]$.

    Then we define {\em the intersection multiplicity of $A$ and $B$ at $W$}, denoted by $\imult(A, B; W)$ as 
    $$ \imult(A, B; W) = \length_{\OO_{\mathfrak p}}(\OO_{\mathfrak p}/I),$$
    where $\OO_{\mathfrak p}$ is the local ring of 
    $\mathfrak p$ and $I$ is the ideal of $\OO_{\mathfrak p}$ generated by $A$ and $B$.
 \end{definition}
 
 Note that if $A$ and $B$ do not both vanish on $W$, then $\imult(A,B;W) = 0$.
 
 \paragraph{Intersections with a hyperplane: }
 One property of $\imult$ that we will need concerns the case where
 $B$ is degree $1$ (i.e., $V(B)$ is a hyperplane $H$). 
 In this case,
 for any irreducible polynomial $U(Z_1, \ldots, Z_{m-1})$ viewed
 as a polynomial on $H$, if we let $W$ denote the $m-2$ dimensional 
 irreducible variety $V(U) \subseteq H$, then we have:
 $$\imult(A, B; W) = \mbox{ the highest power of $U$ that divides $A_H$ }.$$
 Abusing notation, when $B$ is degree $1$ and thus $V(B)$ is a hyperplane $H$, we will sometimes refer to 
 $\imult(A,B ; W)$ as $\imult(A, H ; W)$.
 
 We will also need the property that $\deg(W)$ in this case equals the degree of the polynomial $U(Z_1, \ldots, Z_{m-1})$.

 With the definition of $\imult$ in hand, we can now state Bezout's theorem in the $m$-dimensional setting.

 \begin{theorem}[Bezout's theorem in $m$-dimensions (for $2$ hypersurfaces) {\cite[Theorem I.7.7]{Hartshorne}}]\label{thm:bezout-md}
  Let $A(X_1, \ldots, X_m), B(X_1, \ldots, X_m) \in \F[X_1, \ldots, X_m]$ be relatively prime polynomials, and let $V(A), V(B) \subseteq \F^m$ be their zero sets. 
  Let $W_1, \ldots, W_s$ be the irreducible components of $V(A) \cap V(B)$. 
  Then,
  $$\sum_{j=1}^s \imult(A, B ; W_j) \leq \deg(A) \cdot \deg(B).$$
  
  Furthermore, if the closures $V(A)$ and $V(B)$ in projective space
  $\mathbb{P}^m(\F)$ do not contain any common dimension $m-2$ components, then the inequality above is, in fact, an equality.
 \end{theorem}

 
\subsubsection*{Proof of \autoref{lem:ps-style-lemma-md}}

We can now prove the divisibility lemma for general $m$, \autoref{lem:ps-style-lemma-md}.

\begin{proof}[Proof of \autoref{lem:ps-style-lemma-md}]
The high-level outline of the lemma is similar to that of the two-dimensional case. 
We begin with a high-level sketch.
First, we may assume that $E$ and $P$ are relatively prime;
otherwise we can divide out any common factors and reduce to the relatively prime case. 
Now, suppose $e \geq 1$ -- otherwise $e =0$, and the conclusion holds trivially.

 Our plan is to apply Bezout's theorem to $E$ and $P$. 
 Each $H$ for which $E_H$ divides $P_H$ will account for some irreducible components $W_i$ in the intersection of $V(E)$ and $V(P)$
 which have a large total of the product of their degree and intersection multiplicity. 
 This gives a lower bound, via Bezout's theorem, on $\deg(E) \cdot \deg(P)$, for which we have an apriori upper bound. 
 Resolving this yields the result.
 
As in the $m=2$ case, we need a lemma about the intersection multiplicity of hypersurfaces that share a common tangent hyperplane. We prove this lemma at the end of this section.
 \begin{lemma}\label{lem:md-imult}
  Suppose $E, P$ are relatively prime polynomials in $\F[X_1, \ldots, X_m]$
  and $H$ is a hyperplane in $\F^m$. 
  Let $W$ be an irreducible variety of dimension $m-2$ in $H$, and suppose both $\imult(E, H; W)$ and $\imult(P,H;W)$ are at least $k$. 
  Then,
  $$ \imult(E, P; W) \geq k.$$
 \end{lemma}

 Now, we proceed with the formal proof.
 Let $$\mathcal H' = \{ H \in \mathcal H \mid \deg(E_H) = e \}.$$
 Note that there can be very few $H \in \mathcal H$ which do not lie in $\mathcal H'$. 
 This is a natural generalization of \autoref{clm:no-of-bad-lines}.  
 Indeed, suppose $E(Z_1, Z_2, \ldots, Z_{m-1}, \langle \veca, \mathbf Z \rangle + b)$ has degree strictly less than $e$. 
 Then letting $E^*$ be the highest degree homogeneous part of $E$, we have that $E^*(Z_1, \ldots, Z_{m-1}, \langle \veca, \mathbf Z\rangle) = 0$, and thus 
 $E^*(Z_1, \ldots, Z_{m-1}, Y )$ is divisible by $(Y - \langle \veca, \mathbf Z \rangle)$. 
 Since the $H$ in $\mathcal H$ are in general position and thus correspond to different $\veca$s, we get that each $H \in \mathcal H \setminus \mathcal H'$ yields a distinct factor of $E^*$, and so there can be at most $e$ such $H$. 
 Thus, we showed,
 $$ | \mathcal H' | \geq |\mathcal H| - e > 2d.$$
 Now let $H \in \mathcal H'$, and let us consider
 the polynomials $E_H$ and $P_H$. 
 Let $\mathcal W_H$ be the set of
 irreducible components of $V(E) \cap H$.
 Since $E_H$ divides $P_H$, then for every irreducible factor
 $U(Z_1, \ldots, Z_{m-1})$ of $E_H$, the highest power of $U$ that
 divides $P_H$ is at least the highest power of $U$ that divides $E_H$.
 Thus $W \in \mathcal W_H$, we have,
 $$ \imult(P, H; W) \geq  \imult(E, H; W).$$
 Now, from \autoref{lem:md-imult}, we get that for every $H \in \mathcal H'$ and every 
 $W \in \mathcal W_H$,
 $$ \imult(E, P; W) \geq \imult(E, H; W).$$
 Thus we have,
  \begin{align*}
    &\sum_{W \in \mathcal W_H} \imult(E, P; W) \deg(W) \\
    &\geq 
    \sum_{W \in \mathcal W_H} \imult(E, H; W) \deg(W) \\
    &= \deg(E_H)\\
    &= e.
  \end{align*}
In words, every $H \in \mathcal H'$ accounts for many of the intersection multiplicity of $E$ and $P$. 
Now consider the expression $\Delta$ defined below.
 \begin{align*}
\Delta &:= \sum_{H \in \mathcal{H}'} \sum_{W \in  \mathcal W_H } \imult(E, P ; W) \deg(W) \\
&\geq 
|\mathcal H'| \cdot e \\
&> 2de,
\end{align*}
where the inequality is strict because of our assumption that $e \geq 1$.
 
 On the other hand, $\Delta$ can be bounded from above by Bezout's theorem. 
 Let $\mathcal F$ be the set of all irreducible components of $V(E) \cap V(P)$. 
 (Note that they are all $m-2$ dimensional by~\cite[Theorem I.7.1]{Hartshorne}.). 
 From \autoref{thm:bezout-md}, we have  
 \begin{align*}
\Delta &= \sum_{H \in \mathcal{H}'} \sum_{W \in  \mathcal W_H } \imult(E, P ; W) \deg(W) \\
&= \sum_{H \in \mathcal{H}'} \sum_{W \in  \mathcal F } 1_{W \subseteq H} \imult(E, P ; W) \deg(W) \\
 &= \sum_{W \in \mathcal{F}} \sum_{H \in \mathcal{H}' } 1_{W \subseteq H} \imult(E,P ; W) \deg(W)\\
 &= \sum_{W \in \mathcal{F}}   \imult(E,P ; W) \deg(W) \left(\sum_{H \in \mathcal{H}' } 1_{W \subseteq H} \right)\\
 &\leq 2 \sum_{W \in \mathcal{F}}   \imult(E,P ; W) \deg(W)\\
 &\leq 2 d e,
 \end{align*}
 where the last line is Bezout's theorem, and the penultimate inequality is because any $(m-2)$-dimensional variety $W$ can be contained in at most $2$ of the hyperplanes $H \in \mathcal H'$  (since they are in general position -- and this can only happen if $W$ is the $(m-2)$-dimensional intersection of two hyperplanes in $\mathcal H'$).

 This gives us a contradiction, and the result follows. 
\end{proof}

 \subsubsection*{Proof of \autoref{lem:md-imult}}

 We now prove \autoref{lem:md-imult}.
 
 \begin{proof}[Proof of \autoref{lem:md-imult}]
  Let us rename the $m$ variables $X_1, \ldots, X_m$ to 
  $Z_1, \ldots, Z_{m-1}, Y$.
  
  By an affine change of variables, we may assume that the
  hyperplane $H$ is defined by $Y = 0$, and $W$, the irreducible
  $m-2$-dimensional subvariety of $H$, is defined by the two equations $ Y = 0,   U(Z_1, \ldots, Z_{m-1}) = 0$, where $U$ is an irreducible polynomial.
  (This uses the fact that every irreducible subvariety that is codimension $1$ in an affine space can be defined by a single irreducible polynomial.)
  Thus the prime ideal of $\F[Z_1, \ldots, Z_{m-1}, Y]$ associated with $W$, $\mathfrak p $,  is generated by $y$ and $U(Z_1, \ldots, Z_{m-1})$.
  
  After the variable change, the polynomial $E_H(Z_1, \ldots, Z_{m-1})$ becomes 
  $E(Z_1, \ldots, Z_{m-1}, 0)$ and 
  $P_H(Z_1, \ldots, Z_{m-1})$ becomes $P(Z_1, \ldots, Z_{m-1}, 0)$. Thus $E(Z_1, \ldots, Z_{m-1}, Y)$ and $P(Z_1, \ldots, Z_{m-1}, Y)$ are of the form:
  $$ E(Z_1, \ldots, Z_{m-1}, Y) = E_H(Z_1, \ldots, Z_{m-1}) + Y \cdot C_E(Z_1, \ldots, Z_{m-1}, Y),$$
  $$ P(Z_1, \ldots, Z_{m-1}, Y) = P_H(Z_1, \ldots, Z_{m-1}) + Y \cdot C_P(Z_1, \ldots, Z_{m-1}, Y),$$
  for some polynomials $C_E$, $C_P$. 
 
 Our hypotheses $\imult(E,H; W) \geq k$ and $\imult(P,H ; W) \geq k$ tell
  us that $E_H(Z_1, \ldots, Z_{m-1})$ and $P_H(Z_1, \ldots, Z_{m-1})$ are both divisible by $U(Z_1, \ldots, Z_{m-1})^k$. 
  
  Thus, the polynomials $E(Z_1, \ldots, Z_{m-1}, Y)$ and $P(Z_1, \ldots, Z_{m-1}, Y)$ both lie in the ideal $J$ of $\F[Z_1, \ldots, Z_{m-1}, Y]$ generated by 
 $U(Z_1,\ldots, Z_{m-1})^k$ and $Y$.
  
 Now we consider $\imult(E, P; W)$. 
 Let $\mathfrak p$ be the  prime ideal of $W$ in $\F[Z_1, Z_2, \ldots, Z_{m-1},Y]$. 
 By definition:
 $$\imult(E,P; W) = \length_{\OO_{\mathfrak p}}( \OO_{\mathfrak p} / I ),$$
 where $I$ is the ideal of $\OO_W$ generated by $E$ and $P$.
 By our previous observation, if we let $J$ be the ideal of 
 $\OO_{\mathfrak p}$ generated by $U(Z_1,\ldots, Z_{m-1})^k$ and $Y$,
 then $I \subseteq J$.
 Thus $$ \imult(E,P ; W) \geq \length_{\OO_{\mathfrak p}}(\OO_{\mathfrak p} / J ).$$
This latter expression simply equals
 $\imult( Y, U(Z_1, \ldots, Z_{m-1})^k ; W )$, which,
 by the general comments about intersection multiplicities of polynomials with hyperplanes, equals $k$. 
 Alternatively, $\length_{\OO_{\mathfrak p}}(\OO_{\mathfrak p} / J)$ can be seen to be at least $k$ because of the length $k$ sequence of $\OO_{\mathfrak p}$ modules
 $$ \{0 \} \subseteq \OO_{\mathfrak p}/ \langle Y, U \rangle \subseteq \ldots \subseteq \OO_{\mathfrak p}/ \langle Y, U^i \rangle \subseteq \ldots \subseteq \OO_{\mathfrak p}/ \langle Y, U^k \rangle = \OO_{\mathfrak p}/ J.$$
This completes the proof.
\end{proof}

\bibliographystyle{alphaurlpp}
\bibliography{crossref,references}

\appendix

\section{$d$-robustness in 2 variables}
\label{appendix:d-robustness-bounds}

In this section, we investigate the problem of finding the smallest set with a given $d$-robustness. This corresponds to the best $R$ vs $\delta$ tradeoff that can be obtained from our generalized Schwartz-Zippel lemma for shapes. Our main results are that in the regime of large $\delta$ (close to $1$) and small $\delta$ (close to $0$), the grid and the simplex are essentially optimal. As we already saw, for intermediate $\delta$, there are shapes that outperform both of these.

We first show the grid is optimal for relative $d$-robustness greater than $1/2$.

\subsection{High robustness regime}

First, we show that any set with relative $d$-robustness greater than $1/2$ must contain the point $d, d$.

\begin{lemma}\label{lem:large-distance-contains-dd}
    Let $\delta > 1/2$. Suppose $S \subset \N^2$ is such that $\pi_d(S) \geq \delta$. Then $S$ contains $(d, d)$.
\end{lemma}

\begin{proof}
    By contradiction, suppose $(d, d) \notin S$. 
    Define the following partition of $S$. 
    $S_x = S \cap H(d, 0)$, $S_y = S\cap H(0, d)$, and $S_\bot = \left\{ (x, y): x < d \land y < d \right\}$. 
    In words, $S_x$ is the subset of $S$ with $x$ coordinate at least $d$, $S_y$ is the subset of $S$ with $y$ coordinate at least $d$, and $S_\bot$ is the subset where both $x$ and $y$ coordinates are less than $d$. 
    Since $S$ is downward closed, there are no elements of $S$ with both $x$ and $y$ coordinates at least $d$, and thus $(S_x, S_y, S_\bot)$ is indeed a partition of $S$.

    Since $\pi_d(S) \geq \delta$ $d$-robust, we have $|S_x| \geq \delta |S|$, and $|S_y| \geq \delta|S|$. Thus, we have $|S_x| + |S_y| \geq 2 \delta|S|$ which implies $|S| \geq 2 \delta |S|$, so $\delta \leq 1/2$, which is a contradiction.
\end{proof}

We now prove the following theorem, which implies that the grid is optimal for $\delta \geq 1/2$.

\begin{theorem}
    \label{thm:high-distance}
    Let $S \in \N^2$ with $\pi_d(S) > 1/2$.
    Then $$|S| \geq \left( \sqrt{\delta |S| + \frac{d^2}{4}} + \frac{d}{2}  \right)^2.$$
    Furthermore, setting $R = \frac{d^2}{2|S|}$, we have $R \leq (1 - \delta)^2/2$.
\end{theorem}

\begin{proof}
 Let $V_x = S \cap H(d, 0)$.
 Let $V_y = S \cap H(0, d)$.
 Let $U = V_x \cap V_y$.
 By hypothesis, $|V_x|, |V_y| \geq \delta |S|$. Let $c = \frac{|V_x| + |V_y|}{2}$, and note that $c \geq \delta |S|$.
 
 We have:
 \begin{align*}
 |S| &= d^2 + |V_x| + |V_y| - |V_x \cap V_y|\\
 &= d^2 + 2c - |U|
 \end{align*}
 
 We now get an upper bound on $|U|$ in terms of $c$.
 Let $a_x$ be the largest $a$ such that $(d+a, d) \in S$.
 Let $a_y$ be the largest $a$ such that $(d, d+a) \in S$.
 Then $|U| \leq a_x a_y$.
 Furthermore, we have $|V_x| - |U| \geq d a_x$ and $|V_y| - |U| \geq da_y$.
 
 Thus
 \begin{align*}
  |U| \leq \frac{(|V_x| - |U|) (|V_y| - |U|)}{d^2}.
 \end{align*}
 By the AM-GM inequality, 
 \begin{align*}
  |U| \leq \frac{(c- |U|)^2}{d^2}.
 \end{align*}
 
 Simplifying,
 \begin{align*}
  |U|^2  - (2c + d^2) |U| + c^2 \geq 0.
 \end{align*}
 Solving, we get:
 \begin{align*}
  |U| &\leq c + \frac{ d^2}{2} - \sqrt{\left(c + \frac{d^2}{2}\right)^2 - c^2 }\\
  &= c + \frac{d^2}{2} - \sqrt{ \frac{d^2}{2}\left(2c + \frac{d^2}{2} \right)}\\
  &= c + \frac{d^2}{2} - d \sqrt{ c+ \frac{d^2}{4}}.
 \end{align*}

Thus 
\begin{align*}
|S| &\geq d^2 + 2c - \left( c+ \frac{d^2}{2} - d\sqrt{c + \frac{d^2}{4}}\right)\\
&= \frac{d^2}{2} + c + d \sqrt{c + \frac{d^2}{4}} \\
& = \left( \sqrt{ c + \frac{d^2}{4}} + \frac{d}{2} \right)^2\\
&\geq  \left( \sqrt{ \delta |S| + \frac{d^2}{4}} + \frac{d}{2} \right)^2,\\
\end{align*}
as claimed. Then, substituting $|S| = \frac{d^2}{2R}$, and dividing through by $d^2$, we get
$$
\frac{1}{2R} \geq \left(\sqrt{\frac{\delta}{2R} + \frac{1}{4}} + 1/2\right)^2.
$$
Expanding and multiplying through by $2R$, we get
$$
1 \geq \delta + R + \sqrt{2\delta R + R^2}.
$$
Removing the square root, we get 
$$
(1 - \delta - R)^2 - 2\delta R - R^2 \geq 0.
$$
Expanding and simplifying, we get
$$
\delta^2 - 2R - 2\delta  + 1 \geq 0.
$$
Finally, rearranging for $R$, we get
$$
R \leq (1 - \delta)^2/2,
$$
as claimed.
 
\end{proof}

Let $S$ be the grid of side length $\ell$. Recall from the discussion in \autoref{sec:d-robust} that $S$ has relative $d$-robustness $1 - d/\ell$. Then the RHS of the inequality is $\frac{d^2}{2\ell^2} = \frac{d^2}{2|S|} = R$. Hence, the grid is the smallest set with relative $d$-robustness $\delta$ for $\delta$ for $\delta > 1/2$. Next, we study the low-distance regime.

\subsection{Low robustness regime}

\begin{theorem}\label{thm:low-distance}
    Let $\delta > 0$, and $S \subset \N^2$ be such that $\pi_d(S) \geq \delta$.
    Let $N = |S|$, and $R = \frac{d^2}{2N}$. 
    Then we have 
    $$N - \frac{d^2}{2} \geq \sqrt{2\delta N} d,$$
    and
    $$R \leq 1 - 2\sqrt{\delta^2 + \delta} + 2 \delta$$
\end{theorem}

\begin{proof}
Let $U = \Simplex{2, d}$ be the two-dimensional simplex of side length $d$. 
Since $\pi_d(S) \geq \delta > 0$, $U \subset S$ (otherwise $S$ is contained within the first $d$ column or the first $d$ rows).
Let $V = \overline{U} \cap S$.
We will find several subsets $V_i$ of $V$ that are mutually disjoint
Taking the sum of sizes of these $V_i$ will yield our lower bound. 

For convenience, define $X_{< a} = \{(x, y) \in \N^2: x < a\}$. 
Similarly define $X_{\geq a}, Y_{<a}$.

Let $S_1 = V \cap H(0, d)$.
By hypothesis, $|S_1| \geq \delta N$.
Let $b_1$ be the smallest value for which $S_1 \subset X_{< b_1}$. I.e., $b_1$ is the number of columns in $S_1$.
Then, since $S$ is downward closed, we have that the set $X_{<b_1} \cap Y_{< d} \subset S$.
Define $T_1 = X_{<b_1} \cap Y_{< d} \cap \overline{U}$, and note that it is disjoint from $S_1$.
A simple counting argument shows that $|T_1| = \binom{b_1}{2}$.
We take $V_1 = S_1 \cup T_1$, and note that $|U_1| \geq \delta N + \binom{b_1}{2}.$

Now we consider the part of $V$ with $x$ coordinate at least $b_1$.

Let $S_2 = V\cap H(b_1, d - b_1)$
By hypothesis, we have $|S_2| \geq \delta N$.
Suppose $S_2$ has $b_1 + b_2$ columns, i.e. $b_2$ is the smallest number such that $S_2 \subset X_{<b_1 + b_2}$.
By the same logic, we have that 
$T_2 = X_{\geq b_1} \cap X_{< b_2}  \cap Y{< d - b_1} \cap \overline{U} $
is a subset of $V$ that is disjoint from $S_2$.
We take $V_2 = S_2 \cup T_2$, and note that $|V_2| \geq \delta N + \binom{b_2}{2}$
Also, note that $V_2$ is disjoint from $V_1$ since their projections on the $X$ axis are disjoint.

We repeat the process of taking
$$S_k = V\cap H(\sum_{i=1}^k b_i , d - \sum_{i=1}^{k-1} b_i ),$$ 
and 
$$T_k = X_{\geq \sum_{i=1}^{k-1} b_i} \cap X_{< \sum_{i=1}^k b_i} \cap Y_{< d - \sum_{i=1}^{k-1} b_i} \cap \overline{U},$$
until $\sum_{i=1}^k b_k \geq d$. 

Let $\hat{b_k} = d - \sum_{i=1}^{k-1}b_i$, it is not hard to see that $|T_k| \geq \binom{\hat{b_k}}{2}$. Replace $b_k$ with $\hat{b_k}$, so that $\sum_{i=1}^k b_i = d$, and $T_i \geq \binom{b_i}{2}$ for each $i \in [k]$. Thus, $U_i \geq \delta N + \binom{b_i}{2}$ for each $i \in [k]$. We now use these to bound the size of $S$. We have 

\begin{align*}
    N &\geq |U| + k \delta N + \sum_{i=1}^{k}\binom{b_i}{2}\\
    &= \frac{d(d-1)}{2} + k \delta N + \sum_{i=1}^{k} \frac{b_i(b_i - 1)}{2}\\
    &= \frac{d^2}{2} + k \delta N + \sum_{i=1}^{k} \frac{b_i^2}{2} - d\\
    &\geq \frac{d^2}{2} + k \delta N + \frac{d^2}{2k} - d\\
    &\geq \frac{d^2}{2} + d\sqrt{2 \delta N} - d.
\end{align*}
The third line follows from the fact that $\sum_{i=1}^k b_i = d$, the fourth line follows from the Cauchy-Schwartz inequality, and the fifth line follows from the AM-GM inequality. 
Substituting $N = \frac{d^2}{2R}$, and rearranging, we get that 

$$
1 \geq 2\sqrt{R\delta} + R - O(1/d)
$$
Ignoring the $O(1/d)$ term and solving for $R$, we get that $R \leq 1 - 2\sqrt{\delta^2 + \delta} + 2 \delta$.

\end{proof}

\begin{figure}
    \vspace{-5em}
    \begin{center}
        \includegraphics[width=0.85\textwidth]{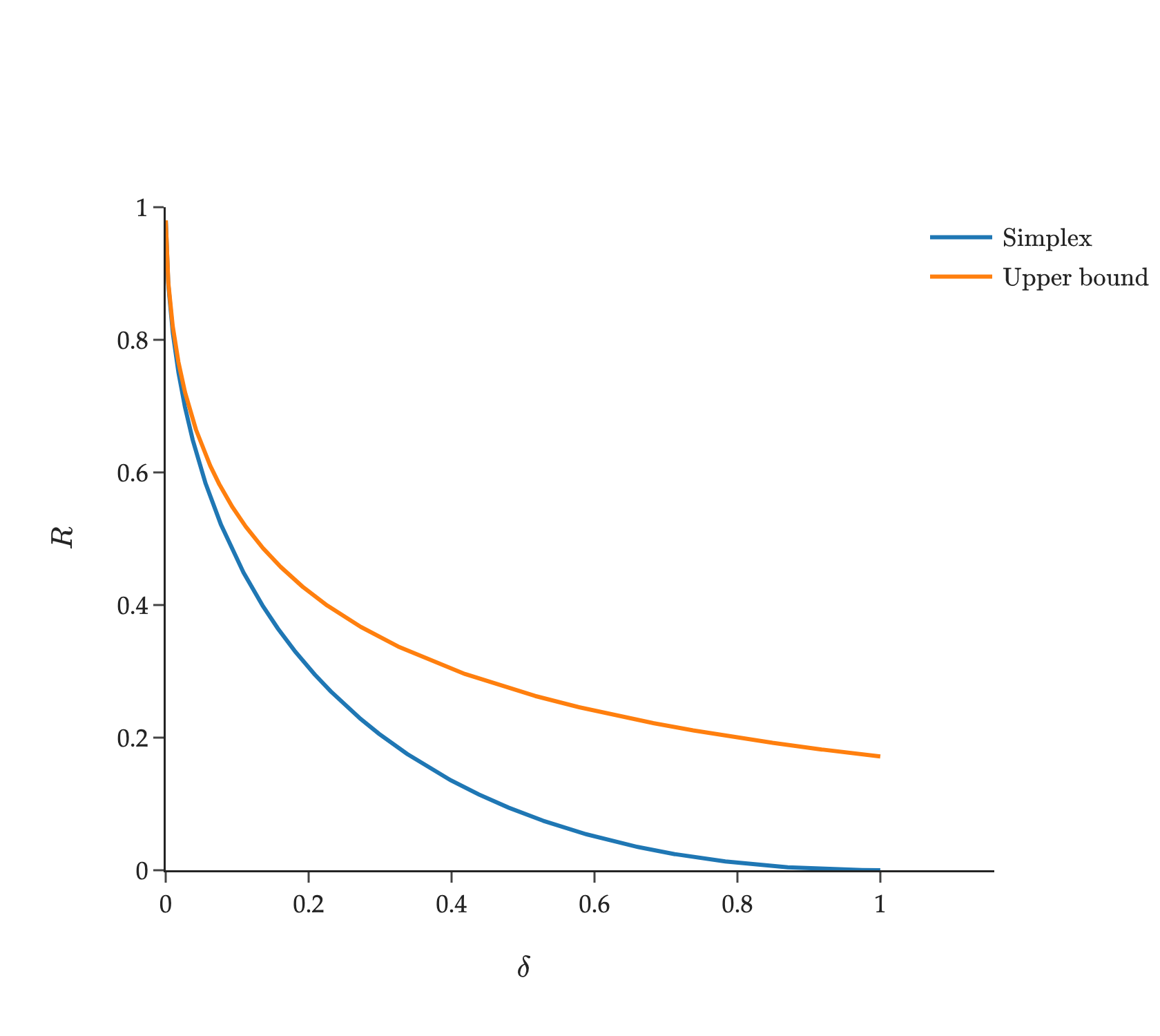}
    \end{center}
    \vspace{-3em}
    \caption{Comparing rate vs. robustness trade-off of the simplex construction to the upper bound}\label{fig:low-distance}
\end{figure}

Recall that the simplex evaluation set obtains a trade-off $R = (1 - \sqrt{\delta})^2 = 1 - 2\sqrt{\delta} + \delta$. 
Taking the Taylor expansion of the expression from \Cref{thm:low-distance}, get $R \leq 1 - 2\sqrt{\delta} + 2\delta + O(\delta^{3/2})$. 
Thus, the simplex has the optimal asymptotic behaviour as $\delta$ approaches $0$, namely, $R = 1 - 2\sqrt{\delta} + O(\delta)$. 
\Cref{fig:low-distance} compares the trade-off obtained by the simplex to the upper bound in \Cref{thm:low-distance}.

\section{Robust local characterization in $m$ dimensions using error-locator polynomials}

In this section we give a proof of a quantitatively weaker robust local characterization (which is completely subsumed by our results in~\autoref{sec:testing}), using an interpolated error-locator polynomial in the spirit of our \autoref{thm:robust-local-consistency-to-global-consistency-2d}. This uses our divisibility lemma for general $m$. We only include this because we find the proof method interesting.

The quantitatively weak aspect is that we cannot take $\frac{d}{t}$ close to $1$ in the statement, and thus it does not apply to high rate GAP codes\footnote{The source of this weakness is the factor $2$ in the divisibility lemma -- however this factor $2$ is tight for the divisibility lemma.}.

\begin{theorem}\label{thm:robust-local-consistency-to-global-consistency}
For all $m, d, t \in \N$ and $\epsilon \in [0,1)$ with $t \geq  12(d+m)$, $\epsilon < \frac{1}{100m^2}$, and for all large enough field $\F$, the following is true. 

Let $\mathcal{H} = \{ H_1, \ldots, H_t\}$ be a set of hyperplanes in general position in $\F^m$. 
Let $\mathcal{S} \subseteq \F^m$ denote the set of all $m$-wise intersection points of the hyperplanes in $\mathcal{H}$. 
For every hyperplane $H$ in $\mathcal{H}$, let $g_{H} : H \to \F$ be a degree $d$ polynomial on $(m-1)$ variables. 

If the polynomials $g_H$ and $g_{H'}$ agree with each other on $H \cap H'$ for at least $(1-\epsilon)\binom{t}{2}$ pairs $\{H, H'\} \subseteq \mathcal{H}$, then there is an $m$-variate degree $d$ polynomial $f : \mathcal{S} \to \F$ and a set $\tilde{\mathcal{H}} \subseteq \mathcal{H}$ of size at least $(1-2\epsilon)t$ such that for every $H \in \tilde{\mathcal{H}}$, the restriction of $f$ on $H$ equals $g_H$. 
\end{theorem}



\begin{proof}[Proof of \autoref{thm:robust-local-consistency-to-global-consistency}]
Let $\mathcal{U}$ be the set of all subspaces of co-dimension two obtained by taking pairwise intersections of distinct hyperplanes in $\mathcal{H}$, and $\mathcal{U}' \subseteq \mathcal{U}$ be the subset of co-dimension two subspaces $A$ such that if $A = H \cap H'$ for $H, H' \subseteq \mathcal{H}$, then $g_{H}$ and $g_{H'}$ agree on $A$, i.e. 
$g_H$ and $g_{H'}$ are consistent with each other. 
From the hypothesis of the theorem, we have that $|\mathcal{U}'| \geq (1-\epsilon) \cdot |\mathcal{U}| = (1-\epsilon) \cdot \binom{t}{2}$. \\

\noindent
\textbf{An error locator polynomial: } We show that there is a non-zero $m$-variate polynomial $E(\vecX)$ of degree at most $\sqrt{\frac{\epsilon}{2}} tm$ that vanishes identically on all subspaces in  the set $\mathcal{U}\setminus\mathcal{U}'$. 
This is again via an interpolation argument. 
We think of the coefficients of $E$ as formal variables of a polynomial of degree $\sqrt{\frac{\epsilon}{2}} tm$ on $m$ variables. 
Now, for any linear subspace $A$ of co-dimension two, we impose the constraint that the restriction of $E$ on $A$, denoted by $E^{A}$ vanishes identically.\footnote{Throughout this proof, we will abuse notation slightly and denote the restriction of polynomials such as $E$ on any subspace $A$ of co-dimension $2$ by $E^A$.} Note that up to an invertible change of basis, $E^A$ is a polynomial of degree at most $\deg(E)$ on $(m-2)$ variables. 
Thus, imposing the constraint that $E_A$ is identically zero corresponds to $\binom{\deg(E) + m-2}{m-2}$ homogeneous linear constraints on the coefficients of $E(\vecX)$. 
Thus, we have a system of $\left(\epsilon \cdot \binom{t}{2} \cdot \binom{\deg(E) + m-2}{m-2}\right)$ homogeneous linear constraints on $\binom{\deg(E) + m-2}{m}$ many variables. 
Thus, if 
\[
\binom{\deg(E) + m}{m} > \epsilon \binom{t}{2} \binom{\deg(E) + m-2}{m-2} \, ,
\]
The system has a non-zero solution. 
For $\deg(E) = \sqrt{\frac{\epsilon}{2}} tm$, it can be verified that this inequality holds. 
Let $E$ be one such non-zero polynomial, and  $e \leq \sqrt{\frac{\epsilon}{2}} tm$ denote its degree. \\

\noindent
\textbf{A global low degree polynomial: }For every $H \in \mathcal{H}$, let $P_{H}(\vecZ)$ be the polynomial 
\[
P_{H}(\vecZ) := g_{H}(\vecZ) \cdot E_{H}(\vecZ) \, ,
\]
where $E_{H}$ is the restriction of $E$ to the hyperplane $H$. 
Clearly, 
\[
\deg(P_{H}) = \deg(g_{H}) + \deg(E_{H}) \leq d + \deg(E_H) \, .
\]
We claim that for every pair of distinct hyperplanes $H, H' \in \mathcal{H}$, the polynomials $P_{H}$ and $P_{H'}$ are consistent with each other, i.e. 
they agree on the subspace $H\cap H'$. 
To argue this, we consider two cases based on whether or not $A = H \cap H’$ is in $\mathcal{U}’$. 
If $H \cap H' \in \mathcal{U}'$, then we have that $g_H$ and $g_{H'}$ are consistent, and hence 
\[
P_H^A = g_H^A \cdot E_H^A = g_{H'}^A \cdot E_{H'}^A  = P_{H'}^A \, .
\]
If $A \notin \mathcal{U}'$, then $E_H^A$ is identically  zero from the definition of $E$, and hence 
\[
P_H^A = P_{H'}^A \equiv 0 \, .
\]
Now, applying the local characterization theorem (\autoref{thm:local-characterization-mD}) to this collection $\{P_{H}: H \in \mathcal{H}\}$ of hyperplane polynomials of degree $d + e$, we get that if $t \geq d + e + m$ (which holds from the lower bound on $t$ in the hypothesis of the theorem), then there is a global $m$-variate polynomial $P$ of degree at most $(d+e)$ such that for every hyperplane $H \in \mathcal{H}$, the restriction of $P$ on $H$ equals $P_{H}$.  

This implies that for every $H \in \mathcal{H}$, the restriction $E_{H}$ of $E$ divides the restriction $P_{H}$ of $P$ on the $H$. 
Since $t > 2(d + e) + e \geq 2\max_{H}\{\deg(P)\} + \deg(E)$, from \autoref{lem:ps-style-lemma-md}, we get that the $m$-variate polynomial $E$ divides the $m$-variate polynomial $P$.


Let $f(\vecX)$ be the resulting quotient $P(\vecX)/E(\vecX)$. 
Clearly, $\deg(f)$ is equal to $\deg(P) - \deg(E)$, which is at most $(d + e) - e = d$. 
To prove the theorem, it now suffices to show that the restriction $f_{H}$ of $f$ on a hyperplane $H$ equals the hyperplane polynomial $g_{H}$ for many hyperplanes $H$ in $\mathcal{H}$. 
Once again, we first prove a weaker bound for this and then amplify it to complete the proof of the theorem. \\

\noindent
\textbf{Agreement with many hyperplane polynomials : }As in the proof of \autoref{thm:robust-local-consistency-to-global-consistency-2d}, we again notice that if $E_H$ is not identically zero for a hyperplane $H \in \mathcal{H}$, then $P_{H}/E_H$ is a polynomial and equals $P_{H}/E_H=g_H=f_H$. 
We now observe that there are at most $\deg(E)$ hyperplanes $H$ in $\mathcal{H}$ such that $E_H$ is identically zero. 
To see this, note that the polynomial $E_H$ can be obtained from the polynomial $E$ by substituting one of the variables of $E$ (w.l.o.g denoted by $X_m$ here) by an affine linear form $L$ in the variables $X_1, X_2, \ldots, X_{m-1}$. 
Thus, $E_H$ is identically zero if and only if $E$ is divisible by $(X_m - L)$. 
So, every hyperplane $H$ such that $E_H$ is identically zero yields a linear factor of $E$, and all these factors are distinct (since the hyperplanes are distinct and, in fact, in general position). 
Thus, $E$ is divisible by the product of all such linear factors, and hence the number of such factors, and therefore, such hyperplanes cannot exceed $\deg(E)$. 
Thus, $f_H$ equals $g_H$ for at least $(t - e)$ of the hyperplanes. 
Now, we amplify this bound further to complete the proof of the theorem.\\

\noindent
\textbf{An improved bound : } This final step also follows the outline of this step for the two-dimensional case. 
For $H \in \mathcal{H}$, let $\mathcal{I}_{H} \subseteq \mathcal{H}$ be the set of hyperplanes $H'$ such that $g_{H}^{H \cap H'} \neq g_{H'}^{H\cap H'}$. 
From the hypothesis of the theorem 
\[
\E_{H \in \mathcal{H}} \left[ |\mathcal{I}_{H}| \right] \leq \epsilon t \, .
\] 
Thus, by Markov's inequality, we get that 
\[
\Pr_{H \in \mathcal{H}} \left[ |\mathcal{I}_{H}| \geq ((1-\sqrt{\epsilon/2} m)t-d-m) \right] \leq \frac{\epsilon t}{(1-\sqrt{\epsilon/2}m)t-d-m} \leq 2\epsilon \, ,
\]
where the last inequality follows from the bounds on $t$ and $\epsilon$ in the theorem. 
Thus, we have that for $(1-2\epsilon)t$ of hyperplanes $H$, the set $\mathcal{I}_{H}$ is of size less than $((1-\sqrt{\epsilon/2}m)t-d-m)$. 
Any such $H$ is consistent with at least $d+m$ of the hyperplanes $H'$ such that  $E_{H'}$ does not vanish, and $f_{H'}$ equals $g_{H'}$, which in-turn equals $g_H$. 
Let $\mathcal{A}$ be this set and let $\mathcal{A}_H = \{H' \cap H : H' \in \mathcal{A}\}$. 
Now, if we focus on the $(m-1)$-dimensional linear space $H$, then $\mathcal{A}_H$ is a set of $(d+m)$ co-dimension one subspaces within $H$, which are also in general position. 
Moreover, $f_H$ and $g_H$ agree on each of these subspaces. 
Thus, they agree on the set of points corresponding to the $(m-1)$-wise intersection of subspaces in $\mathcal{A}_H$. 
Thus, \autoref{thm:geometric-hitting-set}, we get that $g_H$ and $f_H$ must be equal to each other. 

Therefore, for at least $(1-2\epsilon)t$ hyperplanes $H \in \mathcal{H}$, $f_H$ equals $g_H$. 
This completes the proof of the theorem. 
\end{proof}



\section{Intersection points of hyperplanes in general position are an interpolating set}
\label{appendix:GAPsets}

We give a quick proof of Bl\"{a}ser and Pandey's \Cref{thm:geometric-hitting-set}.

\begin{proof}
For each $x \in T$, we will show that there is a polynomial $P_x$ in $\F[X_1, \ldots, X_m]$ of degree at most $d$ which is nonzero at $x$ and zero at $T \setminus\{x\}$. This immediately implies that every function from $T$  to $\F$ can be interpolated by a polynomial of degree at most $d$ -- and since $|T|$ equals
the dimension of the space of all polynomials of degree at most $d$, we get that $T$ is an interpolating set. 

It remains to show the existence of $P_x$. 
Let $H_1, \ldots, H_m \in \mathcal{H}$ be the $m$ hyperplanes whose intersection is $x$. By the general position assumption about $\mathcal{H}$, we have that $x$ does not lie on any other hyperplane in $\mathcal{H}$.
Define
$$P_x(X_1, \ldots, X_m)  = \prod_{H \in \mathcal{H} \setminus\{H_1, \ldots, H_m\} } H(X_1, \ldots, X_m).$$
(Here, abusing notation, we used $H(X_1, \ldots, X_m)$ for a degree $1$ polynomial whose zero set equals the hyperplane $H$).
It is easy to see that $P_x$ does not vanish on $x$ and vanishes on other points of $T$, and that the degree of $P_x$ is at most $D$. This completes the proof.
\end{proof}

\end{document}